%% file: 00_main.tex
\documentclass[trackchanges]{aastex7}

\newcommand{\ieq}{{i_{\rm eq}}}
\newcommand{\jeq}{{j_{\rm eq}}}
\newcommand{\ihost}{{\cal H}}
\newcommand{\imas}{{\cal M}}
\newcommand{\ihms}{{\cal P}}
\newcommand{\igroup}{{\cal G}}
\newcommand{\ianchor}{{\cal A}}
\newcommand{\itype}{{\cal T}}
\newcommand{\icalib}{{\cal C}}
\newcommand{\iobj}{{\cal O}}

\usepackage[normalem]{ulem}
\usepackage{amsmath}
\usepackage{tikz}
\usepackage{amssymb}
\usepackage{adjustbox}
\usepackage{multirow}
\usepackage{blkarray}
\usepackage{mathtools}
\usetikzlibrary{matrix, positioning, calc} 
\definecolor{burntorange}{rgb}{0.8, 0.33, 0.0}
\newcommand{\Hunit}{\ensuremath{\,{\rm km\,s}^{-1}{\rm\,Mpc}^{-1}}}
\newcommand{\Hcst}{\ensuremath{H_0}}
\newcommand{\gaia}{\textit{Gaia}}
\input vardef.tex

\definecolor{newpurple}{rgb}{0.7,0,1}

\newcommand{\DN}{Distance Network}

\begin{document}

\title{The Local Distance Network: a community  consensus report on\\the measurement of the Hubble constant at $\sim$1\% precision}

\collaboration{all}{H0DN Collaboration}

\author[sname='Casertano']{Stefano Casertano}
\affiliation{Space Telescope Science Institute, 3700 San Martin Drive, Baltimore, MD 21218, USA}
\email{stefano@stsci.edu} 

\author[orcid=0000-0002-5259-2314,sname='Anand']{Gagandeep Anand}
\affiliation{Space Telescope Science Institute, 3700 San Martin Drive, Baltimore, MD 21218, USA}
\email{ganand@stsci.edu} 

\author[orcid=0000-0001-8089-4419,sname='Anderson']{Richard I. Anderson}
\affiliation{Institute of Physics, \'Ecole Polytechnique F\'ed\'erale de Lausanne (EPFL), Observatoire de Sauverny, 1290 Versoix, Switzerland}
\email{richard.anderson@epfl.ch} 

\author[orcid=0000-0002-1691-8217,sname='Beaton']{Rachael Beaton}
\affiliation{Space Telescope Science Institute, 3700 San Martin Drive, Baltimore, MD 21218, USA}
\email{rbeaton@stsci.edu} 

\author[orcid=0000-0001-6147-3360,sname='Bhardwaj']{Anupam Bhardwaj}
\affiliation{Inter-University Centre for Astronomy and Astrophysics (IUCAA), Post Bag 4, Ganeshkhind, Pune 411 007, India}
\email{anupam.bhardwaj@iucaa.in} 

\author[orcid=0000-0002-5213-3548,sname='Blakeslee']{John P.~Blakeslee}
\affiliation{NSF NOIRLab, 950 N Cherry Ave, Tucson AZ 85719}
\email{john.blakeslee@noirlab.edu} 

\author[orcid=0009-0003-7937-9526,sname='Boubel']{Paula Boubel}
\affiliation{Research School of Astronomy \& Astrophysics, Australian National University, Cotter Road, Weston Creek ACT 2611, Australia}
\email{paula.boubel@anu.edu.au} 

\author[orcid=0000-0003-3889-7709,sname='Breuval']{Louise Breuval}
\affiliation{European Space Agency (ESA), ESA Office, Space Telescope Science Institute, 3700 San Martin Drive, Baltimore, MD 21218, USA}
\email{lbreuval@stsci.edu} 

\author[orcid=0000-0001-5201-8374,sname='Brout']{Dillon Brout}
\affiliation{Boston University Departments of Astronomy and Physics, 725 Commonwealth Ave, Boston USA}
\email{dbrout@bu.edu} 

\author[orcid=0000-0003-2072-384X,sname='Cantiello']{Michele Cantiello}
\affiliation{INAF -- Astronomical Observatory of Abruzzo, Via Maggini, 64100 Teramo, Italy}
\email{michele.cantiello@inaf.it} 

\author[orcid=0000-0003-2443-173X,sname='Cruz Reyes']{Mauricio Cruz Reyes}
\affiliation{Institute of Physics, \'Ecole Polytechnique F\'ed\'erale de Lausanne (EPFL), Observatoire de Sauverny, 1290 Versoix, Switzerland}
\email{mauricio.cruzreyes@epfl.ch } 

\author[orcid=0000-0001-9210-9860,sname='Cs\"ornyei']{Geza Cs\"ornyei}
\affiliation{Max-Planck-Institute for Astrophysics, Karl-Schwarzschild-Str.~1, 85741 Garching, Germany}
\email{csogeza@mpa-garching.mpg.de} 

\author[orcid=0000-0001-6069-1139,sname='de Jaeger']{Thomas de Jaeger}
\affiliation{Sorbonne Université, CNRS, Laboratoire de Physique Nucléaire et de Hautes Energies, 75252 Paris, France}
\email{thomas.dejaeger@lpnhe.in2p3.fr} 

\author[orcid=0000-0002-2376-6979,sname='Dhawan']{Suhail Dhawan}
\affiliation{School of Physics and Astronomy, University of Birmingham, Edgbaston, Birmingham B15 2TT, UK}
\email{s.dhawan@bham.ac.uk} 

\author[orcid=0000-0001-8408-6961,sname='Di Valentino']{Eleonora Di Valentino}
\affiliation{School of Mathematical and Physical Sciences, University of Sheffield, Hounsfield Road, Sheffield S3 7RH, United Kingdom}
\email{e.divalentino@sheffield.ac.uk} 

\author[orcid=0000-0002-1296-6887,sname='Galbany']{Llu\'is Galbany}
\affiliation{Institute of Space Sciences (ICE-CSIC), Campus UAB, Carrer de Can Magrans, s/n, E-08193 Barcelona, Spain.}
\affiliation{Institut d'Estudis Espacials de Catalunya (IEEC), 08860 Castelldefels (Barcelona), Spain}
\email{l.g@csic.es} 

\author[orcid=0000-0003-0265-6217,sname='Gil-Mar\'in']{Héctor Gil-Mar\'in}
\affiliation{Institut de Ciències del Cosmos (ICCUB), Universitat de Barcelona (UB), c. Martí i Franquès, 1, 08028 Barcelona, Spain}
\affiliation{Departament de Física Quàntica i Astrofísica, Universitat de Barcelona, Martí i Franquès 1, E08028 Barcelona, Spain}
\affiliation{Institut d'Estudis Espacials de Catalunya (IEEC), 08860 Castelldefels (Barcelona), Spain}
\email{hectorgil@icc.ub.edu} 

\author[orcid=0000-0002-7355-9775,sname='Graczyk']{Dariusz Graczyk}
\affiliation{Polish Academy of Sciences, Nicolaus Copernicus Astronomical Center, Department of Astrophysics,\\ul.~Rabiańska 8, 87-100 Toruń, Poland}
\email{darek@ncac.torun.pl} 

\author[orcid=0000-0001-6169-8586,sname='Huang']{Caroline Huang}
\affiliation{Center for Astrophysics $|$ Harvard \& Smithsonian, 60 Garden Street, Cambridge, MA 02138, USA}
\email{caroline.huang@cfa.harvard.edu} 

\author[orcid=0000-0001-8762-8906,sname='Jensen']{Joseph B.~Jensen}
\affiliation{Department of Physics, Utah Valley University, 800 West University Parkway, Orem, UT 84058, USA }
\email{joseph.jensen@uvu.edu} 

\author[orcid=0000-0003-0626-1749,sname='Kervella']{Pierre Kervella}
\affiliation{LIRA, Observatoire de Paris, Universit\'e PSL, Sorbonne Universit\'e, Université Paris Cit\'e, CY Cergy Paris Universit\'e,\\CNRS, 92190 Meudon, France}
\affiliation{French-Chilean Laboratory for Astronomy, IRL 3386, CNRS and U.~de Chile, Casilla 36-D, Santiago, Chile}
\email{pierre.kervella@obspm.fr} 

\author[orcid=0000-0002-4413-7733,sname='Leibundgut']{Bruno Leibundgut}
\affiliation{European Southern Observatory, Karl-Schwarzschild-Strasse 2, 85748 Garching, Germany}
\email{bleibund@eso.org} 

\author[orcid=0009-0007-8211-8262,sname='Lengen']{Bastian Lengen}
\affiliation{Institute of Physics, \'Ecole Polytechnique F\'ed\'erale de Lausanne (EPFL), Observatoire de Sauverny, 1290 Versoix, Switzerland}
\email{bastian.lengen@epfl.ch} 

\author[orcid=0000-0002-8623-1082,sname='Li']{Siyang Li}
\affiliation{Department of Physics \& Astronomy, Johns Hopkins University, Baltimore, MD 21218}
\affiliation{Department of Astronomy, University of California, Berkeley}
\email{seanli@berkeley.edu} 

\author[orcid=0000-0002-1775-4859,sname='Macri']{Lucas Macri}
\affiliation {Department of Physics \& Astronomy, College of Sciences, University of Texas Rio Grande Valley,\\1201 W University Blvd, Edinburg TX 78539}
\email{lucas.macri@utrgv.edu}

\author[orcid=0000-0003-0817-4219,sname='\"{O}z\"{u}lker']{Emre \"{O}z\"{u}lker}
\affiliation{School of Mathematical and Physical Sciences, University of Sheffield, Hounsfield Road, Sheffield S3 7RH, United Kingdom}
\email{e.ozulker@sheffield.ac.uk} 

\author[orcid=0000-0002-5278-9221,sname='Pesce']{Dominic W. Pesce}
\affiliation{Center for Astrophysics $|$ Harvard \& Smithsonian, 60 Garden Street, Cambridge, MA 02138, USA}
\affiliation{Black Hole Initiative at Harvard University, 20 Garden Street, Cambridge, MA 02138, USA}
\email{dpesce@cfa.harvard.edu}

\author[orcid=0000-0002-6124-1196,sname='Riess']{Adam Riess}
\affiliation{Space Telescope Science Institute, 3700 San Martin Drive, Baltimore, MD 21218, USA}
\affiliation{Department of Physics \& Astronomy, Johns Hopkins University, Baltimore, MD 21218}
\email{ariess@stsci.edu}

\author[orcid=0000-0002-5527-6317,sname='Romaniello']{Martino Romaniello}
\affiliation{European Southern Observatory, Karl-Schwarzschild-Strasse 2, 85748 Garching, Germany}
\email{mromanie@eso.org}

\author[orcid=0000-0002-1809-6325,sname='Said']{Khaled Said}
\affiliation{School of Mathematics and Physics, University of Queensland, Brisbane, QLD 4072, Australia}
\email{k.saidahmedsoliman@uq.edu.au}

\author[orcid=0000-0002-7873-0404,sname='Sch\"oneberg']{Nils Sch\"oneberg}
\affiliation{University Observatory, Faculty of Physics, Ludwig-Maximilians-Universität, Scheinerstr.~1, 81677 Munich, Germany}
\affiliation{Excellence Cluster ORIGINS, Boltzmannstr.~2, 85748 Garching, Germany}
\email{nils.science@gmail.com}

\author[orcid=0000-0002-4934-5849,sname='Scolnic']{Dan Scolnic}
\affiliation{Department of Physics, Duke University, Durham, NC 27708, USA}
\email{daniel.scolnic@duke.edu}

\author[orcid=0009-0004-7523-0799,sname='Sicignano']{Teresa Sicignano}
\affiliation{European Southern Observatory, Karl-Schwarzschild-Strasse 2, 85748 Garching, Germany}
\affiliation{ Scuola Superiore Meridionale, Largo S. Marcellino 10, 80138 Napoli, Italy}
\affiliation{ INAF-Osservatorio Astronomico di Capodimonte, Salita Moiariello 16, 80131 Napoli, Italy}
\email{Teresa.Sicignano@eso.org}

\author[orcid=0000-0001-9439-604X,sname='Skowron']{Dorota M. Skowron}
\affiliation{Astronomical Observatory, University of Warsaw, Al.~Ujazdowskie 4, 00-478 Warszawa, Poland}
\email{dszczyg@astrouw.edu.pl}

\author[orcid=0000-0002-9413-4186,sname='Uddin']{Syed A.~Uddin}
\affiliation{ American Public University System, 111 W. Congress St., Charles Town, WV 25414, USA}
\affiliation{Center for Astronomy, Space Science and Astrophysics, Independent University, Bangladesh, Dhaka 1245, Bangladesh}
\email{saushuvo@gmail.com}

\author[orcid=0000-0003-2601-8770,sname='Verde']{Licia Verde}
\affiliation{Institució Catalana de Recerca i Estudis Avançats, Passeig de Lluís Companys, 23, 08010
Barcelona, Spain}
\affiliation{Institut de Ciències del Cosmos (ICCUB), Universitat de Barcelona (UB), c. Martí i Franquès, 1, 08028 Barcelona, Spain}
\email{liciaverde@icc.ub.edu}

\author[orcid=0009-0007-8087-6975,sname='Nota']{Antonella Nota}
\affiliation{International Space Science Institute, Hallerstrasse 6, 3012 Bern, Switzerland}
\email{antonella.nota@issibern.ch} 

\null\vspace{20pt}

\begin{abstract}

\ \par
The direct, empirical determination of the local value of the Hubble constant (\Hcst) has markedly advanced thanks to improved instrumentation, measurement techniques, and distance estimators. However, combining determinations from different estimators is non-trivial, due to their correlated calibrations and different analysis methodologies. Using covariance weighting and leveraging community expertise, we constructed a rigorous and transparent \textit{Distance Network} to find a consensus value and uncertainty for the locally-measured Hubble constant. A broad and comprehensive list of experts across all relevant distance measurement domains were invited to critically review the available data sets, spanning parallaxes, detached eclipsing binaries (DEB), masers, Cepheids, the Tip of the Red Giant Branch (TRGB), Miras, carbon-rich AGB stars (JAGB), Type Ia supernovae (SNe~Ia), Surface Brightness Fluctuations (SBF), Type II supernovae (SNe~II), the Fundamental Plane, and Tully-Fisher relations. Before any calculations, the group voted for first-rank indicators to define a `baseline' \DN. Other indicators were included to assess the robustness and sensitivity of the results.
We provide open-source software and data products to support full transparency and future extensions of this effort. Our key conclusions are:
1) The local $H_0$ is robustly determined, with first-rank indicators internally consistent within their uncertainties;
2) A covariance-weighted combination yields a relative uncertainty of 1.1\% (baseline) or 0.9\% (all estimators);
3) The contribution from SNe~Ia is consistent across four current compilations of optical magnitudes or using NIR-only magnitudes;
4) Removing either Cepheids or TRGB has minimal effect on the central value of \Hcst;
5) Replacing SNe~Ia with galaxy-based indicators changes {\Hcst} by less than 0.1 {\Hunit}, while doubling its uncertainty; % \nilsc{Number missing};
6) The baseline result is $H_0$=$\Hvaluebase \pm \Herrorbase \Hunit $.  Compared to early Universe+$\Lambda$CDM results, our result differs by $\Hsigmabase\sigma$ from the CMB anisotropies within flat $\Lambda$CDM, $ \Hvaluecmb \pm \Herrorcmb \Hunit $ from Planck+SPT+ACT \citep[Eq.~(54)~of][]{2025arXiv250620707C} and $5.0\sigma$ from BBN+BAO within flat $\Lambda$CDM from DESI DR2 \citep[$68.51 \pm 0.58 \Hunit$; Tab.~V~of][]{DESI:2025zgx}. 
A networked approach, such as presented here, is invaluable for enabling further progress in Hubble constant measurements, providing much needed advances in accuracy and precision without overreliance on any single method, sample or group.

\end{abstract}

\keywords{\uat{Hubble Constant}{758} -- \uat{Cosmology}{343} -- \uat{Distance measure}{395}}

\section{Introduction} \label{sec:intro}
\input{01_intro}

\section{Methodology}  \label{sec:meth}
\input{02_method}

\section{Datasets description} \label{sec:data_summary}
\input{03_data_summary}

\section  {Baseline Results} \label{sec:results}
\input{04_results}

\section {Variants} \label{sec:variants}
\input{05_variants}

\section{Conclusions and Discussion} \label{sec:conc}
\input{06_discussion_conclusions}

\input{08_acknowledgments}

\bibliography{bibfile}{}
\bibliographystyle{aasjournal}

\appendix

\input{Appendices/appendix}

\end{document}

%% file: vardef.tex
\newcommand{\Hvaluecmb}{67.24}
\newcommand{\Herrorcmb}{0.35}
\newcommand{\Hvaluebase}{73.50}
\newcommand{\Herrorbase}{0.81}
\newcommand{\Hsigmabase}{7.1}

\newcommand{\redchisqbase}{0.9879}

\newcommand{\Hrelerrorbase}{$1.09\%$}
\newcommand{\Herrorepm}{0.74}

\newcommand{\redchisqtf}{1.4712}

\newcommand{\Hvaluenoceph}{72.51}
\newcommand{\Herrornoceph}{1.30}

\newcommand{\Herrorcustbasenoceph}{0.87}

\newcommand{\Hvaluenoftfe}{73.08}

\newcommand{\Herrornosnia}{1.79}

\newcommand{\Hvaluenohst}{73.75}
\newcommand{\Herrornohst}{1.33}

\newcommand{\Hvaluenojwst}{73.00}
\newcommand{\Herrornojwst}{0.86}

\newcommand{\Hvalueusecmb}{73.06}
\newcommand{\Herrorusecmb}{0.80}

\newcommand{\Hvaluehfnosnia}{75.29}
\newcommand{\Herrorhfnosnia}{0.93}
\newcommand{\Hsigmahfnosnia}{8.1}

\newcommand{\Hvalueeverything}{73.99}
\newcommand{\Herroreverything}{0.70}
\newcommand{\Hsigmaeverything}{8.7}

\newcommand{\redchisqeverything}{1.3078}

\newcommand{\Hvalueeverythingnotf}{73.66}
\newcommand{\Herroreverythingnotf}{0.71}
\newcommand{\Hsigmaeverythingnotf}{8.1}

\newcommand{\redchisqeverythingnotf}{0.8943}

%% file: 01_intro.tex
The current expansion rate of the Universe, quantified by the Hubble constant ($H_0$), is a cornerstone of modern cosmology \citep[see reviews by][and references therein]{1992PASP..104..599J,2010ARA&A..48..673F}. Over the past decade, increasingly precise measurements of $H_0$ have revealed a striking and persistent discrepancy between its value inferred from observations of the early Universe, such as the cosmic microwave background (CMB), and its value measured directly in the local Universe using distance ladder methods. This disagreement, known as the ``Hubble Tension'', has persisted for a decade, exceeds the threshold for a statistical fluctuation, and has withstood extensive scrutiny of both observational data and analysis techniques \citep[see][for recent reviews]{2019NatAs...3..891V, 2021A&ARv..29....9S,2023ARNPS..73..153K,Verde2024}. 
As such, it poses a major challenge to the standard $\Lambda$CDM cosmological model and may point to new physics \citep[see][]{2021CQGra..38o3001D,CosmoVerse:2025txj}, barring the increasingly observationally disfavored possibility of multiple, independent, and unrecognized systematics. While many distance indicators have been used to measure the local value of {\Hcst}, few of these studies attempt to optimally combine these measures, which would require properly accounting for their correlations. In addition, correlations and/or redundant information offer a key advantage to ensure robustness.  Rather than a single distance ladder, or several parallel, partially correlated ladders, we show below that these measurements constitute a (stable) network: the Local Distance Network.  The study presented here is a comprehensive, community-wide effort in 2025 to construct this Local Distance Network via a broad, collaborative effort.

\subsection{The ISSI Bern Workshop setup and philosophy\label{sec:philosophy}}

Building a Local Distance Network requires expert knowledge across diverse astronomical disciplines. There is a wide range of distance indicators with varying levels of maturity, confidence, and uncertainty, necessitating careful consideration before employing them jointly. The various subsets of the astronomical community working on aspects related to the Hubble Tension have interacted at different junctures and have a general understanding of each other's methodologies. However, accurately and reliably combining results while considering all inter-dependencies requires a hands-on collaborative approach and a careful and thorough treatment, rooted in transparency, engagement, and scientific discourse. 

\begin{figure*}
    \centering
    \includegraphics[]{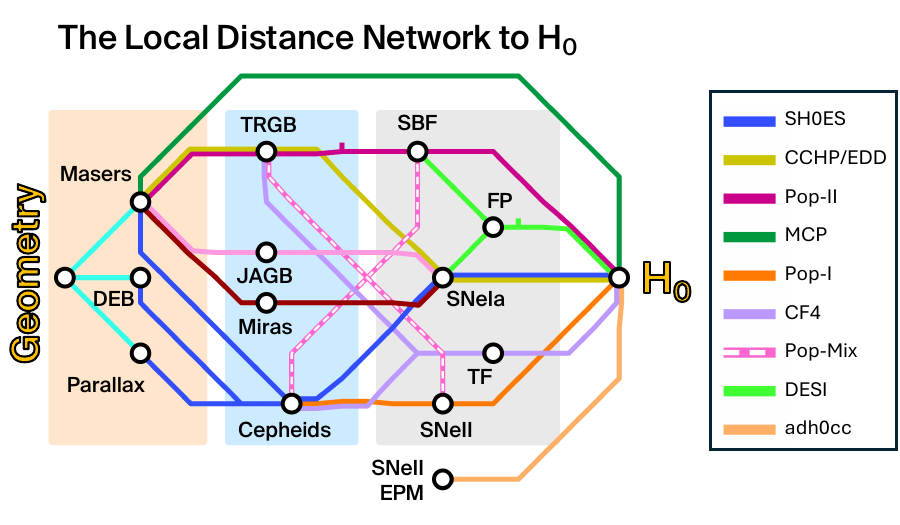}
    \caption{Conceptual overview of the Local Distance Network, a many routes approach. Different methods for distance determination may connect the absolute scale determined by geometric means to \Hcst. A non-exhaustive list of baseline linkages discussed in the literature or the paper is labeled on the right. Links to geometric distances provided by Masers, DEB, and Parallax are indicated as available in our analysis. Background rectangles in orange, light blue, and gray indicate where Rung 1, Rung 2, and Rung 3 of a traditional distance ladder would fall. Unlabeled tickmarks represent Groups (Fornax \& Virgo for the TRGB to SBF, Coma for FP). Example references: 
    \citet[SH0ES]{2022ApJ...934L...7R}, \citet[CCHP]{2025ApJ...985..203F}, \citet[EDD]{2021AJ....162...80A}, \citet[Pop-II]{2024ApJ...973...83A}, \citet[MCP]{2020ApJ...890..118P}, \citet[Pop-I]{2022MNRAS.514.4620D}, \citet[CF4]{2020ApJ...902..145K}, \citet[DESI]{2025MNRAS.539.3627S}, \citet[adh0cc]{2025AA...702A..41V}. Appendix\,\ref{app:replication} replicates a subset of these routes.}
    \label{fig:Metro}
\end{figure*}

This was the underlying motivation and raison d'\^etre for the workshop\footnote{\url{https://workshops.issibern.ch/hubble-constant/}} ``What's under the H0od?'' held at the International Space Sciences Institute (ISSI) in Bern, Switzerland, in March 2025. The goal of this workshop was to arrive at a consensus set of `baseline' and `variant' datasets to include, to define statistically rigorous analysis procedures that account for dataset covariance, and to begin developing the open access tools required to measure {\Hcst} within a networked formalism. The Local Distance Network (Fig.~\ref{fig:Metro}) extends the distance ladder concept ``horizontally'' by linking multiple, overlapping calibration paths. It combines the statistical advantages of consistently averaging the contributions from multiple probes with the robustness to allow for the omission of any single probe. This is possible because there are multiple indicators that can serve the same methodological role (e.g., different anchors, different intermediate calibrators, different tracers of the Hubble flow, cf. Sect.\,\ref{sec:nomenclature}), as well as some parts of the distance network that require fewer connections or steps (megamaser distances, type II supernova modeling).  This ambitious program required assembling the leading experts in each of the relevant reserach areas.
The workshop conveners sought to leverage worldwide expertise across multiple tools as much as possible in selecting the $\sim$40 in-person attendees invited to participate in the ISSI workshop (see Section~\ref{expertise} for a list of attendees and their respective areas of expertise). Attention was given to inviting representatives from the most active groups in the field, including competing groups using similar or identical methods, in order to build consensus on how to consistently incorporate methods within the network and evaluate the level of agreement, especially in cases considered contentious in the literature.\footnote{Though the overall acceptance rate was high ($>$ 90\%), several invitees were unable to attend or declined participation in the workshop. Following the inclusive spirit of the workshop, opportunities for remote participation via Zoom and asynchronous participation by e-mail were offered in such cases. In the case where a group declined to participate in any format, their published data were still included to preserve their scientific contributions.} The group thus assembled (i.e., we, the H${}_{0}$DN collaboration) placed emphasis on the methodology of combining datasets (i.e., the ``how-to''), rather than on the results themselves, which were understood to be subject to quality and consistency checks (e.g., $\chi^2$ or residuals) and to be published irrespective of the resulting value of \Hcst.

%% file: 02_method.tex
\subsection{The Local Distance Network}\label{sec:dn_definition}

The goal of measuring accurate and precise distances well into the Hubble flow 
($D \gtrsim 100 $ Mpc) to directly and empirically determine {\Hcst} requires the overlapping use of multiple techniques, a combination traditionally referred to as a ``distance ladder.''  The primary need for this approach is that geometric distance measurements have been limited in range ($D \lesssim 10$~Mpc, and most often $D \lesssim 100$~kpc), while long-range indicators, such as type-Ia supernovae (SNe~Ia), are too rare to provide enough examples within reach of geometric distances. 
Historically, measuring {\Hcst} thus required a rather large number of methods and combinations of inhomogeneous data sets \citep[e.g.,][]{2001ApJ...553...47F,2006ApJ...653..843S}. Significant improvements in precision and accuracy have since been achieved by streamlining distance ladder setups, focusing on high-quality and maximally homogeneous data sets while maintaining tight control of systematic errors by using the same instrument across rungs to cancel flux calibration offsets.
An example is the three-rung ``SH0ES'' distance ladder \citet{2009ApJ...699..539R}, which calibrates classical Cepheids (henceforth: Cepheids) as standard candles using distances measured by geometrical methods (i.e., higher precision MW parallaxes, detached eclipsing binaries in the Magellanic Clouds,the megamaser distance to NGC 4258), and, in turn, calibrates SNe~Ia luminosity using distances to their host galaxies measured using Cepheids observed exclusively using the {Hubble} Space Telescope (HST). Finally, SNe~Ia luminosity distances in the Hubble flow and their redshifts determine {\Hcst} \citep[see Fig.~10 in][]{2016ApJ...826...56R}. Such streamlined approaches provided significant benefits in terms of robustness and precision, in part because they avoided potentially correlated systematics, ultimately leading to the Hubble tension as discussed today.

In recognition that the tension may be indicating something profound, greater reliability has been sought through an increasing number of different methods, sources, and measurements. These provide multiple, interrelated constraints for the same or different astrophysical sources, resulting in partially independent paths to {\Hcst}. Many of these approaches replace certain steps with alternative methods, sources, or calibrators. For this reason, we believe it is appropriate to consider these tools in aggregate to comprise a ``{\DN},'' illustrated schematically in Fig.~\ref{fig:Metro}, to better convey the interdependence of these methods. While such a goal was considered ``lofty'' and potentially unreachable more than a decade ago \citep[cf.][and Fig.\,1 therein, originally credited to \citealt{2006pnbm.conf...79C}]{2013IAUS..289..351D}, the improvements to the systematics of several distance measurement techniques\hbox{---}inspired not least by the Hubble tension\hbox{---}now provide a wealth of robust information that allow to again ``diversify'' the base on which the local measurement of \Hcst{} rests.

The {\DN} provides two critical advantages on the path to a more accurate measurement of {\Hcst}: robustness (to reduce systematic errors) and statistical advantage (to reduce statistical uncertainties). Systematic errors can be recognized by the redundancy of methods allowing for analyses that ``leave one out.''  At the same time, redundancy offers the means to reduce statistical fluctuations through covariance-weighted averaging.  Informally, method combination and robustness has been evaluated through the display of ``whisker diagrams" which separate measurements of {\Hcst} by the combination of techniques they employ.  Intermediate measure comparisons have also been made, most importantly by comparing multiple ways to measure distances to specific SN Ia hosts, each calibrated by the same geometric source (e.g., Cepheids, TRGB, Miras, and JAGB), though these involve only a fraction of the available data \citep{2024ApJ...977..120R}. Formal covariance-weighting and combining has been attempted for only a limited set of indicators \citep{2022ApJ...934L...7R}. 

In this paper, we introduce an approach to the {\DN} that obviates these shortcomings.  By combining existing measurements \textit{at the component level}, rather than in terms of the resulting value of \Hcst, into a common, statistically rigorous framework encompassing a broad range of methods,  this approach yields a combined value of {\Hcst} with an uncertainty that reflects all available information.  Within this framework, we will also be able to include or exclude different subsets of measurements, thus identifying possible outliers. We will be able to inspect residuals at different levels, verifying whether they are consistent with their stated accuracies.

\subsubsection{Nomenclature and definitions\label{sec:nomenclature}}

Given the intricate interrelations between different methods and measurements, and the complexity of the resulting framework, we define at the outset a set of terms, following past usage as closely as possible, that we will use to formulate our approach. The principles underlying the various methods alongside the datasets used are presented in Sect.\,\ref{sec:data_summary} and App.\,\ref{sec:data_appendix}.

\begin {description}
\item[Anchor] Any object, or collection of objects, whose distance is directly determined by geometric means, such as parallax or measurements of orbiting systems, and is used to calibrate the distances of other indicators. Anchors set the absolute scale of the \DN{}; all other distances, with a few exceptions (Section~\ref{sec:data_summary}), are measured {\it relative} to this scale.  Anchors used in this analysis include NGC 4258 (through Keplerian motion of circumnuclear masers), the Magellanic Clouds (through detached eclipsing binaries, DEBs), and the collection of Milky Way Cepheids (through trigonometric parallaxes).  The Milky Way uniquely provides distances to individual objects rather than a single extragalactic system, mostly measured by the ESA {\it Gaia} mission, which are then combined into a single calibration of the Leavitt Law for Galactic Cepheids.  To a lesser extent, depth effects are also present in the Large Magellanic Cloud (LMC) and Small Magellanic Cloud (SMC), but may be corrected through the use of empirical geometric modeling fit to the collection of DEBs. 
In the traditional distance ladder, these anchors of geometric measures are often referred as the first rung.

\item [Primary Distance Indicator]  An astronomical feature that can be measured or calibrated {\it directly} using the aforementioned geometric means.  Examples include the luminosity of the intercepts of the Leavitt Law of Cepheids or oxygen-rich Mira variables, as well as the luminosities of the Tip of the Red Giant Branch (TRGB) or the J-region of the Asymptotic Giant Branch (JAGB).

\item[Host] An object, typically a galaxy, whose distance can be estimated from its properties via a Primary Distance Indicator.  Relevant hosts include one or more Secondary Distance Indicators and by means of their identical distance, enable absolute calibration of the secondary indicator (i.e., converting relative distance to true distance).  In the distance ladder, these are often referred to as the {second rung}.

\item [Secondary Distance Indicator] An astronomical feature that can be measured in more distant systems (``hosts'') and ideally reach out to Hubble flow systems.  Secondary distance indicators used here include the luminosity of Type Ia and Type II supernovae (SNe~Ia, SNe~II), based on the measurement of objects in nearby hosts; the Tully-Fisher (TF) relation, which relates a spiral galaxy's luminosity to its velocity width; the standardized luminosity of Surface Brightness Fluctuations (SBF); and the Fundamental Plane (FP) of elliptical galaxies, calibrated using its properties in the Coma cluster. 

\item [Calibrator] An astronomical object used in the calibration of a Secondary Distance Indicator, such as SNe~Ia, SNe~II, and galaxies with luminosity estimated via TF or SBF relations.  A Calibrator is in a host (or \textit{is} a host, when the Secondary Distance Indicator is based on the whole galaxy), and its distance is constrained by the host distance.

\item [Group] A grouping (group or cluster) of galaxies that are assumed to be at a common distance, with appropriate dispersion due to depth effects.  Group membership is used to obtain a distance estimate for some Calibrators, including SBF calibrators in Virgo and Fornax and FP calibrators in Coma.

\item [Direct Distance Determination] A method that yields the distance to an astronomical object directly, without relying on intermediate calibration steps.  Such methods are used to determine the distance of anchors, via trigonometric parallaxes (in the Milky Way), the orbital signatures of red giant stars in DEB (LMC, SMC), or the relation between line-of-sight and angular velocity fields in a system of masers (NGC~4258). However, direct distance determinations can sometimes be applied to astronomical objects in the Hubble Flow (see next item) and thus provide a direct constraint on {\Hcst}. Examples include maser systems similar to that in NGC 4258, which yield an angular diameter distance by modeling their recession and angular velocity fields, and SNe~II calibrated via the Expanding Photosphere Method (EPM), which determines distance by comparing angular size measured from surface brightness and color temperature to physical size determined from the evolution of the velocity profile. 

\item [Hubble Flow System (also called Tracer)] An object at sufficiently large distance such that its cosmological redshift can be determined with good accuracy from its measured velocity; together with an angular or luminosity distance determination, such objects provide constraints on the value of $ H_0 $. Tracers used here include SNe~Ia, SNe~II, galaxies with luminosity distances estimated from TF or SBF relations, megamasers, and elliptical galaxies via FP.  The angular or luminosity distance can be determined using a Primary or Secondary Distance Indicator, or through a direct distance determination. Consistent with most determinations of the local value of {\Hcst}, we limit our analysis to redshifts large enough to reduce the impact of correlated flows due to large-scale structure, generally $z>0.01$ or $z>0.023$ and small enough ($ z \lesssim 0.15 $) that a simple, kinematic form of the redshift-distance relation suffices.  In practice, the effective redshift range for Hubble flow systems follows the relevant literature and depends on the tracer; tracers that require resolved galaxy images, such as SBF, typically occupy a lower redshift range than SNe~Ia.

\item [Use of Variance and Covariance] For each measured quantity, the original sources often define an uncertainty, which can be the combination (in quadrature) of several terms.  For example, the uncertainty in a TRGB-based measurement of the distance to a host galaxy (calibrator) may combine at least three terms: the uncertainty in the geometric distance to the anchor, such as NGC~4258; the uncertainty in the apparent magnitude of the TRGB in the anchor; and the uncertainty in the TRGB of the calibrator itself. However, several of these terms may be in common with other data. For example, all distance estimates anchored to NGC 4258 share the variance associated with its geometric distance measurements, and all TRGB distances estimated relative to NGC~4258 share the uncertainty of the apparent TRGB magnitude determination in this galaxy.
In order to make use of more than one such measurement, these terms must be included as \textit{covariances} between the relevant data, and thus added as off-diagonal elements to a full covariance matrix of the data system.  More details are given in the description of the equations in Section~\ref{app:equations}.

Based on the need to define covariance and successfully leverage a broad range of distance indicators, it is necessary to restrict the use of measurements to those with {\it direct} traceability to well-defined sources.  Likewise, direct linkages assume consistent measurements between sources (e.g., Cepheids in an anchor and an SN Ia host), a reasonable assumption when both employ the same telescope and instrument.  In contrast, the combination of measures from different telescopes and instruments involves substantial covariance (most certainly from different zeropoints) whose characterization is rarely provided in publications and is beyond the scope of this work to define.  The above serve as ``quality cuts'' for the inclusion of data in the distance network.  As an example, we make use of TRGB measurements in hosts calibrated with NGC 4258, where both are identically measured with HST (or JWST). However, we do not combine ground-based and space-based calibrations.

\end{description}

\begin{figure*}
    \centering
    \includegraphics[width=\linewidth]{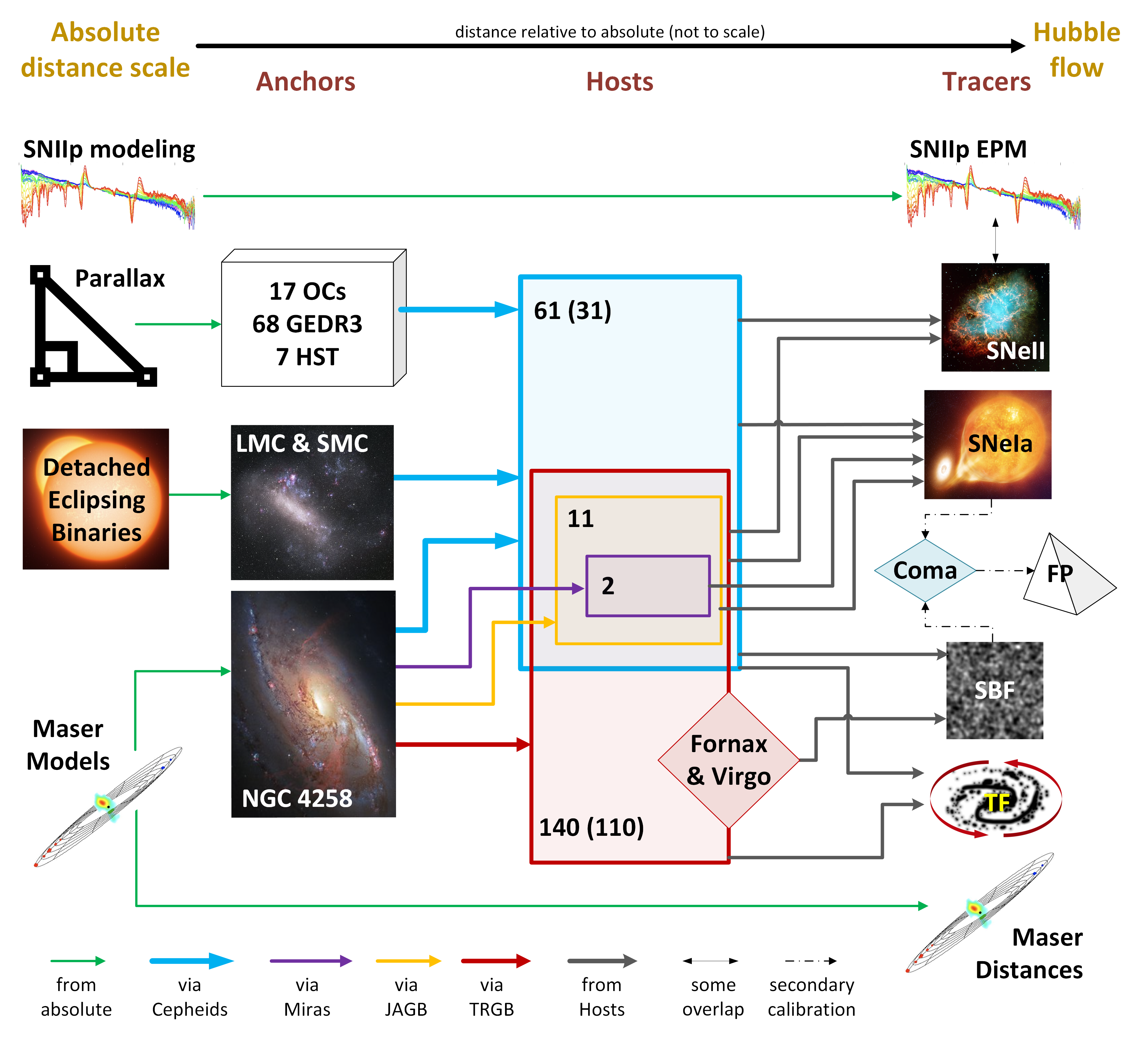}
    \caption{The complete {\DN}, with all possible pathways illustrated. Anchors are objects that establish an absolute scale based on the methods shown to their left. Primary distance indicators (Cepheids, TRGB, Miras, JAGB) transfer the absolute scale to hosts (i.e., galaxies), the ensemble of which calibrates secondary distance indicators in the Hubble flow (tracers).  Exceptions are Megamasers and astrophysically modeled SNe~II, both of which serve as primary distance indicators and are capable of reaching the Hubble flow without intermediate steps. Green arrows illustrate direct connections between anchors or tracers and the method used to determine the absolute scale. Blue, violet, yellow, and red arrows show which calibrators constrain host distances; line width qualitatively distinguishes the attainable precision. Among hosts, rectangles qualitatively indicate overlap among objects measured via multiple methods.
    Diamond shapes represent groups. Dark gray arrows tie subsets of hosts whose distance is constrained by different calibrators to tracers. Any given arrow may represent multiple data sets, e.g., HST or JWST photometry of Cepheids, or optical vs.\ infrared photometry of SNe~Ia. The number of hosts is labeled for Cepheids, TRGB, JAGB, and Miras, with the number of hosts exclusively available to each method shown in parentheses.}
    \label{fig:DN_full}
\end{figure*}

\subsubsection {Distance Network Architecture}

Given very few assumptions,\footnote{We make conventional assumptions common to distance ladders: that the cosmological principle holds so that we do not live in a special place (i.e., there is no intrinsic difference in objects solely based on their separation from us); that evolution over short lookback times, of order a few hundred~Myr, is negligible; and that the expansion history for $z<0.15$ can be approximated with an expansion of a few kinematic terms, namely {\Hcst}\,, $q_0$\,, and $j_0$\,.} the measures and concepts given above may be linked together as a system of equations, as we will show below and in Appendix~\ref{app:equations}, which we refer to as the {\DN}. The multiplicity of data (see Sect.~\ref{sec:data_summary} and Appendix~\ref{sec:data_appendix}) means we can optimize the system by introducing free global parameters.   The wealth of sources, methodologies, and distance indicators also provide important additional and complementary  information which can help strengthen the distance ladder or can contribute to constrain {\Hcst}.  For example, the distance to a given supernova host can be measured via different methods (TRGB, Cepheids, JAGB, Miras) depending on which observations are available for that galaxy, and each method can be related to a different anchor.  However, these determinations are not independent, as explained under ``Use of Variance and Covariance'' in Sect.\,\ref{sec:nomenclature}. The full {\DN} is illustrated in Fig.~\ref{fig:DN_full}.

Within the approximations of the present analysis, all equations are linear (or linearized), and all probability distributions are Gaussian in magnitude, distance modulus, or $\log_{10}{(H_0)}$.
Due to shared calibration sources among primary or secondary distance indicators, a covariance matrix with off-diagonal elements provides a useful method for proper accounting and weighting. The optimization procedure is that of a generalized linear least squares problem. The global solution to this problem provides best-fit values for the underlying parameters of the {\DN} as well as their uncertainties, most notably the distance moduli of included host galaxies, absolute magnitude calibrations, and the Hubble constant. These parameters include $\log_{10}{(H_0)}$, the distance moduli to host systems, $\mu_{\mathrm{host},\mathcal{H}}$, and groups included in the host data, $\mu_{\mathcal{G}}$, as well as the reference magnitudes for each type of calibrator, $M_{\mathrm{ref},\mathcal{T}}$. % \nilsc{I don't know what is the added value by showing symbols here.}
The full description of the system of equations and the covariance matrix is given in Appendix~\ref{app:equations}. The employed data sets are listed in Table~\ref{tb:datasets}, described briefly in Section~\ref{sec:data_summary}, and in greater detail in Appendix~\ref{sec:data_appendix}.  

\begin{deluxetable*}{lll}[t]
\tablewidth{0pc}
\tabletypesize{\scriptsize}
\tablecaption{Distance Network Datasets. Note that for all references we have made sure to decompose the listed uncertainties into those coming from the anchor (which the {\DN} treats separately) and those coming from the actual measurement of the distance indicator (separated to that within and without the anchor).
\label{tb:datasets}}
\tablehead{\colhead{Measure} & \colhead{Reference} & \colhead{Notes}}
\startdata
\hline
Milky Way (MW)& \citet{2018ApJ...855..136R, 2021ApJ...908L...6R,2022ApJ...938...36R} & (a)\\
LMC& \citet{2019Natur.567..200P} &$\mu_0$=18.477$\pm$0.024; converted from distance in kpc\\
SMC& \citet{2020ApJ...904...13G} & $\mu_0$=18.977$\pm$0.032; converted from distance in kpc\\
NGC 4258 Masers& \citet{2019ApJ...886L..27R} & $\mu_0$=29.397$\pm$0.032\\
\hline
\multicolumn{3}{c}{Primary distance indicators} \\[-0.015cm]
\hline
HST Cepheids & \citet{2022ApJ...934L...7R} & SN hosts, separate measures, N4258, MW, LMC (R22a fits 9,10,11) \\
& \citet{2022ApJ...938...36R} & MW cluster Cepheid photometry \\
& \citet{2019ApJ...876...85R} &  LMC  \\
 & \citet{2022ApJ...940...64Y} &  NGC 4258  \\
  & \citet{2024ApJ...973...30B}  &  SMC  \\
JWST Cepheids & \citet{2024ApJ...977..120R,2025arXiv250901667R}  & \\
JWST TRGB & \citet{2024ApJ...966...89A}   & \\
& \citet{2024ApJ...976..177L, 2025arXiv250408921L}  & \\
JWST TRGB & \citet{2025arXiv250311769H}  & \\
JWST TRGB & \citet{2024ApJ...973...83A,2025ApJ...982...26A} &  \\
HST TRGB & \citet{2024ApJ...966...89A}  &  \\
HST TRGB & \citet{2025ApJ...985..203F} &  \\
HST TRGB & \citet{2022MNRAS.514.4620D}  & and contained references \\
JAGB & \citet{2025ApJ...988...97L} & \\
JAGB &  \citet{2025ApJ...985..203F}  & \\
Miras & \citet{2024ApJ...963...83H} & Modified in correspondence with C. Huang to N4258-only values \\
\hline
\multicolumn{3}{c}{Secondary distance indicators} \\[-0.015cm]
\hline
SNIa Pantheon+ & \citet{2022ApJ...938..110B}  & \url{https://github.com/PantheonPlusSH0ES/DataRelease}\\
SNIa SNooPy (post v2.7) & C. Burns 2025, priv.\ comm.  &  Re-fitting the official data from \citet{2024ApJ...970...72U} \\
SNIa SNooPy (pre v2.7) & \citet{2024ApJ...970...72U}  & \url{https://github.com/syeduddin/h0csp/tree/main/data/working} \\
SNIa BayesSN & \citet{2023MNRAS.524..235D} & also E.~Hayes 2025, priv.\ comm.\ for additional fits\\
SNIa Salt3 & \citet{2021ApJ...923..265K}  &  \\
SNIa IR & \citet{2023AA...679A..95G}  & J, H bands separately\\
SNII (SCM) & \citet{2020MNRAS.496.3402D}  & T. de Jaeger 2025, priv.\ comm.\\
TF & \citet{2024MNRAS.531...84B} & Cosmicflows-4 catalog; data curated by the authors\\
\hline
\multicolumn{3}{c}{Other} \\[-0.015cm]
\hline
SNII EPM & \citet{2025AA...702A..41V},  & Removed galaxies (M61 and N6946), where peculiar \\
 & \citet{2023AA...672A.129C} &  velocities could not be meaningfully established.\\
% &   &  & \\
Megamasers & \citet[Tab.~1]{2020ApJ...891L...1P} & We also adopt our own peculiar velocity treatments\\
Coma FP & \citet{2025MNRAS.539.3627S}  & See also Appendix \ref{sec:eq_special}\\
SBF & \citet{2025ApJ...987...87J}  & Ancillary info from priv.\ comm.\ with J.~B.~Jensen.\\
\enddata
\tablecomments{(a) For the measurement of calibrators within the Milky Way galaxy, its extent is quite relevant --- therefore we adopt the individual measurements standardized to a 1~kpc reference distance.} 
\end{deluxetable*}

\subsubsection{Selecting the baseline}\label{sec:define_baseline}

There was strong consensus among the workshop participants, which is also shared more broadly within the community, that not all data, measurements, and tools are viewed with equal confidence.  For this reason, we decided to produce a ``baseline'' result based on all methods which enjoy the broadest and highest level of confidence (sometimes referred to in the literature as ``gold standards''), along with variants to the baseline that incorporate additional methods or datasets, or exclude certain data in order to reflect a broad range of minority viewpoints and explore the robustness of the results. To reach a true consensus for the baseline and variants, we adopted a collaborative, methodologically inclusive approach during the ISSI workshop ``What’s under the H0od.'' Prior to analyzing any combined fits, we engaged in extensive discussion of the various available methods, the choice of data sets (for instance, Cepheids from HST, TRGB distances from JWST, etc.), distance calibrators, and the systematic uncertainties involved in the determination of local distances and \Hcst\,. These included both classical and emerging techniques, ranging from geometric anchors (e.g., megamasers, {\it Gaia} parallaxes, detached eclipsing binaries) to stellar standard candles (e.g., Cepheids, TRGB, Miras, JAGB), as well as other methodologies such as Surface Brightness Fluctuations from HST and JWST and the Fundamental Plane from DESI. We also carefully reviewed the use of Type~Ia and Type~II supernovae, which extend the distance ladder into the Hubble Flow. In particular, we examined various SNe~Ia data sets (e.g., Pantheon+, CSP, NIR-only samples) as well as SNe~II determinations based on both the standard candle method and spectral modeling.

\begin{deluxetable}{llcccc}[h]
\tablewidth{\textwidth}
\tablecaption{Result from anonymous vote on ``technical readiness level" of distance calibrators. Above the line the consensus is that these  will be included in the baseline, below the line will be included  as variations.}
\label{table:rank1}
\tablehead{\colhead{Rank} & \colhead{Distance calib}& \colhead{in baseline}&\colhead{in variants} & \colhead{excluded} & \colhead{abstain}}
\startdata
\hline
1&Cepheids &  {\bf 28} & 1 &0&0 \\
2&DEB     & {\bf 26}   &3  &0 &0\\
3& TRGB   &  {\bf 26}&  2   &0 &1\\
3& NGC4258 & {\bf 26}&   2&  1&0\\
4& {\it Gaia} parallaxes & {\bf 24}& 3&2&0\\
\hline
5& Miras & 8& {\bf 15}&1&3\\
6& JAGB & 5& {\bf 15}&3&3\\
\hline
\enddata
\end{deluxetable}

\begin{deluxetable}{llcccc}[h]
\tablewidth{\textwidth}
\tablecaption{Result from anonymous vote on the method and secondary distance indicators.  Above the line the consensus is that these  will be included in the baseline, below the line will be included  as variations.}
\label{table:rank2}
\tablehead{\colhead{Rank} & \colhead {Method} & \colhead{in baseline}& \colhead{in variants} & \colhead{excluded} & \colhead{abstain}}
\startdata
\hline
1&SNIa &  {\bf 32} & 0 &0&0 \\
2&SBF     & {\bf 23}   &8  &0 &1\\
3& Masers in Hubble flow   &  {\bf 19}&  12   &0 &1\\
\hline
4& SNe~II, empirically standarized &15&   {\bf 14}&  0&3\\
5&SNe~II, astrophysically calibrated & 1& {\bf 22}&3&6\\
6& DESI FP (calibrated to Coma) & 11& {\bf 20}&0&1\\
7& TF from CF4 & 8& {\bf 21}&1&2\\
\hline
\enddata
\end{deluxetable}

A key element of this process was a series of open, expert-led discussions in a plenary setting, during which participants evaluated the strengths, systematics, data availability, and limitations of each method and dataset. Following these in-depth assessments, we held anonymous ballot votes to determine the configuration of the \textbf{baseline} analysis and well-motivated \textbf{variants} thereof, as well as configurations that should be excluded entirely.

The first vote identified the primary and secondary distance indicators and methods considered most technically mature and broadly supported, cf.  Table~\ref{table:rank1}. In a second vote, we decided on specific combinations, methodological choices, and alternative calibrations, resulting in a comprehensive suite of variants, see Table~\ref{table:rank2}. Methods were included in the baseline if they garnered more than half the total votes, including abstentions, as indicated by the horizontal lines in Tables~\ref{table:rank1} and \ref{table:rank2}.  All votes were taken prior to deriving a value of {\Hcst}, and their outcomes were considered binding. The datasets included in the baseline configuration are listed in Table~\ref{tb:datasets} and discussed in detail in Appendix~\ref{sec:data_appendix}. Where multiple datasets were available in principle, preference was given to datasets that had been fully published 
and/or have been most widely adopted. Alternative datasets, notably of SNe~Ia (i.e., Pantheon$+$ vs. CSP vs. BayesSN), were included as variants.

This process aimed at identifying the most widely supported and robust measurement paths based on current data and understanding, not at enforcing a single value of \Hcst. The resulting baseline, variants, datasets and code provide a flexible framework to assess the stability and reliability of the derived consensus {\Hcst} value under well-motivated methodological changes. This structured approach reflects the collective judgment of the experts present at the workshop, and provides a reproducible path forward for future analyses. Hence, ``consensus'' here refers to an a priori agreement on \emph{how} a {\DN} should be constructed in terms of methods, datasets, and their combinations, rather than on the outcome, which was not known when the process was fixed.

In summary, the baseline analysis reflects the scientific consensus among workshop participants concerning the optimal combination of datasets and methodologies as established prior to obtaining results. Variants of the baseline analysis were performed to test the robustness of the result to individual methods, to different datasets corresponding to identical methodologies, and to quantitatively explore ``hypothesis-driven'' setups that combine multiple elements deviating from the baseline. Further detail on the process of selecting variants is provided in Sect.~\ref{sec:select_variants} below.

\subsubsection{Selecting variants and other methodological choices }\label{sec:select_variants}

We defined analysis \emph{variants} to explore the sensitivity of the {\Hcst} determination to {\it well-motivated} modifications in the analysis setup. Possible variants were identified through brainstorming and subsequently discussed for their merits and weaknesses. These variants serve as internal consistency tests (or ``null tests''), helping to assess the robustness of the baseline result under plausible changes to inputs or methodology. Considerable discussion was involved in ensuring that variants were well-motivated prior to knowing their impact on \Hcst{}. We specifically avoided combinations targeting maximal changes in \Hcst{}, as these would require {\it a posteriori} information or intermediate outcomes, potentially becoming prone to confirmation bias. Our process described in Sections~\ref{sec:philosophy} and \ref{sec:define_baseline} was key to taking these decisions well informed by broad interdisciplinary and expert knowledge. In each variant, we furthermore aimed to adequately reflect the current state-of-the-art in the various disciplines by including all data sets sufficiently precise to carry significant weight in the determination of \Hcst{}. In other words, we omitted data sets or methods only when motivated by a specific hypothesis in order to avoid artificially inflated uncertainties (i.e., caused merely by insufficient data) that would reduce the Hubble tension in an uninformative way not representative of the state of knowledge.

We reiterate that most variants are effectively ``null tests.''  Their consistency (or lack thereof) serve as a test of robustness of the baseline.  In general, they are not fair substitutes for the baseline solution, and should always be referenced together with a description of their underlying choices.  It is also critical to note that the values of {\Hcst} obtained with all variants are \textit{highly correlated}.  They share a large fraction of the underlying data, and are therefore \textit{expected} to differ from each other by much less than the nominal uncertainty.  Therefore, quantifying if individual datasets have anomalous pulls on the final result requires a more careful analysis; see Section~\ref{sec:variants} for further discussion. 

The variants discussed in this paper can be organized into different categories:

\paragraph{Add-one-in variants} include additional information that was not voted for inclusion in the baseline, introduced through a step-by-step process. This category can  introduce additional anchors (i.e., the SMC), as well as primary or secondary distance indicators. For example, V01 (Baseline+JAGB) adds JWST/NIRCam distances of the JAGB in supernova hosts calibrated to that in NGC~4258. The addition of calibrators to the baseline configuration serves to explore whether these supplementary methods shift or reinforce the baseline results. Furthermore, this category includes the incorporation of alternative distance indicators in the Hubble flow, such as SNe~II calibrated via the standard candle method or spectral modeling, the Fundamental Plane from DESI anchored to Coma, and the Tully-Fisher relation as implemented in the Cosmicflows-4 catalog \citep[CF4;][]{2020ApJ...902..145K}. 

\paragraph{Leave-one-out variants} involve the exclusion of individual methods or calibrators from the baseline configuration, motivated by the hypothesis that there is an undiscovered error in a type of measurement. Each variant is constructed by removing one key element at a time—such as Cepheids, SNe~Ia, TRGB, {\it Gaia} parallaxes, or NGC~4258—and observing the resulting impact on the derived {\Hcst} value. These tests help identify whether any individual dataset or calibrator is significantly impacting the result. This category also includes the exclusion of specific indicators in the third rung, such as SBF, SNe~Ia, and masers in the Hubble flow.  Since leaving out a given element can implicitly remove other elements (e.g., specific SNe~Ia calibrators), we have constructed custom baselines (cf.~below) to separately assess the effect of the implicit (undesired) modification relative to the baseline.

\paragraph{Hypothesis-driven variants} involve compound 
configurations motivated by specific physical or observational hypotheses. These include the use of alternative calibrations (e.g., CSP/SNooPy instead of Pantheon+/SALT2 for SNe~Ia), different treatments of systematics (e.g., inclusion or exclusion of peculiar velocity corrections, metallicity corrections, or near-infrared SN data), and restricted subsets of the data (e.g., only modern SNe~Ia, or cutting the Hubble flow sample at $z > 0.06$). Additionally, we included a variant that considered the Cepheid Leavitt law to be independent of chemical composition.

\paragraph{Instrument-suspicious variants} explore the impact of removing specific observatories. In particular, we tested the exclusion of all HST or JWST-based observations, all {\it Gaia} measurements, etc.

\paragraph{Custom baselines} The exclusion of certain measurements can also cause the exclusion of some SNe~Ia calibrators; for example, when Cepheid measurements are excluded, only 35 of the 55 SNe~Ia calibrators in the baseline can be included, since the other 20 only have determinations of the host distances through Cepheids. The resulting change in the {\Hcst} value and uncertainty is due to both the exclusion of the measurements and the change in the SNe~Ia calibrator sample. To cleanly separate the two effects, we define a custom baseline that includes those measurements, but is restricted to only the SNe~Ia calibrators for which other measurements are available. Custom baseline versions are not variants {\it per se}; they are only intended to facilitate the interpretation of the results for the corresponding variants.  

\paragraph {Include Everything}
We also consider a variant in which all independent methods and data sets are included, to illustrate the accuracy that can be achieved with presently available measurements if systematic effects and other issues are resolved.  A secondary variant in this class uses all independent methods \textit{except} Tully-Fisher, for which the current data set has excess dispersion.  In either version, this variant includes only one set of measurements for SNe~Ia, since different measurements are likely correlated (because of astrophysical variance) to a degree that has not been sufficiently quantified in the literature.  

\paragraph {Additional solutions}  In addition to the variant solutions described above, we also consider special solutions designed to test our methodology. These are not ``variants'' in the same sense as those listed previously, and do not appear in Table~\ref{tab:variants}; they are discussed in mode detail in Sections~\ref{sec:orthogonal_paths}, \ref{sec:consistency_checks}, and Appendix~\ref{sec:comparisons}.  They include \textit{independent grouping solutions}, \textit{consistency checks}, and \textit {emulator solutions}.  Independent grouping solutions consist of configurations in which two or more fully independent paths through the distance ladder are identified, sharing no common calibrators or intermediate steps. These allow for the construction of entirely separate determinations of {\Hcst}, such that agreement between them provides a strong consistency check and minimizes the risk of shared systematic effects. They are the analogous to splitting the data in uncorrelated halves or thirds. Two such variants are discussed in Section~\ref{sec:orthogonal_paths}.  Consistency checks are employed to make sure that the network results are statistically self-consistent for different paths through it and are further discussed in Section~\ref{sec:consistency_checks}.  Emulator solutions involve a set of \textit{emulators} designed to reproduce key published results, such as SH0ES, CCHP, and recent SBF analyses, within our unified framework and dataset handling. These configurations serve both as validation tests of our pipeline and as transparent benchmarks for comparison with previous literature, and they may also help explain sources of differences. These are discussed in Appendix~\ref{sec:comparisons}.

%% file: 03_data_summary.tex
The analysis presented in this work is based on a comprehensive set of local distance measurements and their corresponding calibrators, as detailed in Section~\ref{sec:meth} and conceptually shown in Fig.\,\ref{fig:Metro}. The dataset includes a broad array of distance indicators, spanning geometric anchors, calibrators, and tracers. These are combined using a statistically rigorous framework that accounts for shared systematics and covariances among measurements.
To maintain clarity and readability in the main text, we provide here only a brief summary of the datasets used to construct the \DN, summarized also in Table~\ref{tb:datasets}. Methods employed in the baseline are marked in bold. A more detailed description of the individual measurements, sources, and associated assumptions is given in Appendix~\ref{sec:data_appendix}. We employ peculiar velocity corrections as described in Appendix~\ref{app:sec:pec_vel}.

%\nilsc{Honestly, I think this section should be much more explanative about HOW these methods work, with just a tiny bit of info on which we use precisely, with the big info on WHAT data we use mostly already found in App. A. For example, for the JABG we say nothing of how it works, just which PI's have observed it in which programs, which IMHO is irrelevant information for any non-expert.}

% \riacomm{In keeping with the below listing methods (not objects), I have slightly rearranged and relabeled the first three paragraphs} \nilsc{I found the previous labeling more intuitive, since it separated out rungs.}

\paragraph{\textbf{Parallaxes}}
%Trigonometric parallaxes of Milky Way (MW) stars}}
We adopted trigonometric wide-angle parallaxes (henceforth: parallaxes) of Milky Way (MW) stars from the early third data release of the ESA {\it Gaia} mission \citep[GEDR3;][]{2016A&A...595A...1G,2021A&A...649A...1G}. GEDR3 parallaxes were corrected for systematics following \citet{2021A&A...649A...4L} and for additional residual offsets as determined for field Cepheids \citep{2021ApJ...908L...6R}. GEDR3 parallaxes of open stars clusters were determined using non-variable member stars \citep{2022ApJ...938...36R,2023A&A...672A..85C}. An additional seven narrow-angle parallaxes of Cepheids based on the HST/WFC3 spatial scanning technique were included \citep{2018ApJ...855..136R}. 
% \citep[7 Cepheids]{} 
% We adopt the trigonometric parallaxes of individual Cepheids within the Milky way, taking into account its extended nature (as opposed to other anchors, whose extent relative to their distance is often negligble). The distances are measured with high signal-to-noise ratio by the ESA {\it Gaia} mission~\citep{2016A&A...595A...1G}.

\paragraph{\textbf{Detached Eclipsing Binary (DEB) distances}} % to the Magellanic Clouds}}
The distances to both Magellanic Clouds have been determined using helium burning red giants in detached eclipsing binary systems (henceforth: DEBs). These distances are determined geometrically as the ratio of the physical stellar radii determined from full orbital solutions to the angular diameters determined by surface-brightness-color relations calibrated empirically using long-baseline interferometry. 
% solved orbits \nils{. This method uses the fact that the orbits closely determine the (eclipsed) star's linear radius to extreme precision. Estimating then the angular size of the star from the measured flux as well as the surface brightness (calibrated using an emprical surface-brightness-color relation) allows the determination of the distance}. 
% For the present work, we assume a 
The distance to the LMC from \cite{2019Natur.567..200P} contributes to the Cepheid calibration in the baseline analysis. An analogous distance to the SMC from \cite{2020ApJ...904...13G} was considered as part of a variant. 
% In our analysis, both galaxies are used as geometric anchors to calibrate the period-luminosity relation of classical Cepheids. 

\paragraph{\textbf{Megamaser distances}}
We adopted geometric distance measurements to the anchor galaxy NGC$\,$4258 and five additional galaxies out to the Hubble flow. The distances were determined via very long-baseline interferometric (VLBI) radio-wavelength observations of water megamasers in accretion disks surrounding central SMBHs of their host galaxies. Distances were determined by comparing the physical scale of the megamasers determined from a Keplerian disk model to the angular extent of the features \citep{2015IAUGA..2255730B,2019ApJ...886L..27R,2020ApJ...892...18K}. 

%% merged the following two together since it's the same method
% \paragraph{\textbf{NGC$\,$4258}}
% We adopt the geometric distance measurement to NGC$\,$4258 from~\citet{2019ApJ...886L..27R}, who used VLBI observations of water masers in the accretion disk around the central SMBH to model their orbital motion and determine the distance to the system.  In our analysis, NGC$\,$4258 is used as an anchor to calibrate the Cepheid variable, TRGB, JAGB, and Mira variable primary distance indicators.
% \paragraph{\textbf{Megamaser distances}}
% We use a sample of 5 megamaser-hosting galaxies as single-step tracers in the Hubble flow.  The distances to these galaxies were compiled by the Megamaser Cosmology Project  \citep[][MCP]{2015IAUGA..2255730B,2020ApJ...892...18K}, which used spectral monitoring and VLBI mapping observations to measure the orbital motions of the masers and thereby determine geometric distance measurements to each system.

\paragraph{\textbf{Cepheids}}
The Leavitt Law, or Period-Luminosity relation of classical Cepheids (throughout this work: Cepheids) is a well-understood consequence of the dependence of acoustic oscillations on stellar density, the relation between mass and luminosity, and the Stefan-Boltzmann law. The periods of their observed brightness variations link directly to their intrinsic luminosity  \citep{1912HarCi.173....1L} with particularly small scatter in the infrared, and this is well described by stellar evolution models \citep[e.g.,][]{2025arXiv250522512K}. Distances to Cepheids pulsating in the fundamental mode were measured using the reddening-free near-infrared Wesenheit magnitudes based on HST photometry published by \citet{2018ApJ...855..136R, 2021ApJ...908L...6R, 2022ApJ...938...36R, 2019ApJ...876...85R, 2022ApJ...934L...7R, 2022ApJ...940...64Y, 2024ApJ...973...30B}, and more recently with JWST by \citet{2023ApJ...956L..18R, 2024ApJ...977..120R, 2025arXiv250901667R}. The Cepheid Period-Luminosity relation was calibrated in anchor galaxies using geometric distances in the Milky Way, Magellanic Clouds, and NGC$\,$4258.

\paragraph{\textbf{Tip of the Red Giant Branch (TRGB)}}
The TRGB represents a recognizable feature in the color-magnitude diagram of galaxies that is caused by the nearly constant luminosity of the Helium flash of first-ascent red giant stars, which continue along their evolution towards the Horizontal Branch.
%sudden drop in luminosity once Helium burning starts, whose intrinsic luminosity is universal.}
The TRGB distances were measured using HST and JWST observations of resolved stars in nearby (D$<$30 Mpc) galaxies. These distances were measured by the Carnegie-Chicago Hubble Program (CCHP; \citealt{2019ApJ...882...34F, 2021ApJ...915...34H, 2025arXiv250311769H}), the Extragalactic Distance Database (EDD; \citealt{2009AJ....138..323T, 2021AJ....162...80A}), the SH0ES team \citep{2024ApJ...966...89A,2024ApJ...976..177L}, and the TRGB-SBF project team \citep{2024ApJ...973...83A, 2025ApJ...982...26A}. Each team used somewhat different reduction, analysis, and calibration techniques, but we find that the resulting distances are generally consistent across groups, as shown in Section~\ref{sec:results}.

\paragraph{J-region Asymptotic Giant Branch (JAGB)} The JAGB is a recognizable feature in the near-infrared color-magnitude diagram of galaxies attributed to post-third dredge-up carbon-rich intermediate-mass AGB stars. The JAGB distances used here are derived from data taken with {JWST} NIRCam as published by \citep[][SH0ES]{2024ApJ...966...20L,2025arXiv250205259L} and \citep[][CCHP]{2025ApJ...985..203F}; see also Appendix~\ref{sec:data_appendix}. In its current implementation, the JAGB method assumes a homogeneous stellar population with a constant average luminosity. Systematic uncertainties, e.g., asymmetric luminosity functions or metallicity effects, were considered following  \citet{2024ApJ...966...20L} and remain a subject of research \citep[e.g.,][]{2021ApJ...916...19Z,2024AA...691A.350M,2025ApJ...985..182L}.
% The JAGB method currently assumes a homogeneous stellar population in the J-region; while this may not be the case \nils{and there could be some} inhomogeneity, which can manifest as asymmetry of the J-region luminosity function, \nils{it} is accounted for as a systematic uncertainty \nils{determined} via measurement variants, for example as described in \cite{2024ApJ...966...20L}.

\paragraph{Mira distances}
Miras are high-amplitude long-period variable asymptotic giant branch stars that obey period-luminosity relations.
We use a sample of 3 Mira host galaxies consisting of the geometric anchor NGC$\,$4258 \citep{2018ApJ...857...67H} and two SNe~Ia calibrator galaxies, NGC$\,$1559 and M101 \citep{2020ApJ...889....5H, 2024ApJ...963...83H}, hosts of SN~2005df and SN~2011fe, respectively. All observations consisted of at least 10 epochs of HST WFC3/IR time series photometry spanning a minimum of one year, and distances were obtained using the Mira period-luminosity relation in the {F160W} bandpass.

\paragraph{\textbf{Type Ia Supernovae (SNe~Ia)}}
SNe~Ia are standardizable candles whose light curve information (e.g., the duration) can be related to their intrinsic magnitude.
We incorporate five different SNe~Ia datasets, ranging from optical to near-infrared (NIR) wavelengths and spanning different SNe~Ia modeling methodologies (Spectral template PCA with SALT2, SALT3 and SNooPy v2.7, and template-free with BayesSN). Four of these datasets are used in the Hubble Flow: (1) Pantheon+ (optical) with SALT2 \citep{2022ApJ...938..113S,2022ApJ...938..111B}, (2) Carnegie Supernova Project (CSP) I \& II (with SNooPy v2.7; \citealt{2011AJ....141...19B,2024ApJ...970...72U}), (3) template-independent distances in NIR \citep{2023AA...679A..95G}, (4) optical+NIR samples processed with BayesSN \citep{2023MNRAS.524..235D}. A fifth dataset of 13 SNe~Ia is used to calibrate the distance to the Coma cluster \citep{2025ApJ...979L...9S}. % \riacomm{inconsistency: \ref{sec:eq_special:FP} and FP below mention 13 SNe Ia, one of which is used twice}

\paragraph{Type II Supernovae (SNe~II)}
Several methods exist to standardize SN~II magnitudes and derive distances (see \citealt{2024hct..book..177D}). In particular, the Standard Candle Method (SCM; \citealt{2002ApJ...566L..63H}) standardizes SN~II luminosities based on correlations with photospheric velocity (from H$_\beta$) and color.
We use a sample of 89 SNe~II in the Hubble flow at $z > 0.01$ as well as 14 calibrator supernovae from \citet{2020MNRAS.495.4860D,2022MNRAS.514.4620D}, compiled from CSP-I, LOSS, SDSS-II, SNLS, DES-SN, and SSP-HSC \citep{2017ApJ...835..166D,2017MNRAS.472.4233D,2020MNRAS.495.4860D}.

\paragraph{Expanding photosphere method (EPM) of Type IIP supernovae}
We adopted distances of Type IIP supernovae (SNe~IIP) determined by astrophysical modeling via the tailored expanding photosphere method (EPM) from \citet{2025AA...702A..41V}. Such distances relate the angular extent of SNe~IIP measured from light curves to their physical expansion determined by radiative transfer modeling of optical spectra \citep{2020A&A...633A..88V,2023AA...672A.129C}. The EPM can be applied to SNe~IIP in the Hubble flow without requiring additional calibration, although it depends on the accuracy of the underlying astrophysical modeling.

\paragraph{\textbf{Surface Brightness Fluctuations (SBF)}} Spatial fluctuations in the surface brightness of an otherwise smooth galaxy arise from the statistics of the discrete number of stars per pixel. The amplitude of these fluctuations depends inversely on the distance of the galaxy, as well as on age, metallicity, and other properties of the stellar population.
For galaxies with evolved stellar populations, the SBF method can be calibrated through an empirical relation between the intrinsic magnitude of the fluctuations and the galaxy's integrated color, used as a proxy for the stellar population properties. SBF measurements in the near-infrared are particularly useful because the fluctuations are bright at these wavelengths and can be well calibrated using optical colors.
For this study, we use a sample of 61 HST WFC3/IR SBF elliptical galaxy distances reaching out to 100 Mpc \citep{2025ApJ...987...87J,2021ApJS..255...21J,2021ApJ...911...65B}; distances to the 14 calibrators in the Virgo and Fornax galaxy clusters are based on TRGB; see Appendix~\ref{sec:eq_groups}.

\paragraph{Fundamental Plane (FP)}
The distance to an elliptical galaxy can be estimated by means of the fundamental plane relationship between central velocity dispersion, effective radius, and surface brightness.
Fundamental Plane distances were measured using the Dark Energy Spectroscopic Instrument (DESI) Early Data Release, analyzing 4191 early-type galaxies within $0.01 < z < 0.1$ with photometry from the DESI Legacy Imaging Surveys and spectroscopic velocity dispersions from DESI observations \citep{2025MNRAS.539.3627S}. The FP zero-point calibration was established using a collection of galaxies in the Coma cluster, with the distance of the latter constrained by 13 Type Ia supernovae within Coma analyzed by \citet{2025ApJ...979L...9S}, as well as an additional (fixed) constraint from SBF; see Appendix~\ref{sec:eq_special}.
%\riacomm{What does `additional calibration' mean in this case? App.\ref{sec:eq_special:FP} suggests that SNe\,Ia alone set the absolute scale?}

\paragraph{Tully-Fisher (TF)\label{sec:data:TF}}
The Tully-Fisher relation relates the rotation velocity of spiral galaxies (determined with HI line widths) to the total intrinsic luminosity. TF data were obtained from the Cosmicflows-4 catalog \citep{2020ApJ...902..145K}, which compiled HI line widths, redshifts, and photometry for $\sim 10,000$ spiral galaxies across the full sky, out of which we use 3400 galaxies with complete infrared photometry (as recommended by P.~Boubel 2025, priv.\ comm.). The TF relation zero-point has been re-calibrated within the {\DN} using calibrator galaxies identified in \citet{2024MNRAS.533.1550B}, with systematic corrections applied by \citet{2024arXiv241208449S}.

%% file: 04_results.tex
\begin{figure}
    \centering
    \includegraphics[width=\textwidth]{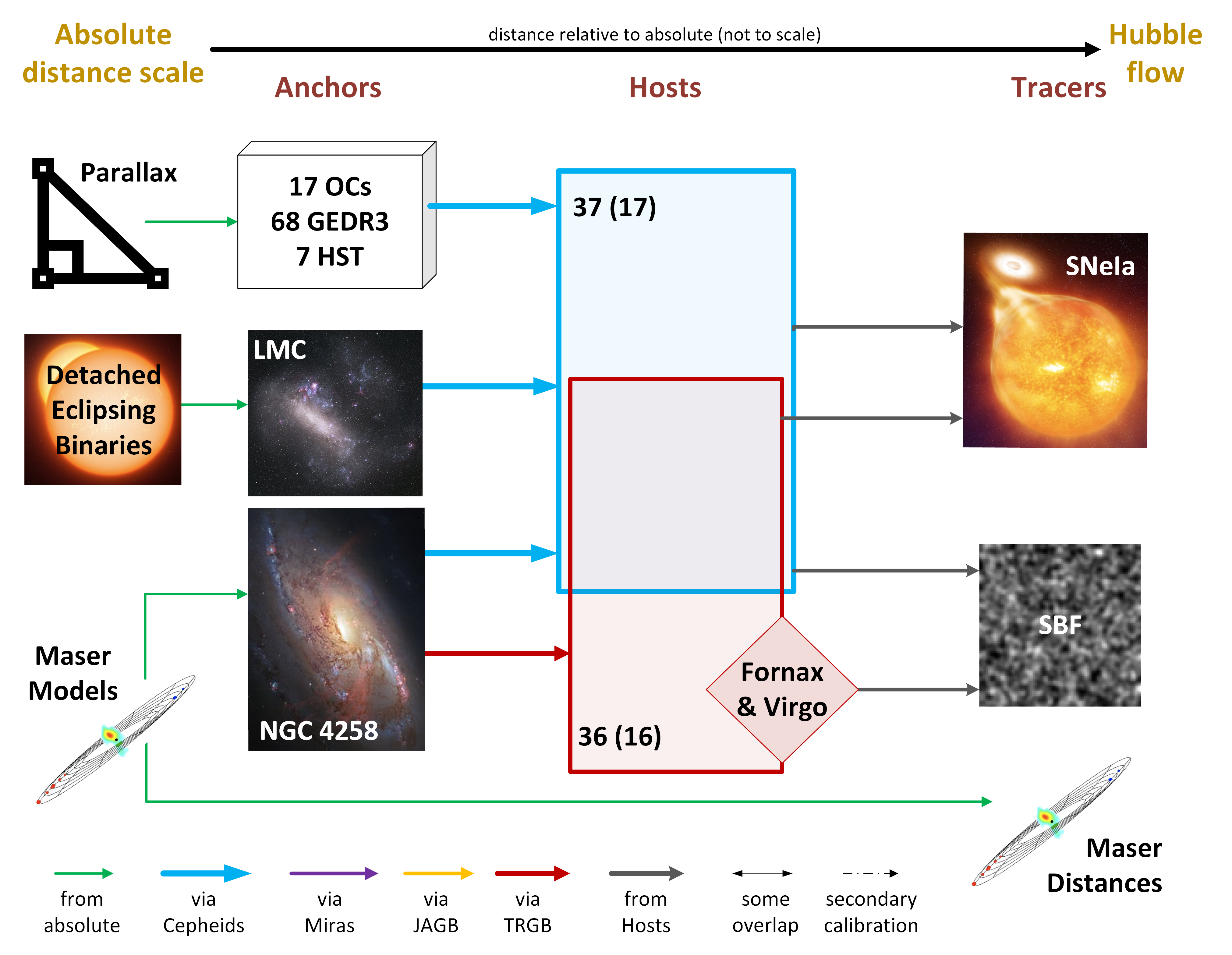}
    \caption{The Baseline Distance Network, illustrated analogously to Fig.\,\ref{fig:DN_full} \label{fig:baseline_DN}}
\end{figure}

The baseline solution for the Distance Network  as illustrated in Figure~\ref{fig:baseline_DN} yields 
\begin{equation}
H_0 = \Hvaluebase \pm \Herrorbase \, \Hunit~.
\end{equation} 
As described in Section~\ref{sec:define_baseline}, this solution includes SNe~Ia, SBF, and megamaser measurements as Tracers. The distances to hosts of SNe~Ia and SBF calibrators are obtained from the solution of the full \DN, incorporating measurements based on TRGB and Cepheids from the sources described in Section~\ref{sec:data_summary}, with the Milky Way, LMC, and NGC$\,$4258 as anchors for Cepheid distances.\footnote{Note that the current version only uses TRGB measurements for the calibration of the SBF method, in accordance with the most recent publication \cite{2025ApJ...987...87J}; in future iterations we intend to allow for arbitrary primary distance indicators for the calibration.} SNe~Ia measurements are from the Pantheon+ sample, and peculiar velocity corrections in the Hubble flow are based on the 2M++ model \citep{2015MNRAS.450..317C}.  This solution has an overall $ \chi^2 $ of {\redchisqbase} per degree of freedom, indicating broad agreement between the estimated uncertainties---all based on original sources---and the statistical properties of the solution.  Note that the value of $ \chi^2 $ does not include degrees of freedom associated with SNe~Ia and SBF tracers in the Hubble flow. In our methodology, we solve separately for the Hubble flow intercept for SNe~Ia and SBF tracers in the Hubble Flow---as well for others that are not included in the baseline solution, such as SNe~II and TF tracers---and only the intercept is included in the final \DN{} solution. It can be shown that, apart from a separation of the $\chi^2$ values, this approach is equivalent to directly including the Hubble tracer equations into the {\DN}; see also Appendix~\ref{app:sec:intercept}.

\begin{figure}
    \centering
    \includegraphics[width=0.9\linewidth]{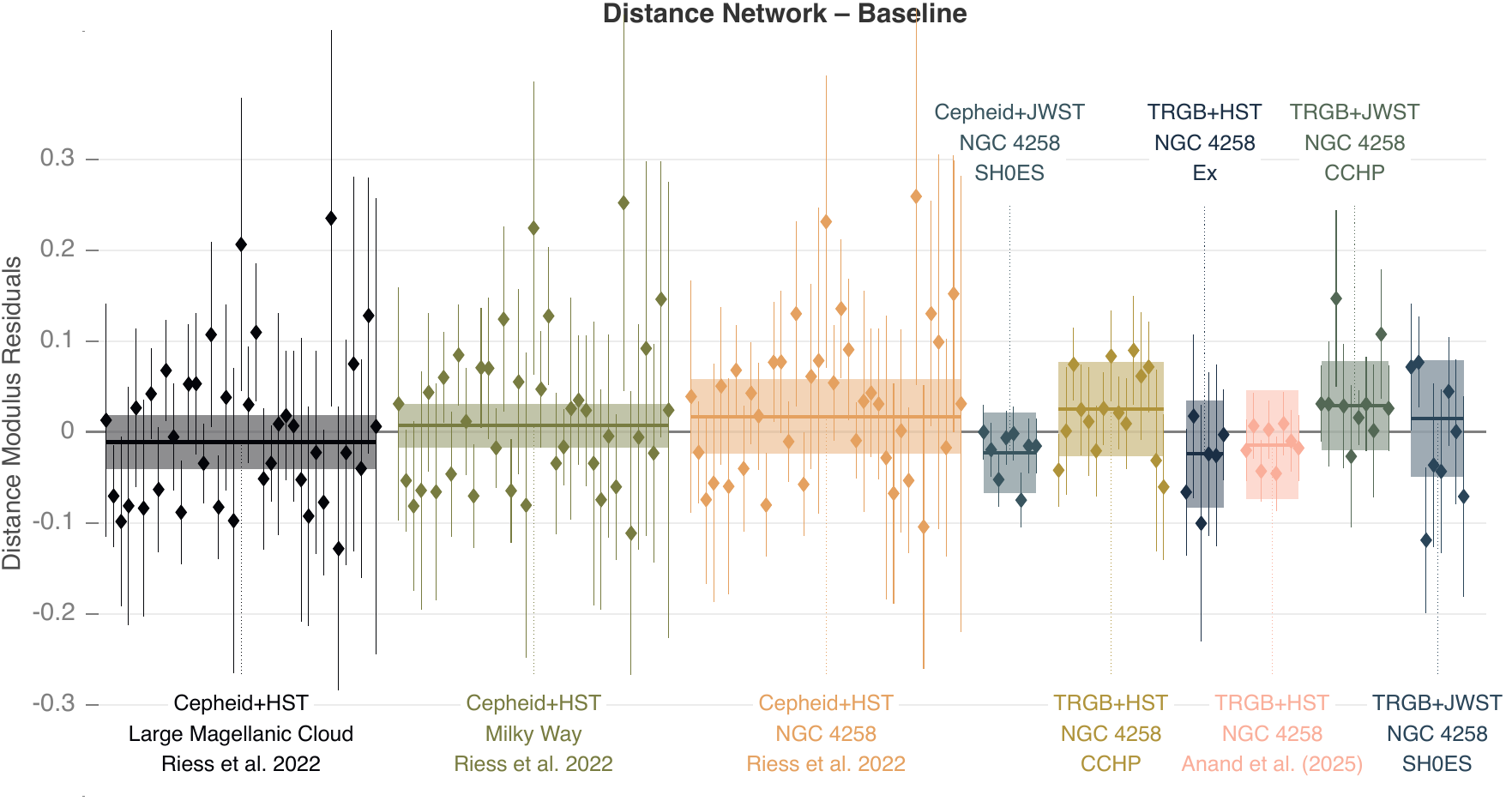}
    \caption{Residuals for each category of host distance measurements from the Baseline solution.  Each panel represents a group of measurements of host distances that share the same method, anchor, and authors, and shows the deviation of those measured host distances from the full distance network value.  Error bars represent the individual uncertainty of each measurement, while the shaded regions for each group shows the common (fully correlated) uncertainty due to the reference system.}  
    \label{fig:baseline_residuals}
\end{figure}

Figure~\ref{fig:baseline_residuals} shows the residuals of host distances, estimated from the network solution, grouped by methodology. 
Each grouping, identified by a different color, refers to a specific combination of primary distance indicator, anchor, and source, as indicated in the labels below the group. The error bars shown correspond to the uncertainty of the specific distance measurement for that host, not including terms that are covariant with the other elements of the group. For example, each Cepheid measurement is shown with the uncertainty of the Period-Luminosity intercept for that galaxy, which reflects the expected scatter of residuals \textit{within that group}.  The color-shaded band for each group shows instead the uncertainty associated with common error terms for that group, i.e., the anchor uncertainty and the uncertainty in the indicator measurement (TRGB magnitude or Cepheid Period-Luminosity intercept) for that anchor, combined in quadrature; this term is indicative of the likely scatter of the \textit{average} of the group---shown by the corresponding horizontal line at the center of the band---with respect to zero.\footnote{This is a slightly simplified description of the uncertainties, as it accounts for most, but not all, covariances in the data. However, it is a useful guide to the properties of residuals, as long as the formal $ \chi^2 $ is consistent with the statistics of the system.}  As can be seen, the distribution of residuals is consistent with expectations, with no significant deviation from the published uncertainties for any subset of data.  

A closer look at the residuals for Cepheids shows clearly that all anchors are consistent with one another \textit{and} with TRGB; the systematic offsets (horizontal bars) for Cepheids anchored to the LMC, the Milky Way, and NGC$\,$4258 differ by 0.02 mag or less from the distance network solution.  Subsets of TRGB measurements have slightly larger typical offsets, about 0.03 mag, justified in part by the smaller number of systems included in each study; but overall they are in very good agreement, within the expected statistical uncertainties. 

\begin{figure}
    \centering
    \includegraphics[width=0.9\linewidth]{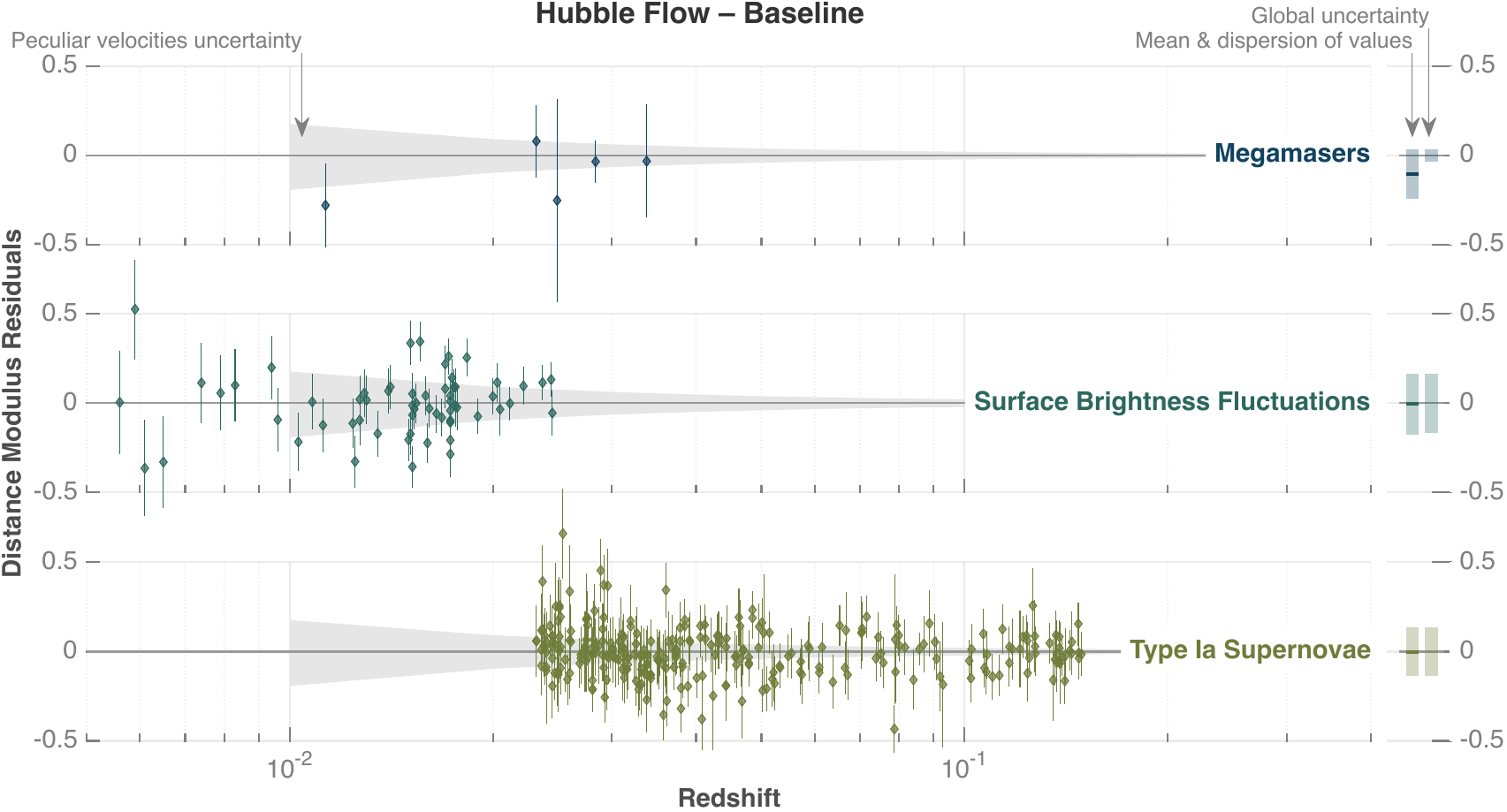}
    \caption{Residuals in distance modulus as a function of redshift for objects in the Hubble flow in the Baseline solution.  Error bars reflect the individual source scatter, without the common calibration uncertainty.  The shaded region in each plot indicates the effect of a velocity uncertainty of $ 250 \, \mathrm{km \, s^{-1}} $.  The bars at the right show the mean and dispersion for each group of sources and the calibration uncertainty, which is a common error mode for all points in each panel.} 
    \label{fig:baseline_hf}

\end{figure}

\begin{figure}
    \centering
    \includegraphics[width=0.9\linewidth]{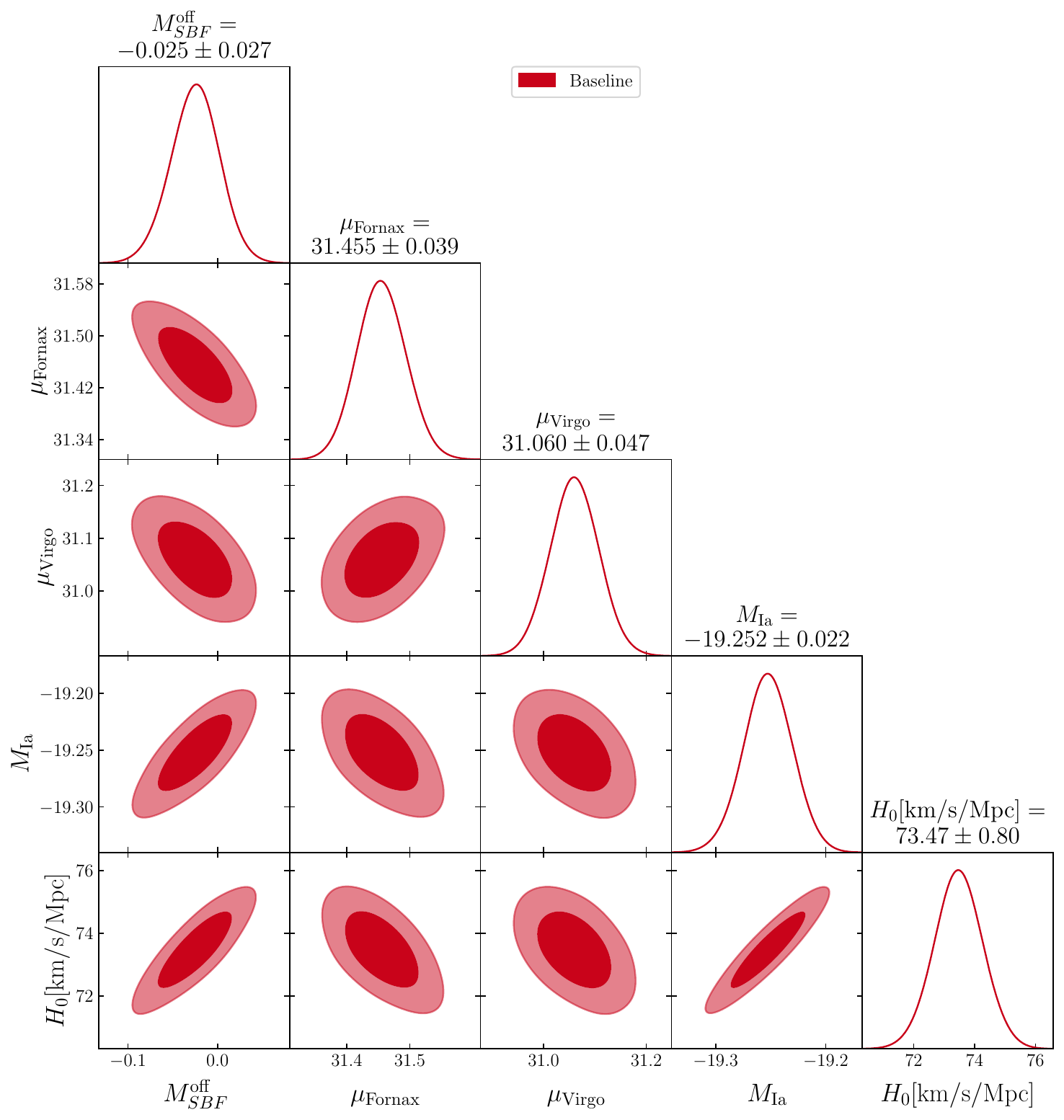}
    \caption{Corner plot for the baseline solution, illustrating the correlations between {\Hcst}, the calibration of SNe~Ia and SBF, and the distance to Fornax and Virgo, which contribute to the SBF calibration. The parameters use the naming convention of Appendix~\ref{sec:optimization_parameters}. Deviations from the analytically calculated result are due to the numerical precision of a finite-length chain.}
    \label{fig:baseline_corner}
\end{figure}

Figure~\ref{fig:baseline_hf} shows the distribution of residuals for objects in the Hubble flow as a function of redshift for three classes of objects.  The ordinate is the offset in $ \log_{10}{(H_0)} $ inferred for that specific object; the error bar reflects a combination of the measurement uncertainty (photometric error, error in standardization, and/or uncertainty in the solution, as applicable), the effect of the dispersion in peculiar velocities, the uncertainty in the peculiar velocity reconstruction, and the intrinsic dispersion in the relevant distance indicator, as determined in the original sources from the dispersion in the calibrators. They do not include the calibration uncertainty, which is determined as part of the distance network calibration solution for each group when applicable.  The shaded bars at the right show the measured mean and dispersion for each of the three sets of data. Again, both the dispersion in individual measurements and the mean for each subset are consistent with expectations, which represent the direct measurement uncertainties for each value---not including calibration uncertainties.  Figure~\ref{fig:baseline_corner} shows the posterior distribution for the global fit parameters for this solution: the absolute magnitude calibration for SNe Ia, the offset in the SBF calibration, the estimated distances to Virgo and Fornax, and the value of {\Hcst}.  These are based on finite-length chains extracted from the probability distribution of the solution, and the values of the parameters may be slightly different from the analytically calculated results.

The solution for the baseline {\DN} yields the most accurate direct {\Hcst} measurement to date, with a relative uncertainty of \Hrelerrorbase, including systematic uncertainties. It agrees to much better than the $1\sigma$ uncertainty with previous determinations of {\Hcst} by the SH0ES team \citep{2022ApJ...934L...7R}.

\begin{figure}
    \centering
    \includegraphics[width=0.8\linewidth]{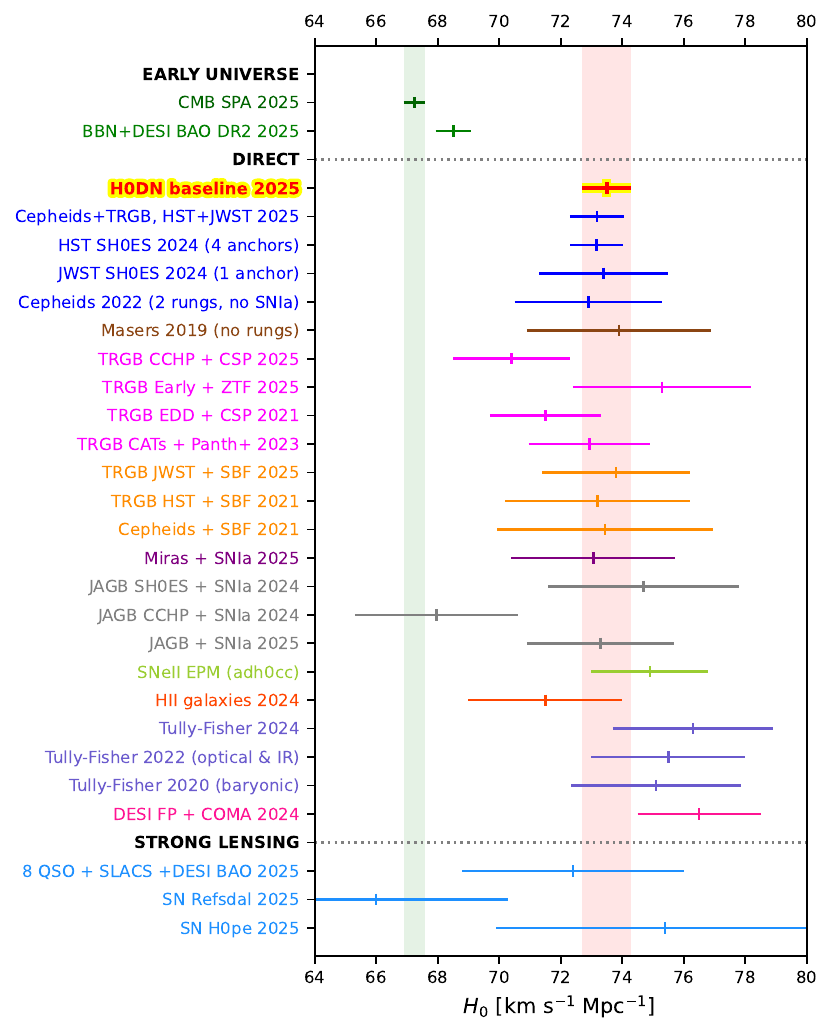}
    \caption{Comparison of the baseline result with  previous literature results, based on early universe indirect inferences using the flat $\Lambda$CDM model, based on direct measurements in the late universe, and based on strong lensing results (which involve additional modeling).
    }
    \label{fig:whiskerCMB}
\end{figure}

The relative uncertainty of this result improves by  $\gtrsim 13\%$ over updates by the SH0ES team based on cluster Cepheids \citep{2022ApJ...938...36R} and sibling SNe~Ia \citep{2023JCAP...11..046M}, and by $\sim 7\%$ compared to the most recent SH0ES update provided by \citet{2024ApJ...973...30B}, which furthermore incorporated the SMC as an additional anchor. As shown by \citet{2022ApJ...934L...7R}, integrating TRGB distances into the fit strengthens the result, but does not cause it to deviate strongly. In the baseline case, this finding extends to the inclusion of distant megamasers and the SBF measurements. More generally, we find excellent agreement with most previous literature results based on direct measurements, as we show in Fig.~\ref{fig:whiskerCMB}---including most of the studies corresponding to the individual distance ladders that can be constructed within the {\DN}. We explicitly investigate the similarities and differences with respect to previous literature results that are part of the {\DN} in Appendix~\ref{sec:comparisons}, where we also showcase what is necessary to obtain the results with lower central values in \Hcst\,. We also find great agreement with the recent TDCOSMO results from \citet{2025arXiv250603023T}, which give $\Hcst = 71.6^{+3.9}_{-3.3}\Hunit$ and are compatible at $\sim0.5 \sigma$.

The consensus \Hcst{} measurement differs from the indirect, $\Lambda$CDM-dependent, measurement based on the CMB anisotropies of $\Hvaluecmb \pm \Herrorcmb \,\Hunit$ \citep[][Eq.~(54)]{2025arXiv250620707C} at a statistical significance of $\Hsigmabase \sigma$. 

%% file: 05_variants.tex
Exploring variants to the baseline solution serves multiple purposes.  First, several available methods and data sets were not included in the baseline solution, for various reasons---primarily because of maturity of analysis or questions about uncertainty estimates. These data can still be used to highlight potential issues with the baseline solution, or point toward future research directions. In some cases, additional observations may eventually justify including these methods in the baseline. Second, it is informative to assess what happens to both the value and uncertainty in the Hubble constant when different categories of data are included or excluded in the solution. Third, some methods\hbox{---}e.g., SNe~Ia\hbox{---}can be included in several different ways; different filters, different light curve analysis, or different redshift selections. Finally, it is useful to consider a solution that includes all available, independent data and methods to assess the potential precision achievable with existing data once the remaining analysis issues are resolved. We stress that our process determined the baseline analysis as the primary solution \textit{priori to} obtaining any results on \Hcst{}. 
Variants in these categories are defined in Section~\ref{sec:variant_description}, and the results discussed in Section~\ref{sec:variant_results}.

Separately, we also consider two additional sets of consistency checks presented in Sect.\,\ref{sec:consistency_both}. One builds fully independent, ``orthogonal'' paths capable of determining \Hcst{} to verify that no single path/method dominates the \DN{}, cf. Section~\ref{sec:orthogonal_paths}.
The other employs a variety of different, albeit not independent, paths through the data to verify that their statistical properties are consistent with expectations, cf.~Section~\ref{sec:consistency_checks}. These checks serve to demonstrate the statistical consistency of the baseline solution with all possible configurations. Replications of results presented in the literature provide additional consistency checks of our methodology and are presented in Appendix~\ref{sec:comparisons}.

Some general considerations pertain to all these analyses and their interpretation. First, the baseline and variants were defined and classified {\it before} considering the resulting values of {\Hcst}. Very minor changes were introduced afterwards if required by updated data availability, albeit without changing the definition of the baseline.  
Second, most ``variants'' should not be regarded as alternative, equally valid solutions. Rather, their role is to illustrate the statistical properties of the data, identify potential problems, and/or reinforce their validity. This is especially true when considering the consistency checks, which are designed specifically so that each path uses only a small fraction of the available data, and thus has by construction a large nominal uncertainty, which is \textit{in no way} representative of the uncertainty in {\Hcst}. Third, the results of all variants are highly correlated because they share large portions of data. An exception are the orthogonal paths in Section~\ref{sec:orthogonal_paths}, which serve as consistency checks. Hence, the scatter between variants is expected to be much smaller than their error estimates. 

\subsection {Description of variants}\label{sec:variant_description}

The following is a short description of the main variants included in Table~\ref{tab:variants}. Specific sources of data products are provided in the text and in Appendix~\ref{sec:data_appendix}.

\begin{itemize}
\item[]\textbf{Baseline}
    \item[V00] Baseline: includes all measurements of SNe~Ia host distances using either Cepheids or TRGB, with the Milky Way via \gaia, the LMC, and NGC$\,$4258 as anchors, as well as SBF calibrated via TRGB and megamasers in the Hubble flow.   Note that TRGB distances are only anchored to NGC$\,$4258, since TRGB calibrations from the Milky Way and LMC do not meet our data requirements (notably photometric homogeneity across anchors and hosts in HST/JWST passbands).  Cepheids use all three anchors.  
    \smallskip\item[]\textbf{Add-one-in variants}
    \item[V01] Baseline + JAGB: includes all baseline methods, plus measurements of SNe~Ia hosts using the JAGB method, calibrated to NGC$\,$4258.
    \item[V02] Baseline + Miras: includes all baseline methods, plus measurements of SNe~Ia hosts using  Mira variables, calibrated to NGC$\,$4258.
    \item[V03] Baseline + Fundamental Plane: includes all baseline methods, plus distances to elliptical galaxies using the Fundamental Plane as measured by DESI, and calibrated to the distance of the Coma cluster measured with SBF and SNe~Ia.
    \item[V04] Baseline + empirically calibrated SNe II: includes all baseline methods and adds SNe~II (standard candle method) both as calibrators and in the Hubble Flow.  SNe~II host distances are determined via the distance network.
    \item[V05] Baseline + SNe~II with Expanding Photosphere Method: includes all baseline methods, plus SNe~II distances measured by the (astrophysical model-dependent) Expanding Photosphere Method.  Excludes a handful of supernovae that are too nearby to have significant weight on the determination of $ H_0 $.
    \item[V06] Baseline + Tully-Fisher: includes all baseline methods and adds calibrators for the Tully-Fisher relation and Tully-Fisher galaxies in the Hubble flow. The calibrators are treated in the same way as SNe Ia hosts, and their individual distances are determined from all available measurements.
    \item[V07] Baseline + SMC as anchor: includes all baseline methods and includes the SMC as an additional anchor for Cepheids.
    \smallskip\item[]\textbf{Leave-one-out variants}
    \item[V08] Baseline without Cepheids: excludes all Cepheid measurements from the distance network.  As a consequence, it only uses NGC$\,$4258 as anchor, since the Milky Way and LMC are currently only used via Cepheids. It furthermore excludes some SNe Ia for which TRGB measurements are not available.
    \item[V09] Baseline without TRGB: excludes all TRGB-based measurements.  As a consequence, it excludes some SNe Ia for which Cepheid measurements are not available, and the SBF method, which in this version it is only calibrated via TRGB.  A future version could include other primary distance indicators to calibrate the SBF method.
    \item[V10] Baseline without parallaxes: excludes parallaxes measured by {\it Gaia} and {HST}. As a consequence, the MW is removed as an anchor.
    \item[V11] Baseline without DEB distances to the LMC: removes the LMC as an anchor.  Note that the SMC is not included in the baseline.
    \item[V12] Baseline without NGC$\,$4258: this excludes all measurements based on NGC 4258 as anchor. As a consequence, all TRGB measurements are removed, since they are exclusively calibrated via NGC$\,$4258. As in V09, this also results in the exclusion of a number of SNe Ia for which no Cepheid measurements are available, as well as the SBF method.
    \item[V13]  Baseline without SNe~Ia: excludes all SNe~Ia, both as calibrators and as Hubble Flow objects.
    \item[V14]  Baseline without SBF: excludes all SBF-measured galaxies in the Hubble Flow.
    \item[V15]  Baseline without masers in the Hubble flow: retains the maser distance to NGC$\,$4258, but excludes other megamaser measurements. 
    \smallskip\item[]\textbf{Instrument-suspicious variants}
    \item[V16]  Exclude HST data: excludes {\it all} HST data from the baseline.  Since JWST anchor measurements are only available for NGC$\,$4258, this version also excludes the Milky Way and the LMC as anchors. 
    \item[V17]  Exclude JWST data: removes all JWST observations from the baseline.
    \item[V18]  Exclude SN~1994D and earlier: excludes 7 SNe~Ia calibrators that do not have modern data, namely SN 1980N, 1981B, 1981D, 1989B, 1990N, 1992A, and 1994D.
    \smallskip\item[]\textbf{Custom baselines}
    \item[V08B] Modified baseline for the no-Cepheid variant V08.  This is identical to the baseline, except that SNe~Ia for which we only have Cepheid distance estimates have been removed, leaving 35 SNe Ia calibrators.  Comparing V08 and V08B allows a direct estimate of the impact of removing Cepheids from the distance network, without the effect of changing the subset of SNe~Ia calibrators.
    \item[V09B] Modified baseline for the no-TRGB variant V09. This is identical to the baseline, except that SNe~Ia for which we only have TRGB distance estimates have been removed, leaving 42 SNe Ia calibrators. Comparing V09 and V09B allows a direct estimate of the impact of removing TRGB from the distance network, without the effect of changing the subset of SNe~Ia calibrators. However, this custom baseline includes SBF, which are excluded from V09 because they are calibrated using TRGB distances; therefore the difference between V09 and V09B is due to both the TRGB contribution to SN calibrator distances and to the omission of SBF.
    \item[V12B] Modified baseline for the no-NGC 4258 variant V12.  This is identical to the baseline, except that SNe~Ia for which we only have  distance estimates based on NGC 4258 have been removed, leaving 42 SNe Ia calibrators.  This variant allows a direct estimate of the impact of removing NGC 4258 from the distance network, without the effect of changing the subset of SNe~Ia calibrators.  Unlike V09B, this custom baseline also excludes SBF measurements;
    therefore the difference between V12 and V12B is only in the contribution of NGC 4258 to the distances to SNe Ia calibrators.  
    \item[V16B] Modified baseline for the no-HST variant V16. This is identical to the baseline, except that SNe~Ia for which we only have HST-based distance estimates have been removed, leaving 24 SNe Ia calibrators. Comparing V16 and V16B allows a direct estimate of the impact of removing HST measurements from the distance network, without the effect of changing the subset of SNe~Ia calibrators.
    \smallskip\item[]\textbf{Hypothesis-driven variants} 
    \item[V19] Modified baseline in the CMB frame: removes the peculiar velocity corrections for SNe~Ia, SBF, and megamasers in the Hubble flow.
    \item[V20] Modified baseline that limits Hubble flow SNe~Ia to $ 0.023 \leq z \leq 0.06 $.
    \item[V21] Modified baseline, with SNeIa in the Hubble flow limited to $ 0.03 \leq z \leq 0.1 $.
    \item[V22] Modified baseline, with SNeIa restricted to the CSP sample fit using SNooPy v2.7 for both calibrators and tracers. 
    \item[V23] Modified baseline with SNe~Ia measurements fitted using BayesSN for both calibrators and Hubble flow objects.
    \item[V24] Modified baseline, wherein SNe\,Ia light curves were fitted using the SALT V3 fitter instead of SALT V2.4 (as in Pantheon+)
    \item[V25] Modified baseline that exclusively uses H-band NIR measurements for both calibrators and Hubble flow SNe~Ia.
    \item[V26] Modified baseline that exclusively uses J-band NIR measurements for both calibrators and Hubble flow SNe~Ia.
    \item[V27] Modified baseline that ignores off-diagonal covariance terms for Hubble flow SNe~Ia.
    \item[V28] Modified baseline without metallicity corrections for Cepheid PL relation
    \smallskip\item[]\textbf{Include everything}
    \item[V99] ``Everything'': All data included in any variant, without repeating Hubble Flow objects; SNe~Ia exclusively from Pantheon+.
   \item[V99a] ``Everything but Tully-Fisher'': as V99, but excludes the Tully-Fisher method, which contributes to the total $ \chi^2 $ beyond expectations, and has high $ \chi^2 $ in the Hubble Flow.
\end{itemize}

\input Tables/vartable.tex

\begin{figure}
    \centering
    \includegraphics[width=\linewidth]{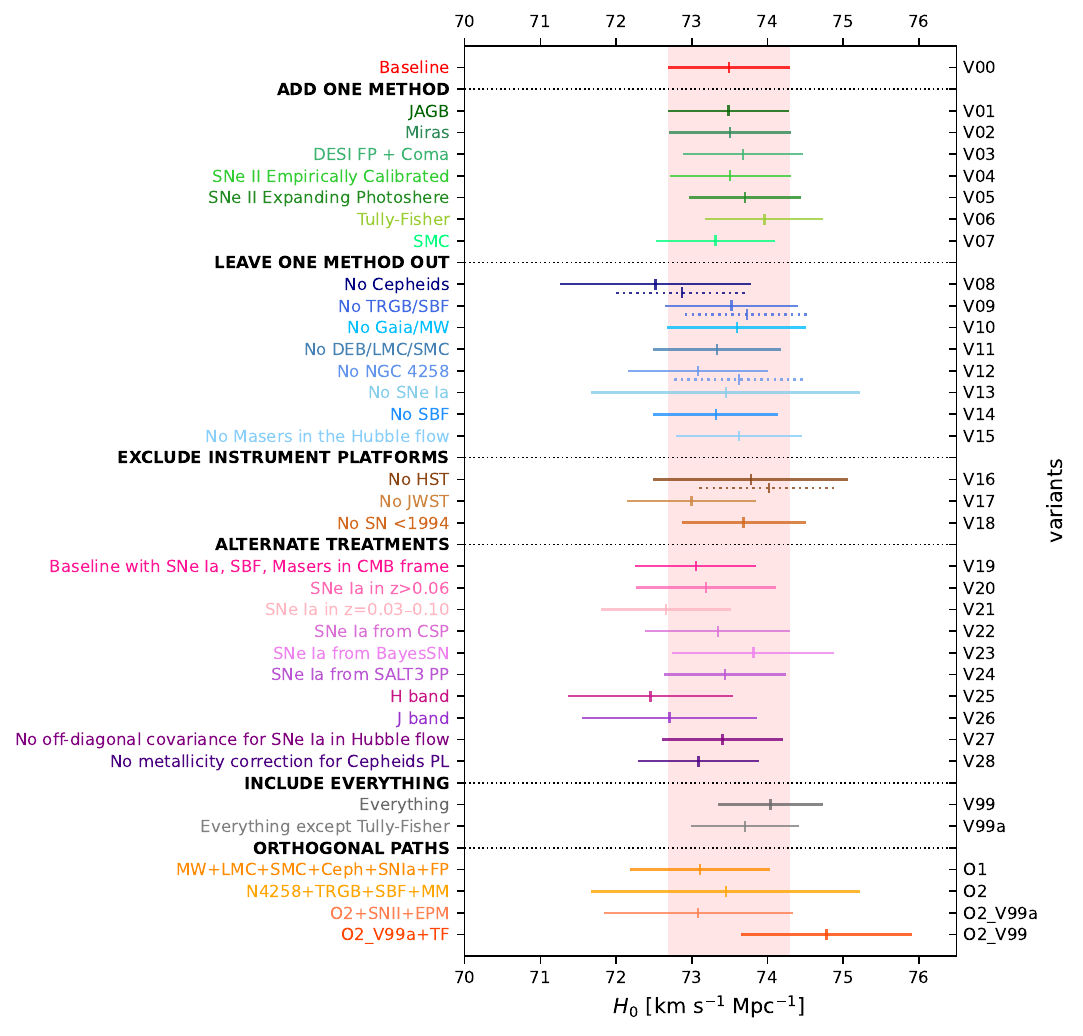}
    \caption{Baseline and variants results from Table~\ref{tab:variants}. For some variants, a labeled change has a secondary consequence (i.e., removing the identified primary distance indicator also results in the loss of calibration to some SN Ia).  In these cases a dashed line immediately below such variants provides a custom baseline--i.e., the same SNe are removed before the removal of the primary indicator--is shown to provide a custom reference of impact for that variant; see text for details.
    }
    \label{fig:whisker}
\end{figure}

\subsection{Overview of results based on variants} \label{sec:variant_results}

The results of the baseline and all variants are illustrated in Figure\ref{fig:whisker} and reported in Table~\ref{tab:variants}. All variants yield results consistent with the baseline, typically varying only slightly from the baseline result; the dispersion of central values is $0.400$\,\Hunit, or $50\%$ of the baseline's $1\sigma$ uncertainty. Since the baseline and its variants are highly correlated due to shared datasets and methodologies, a scatter significantly below $1\sigma$ is expected. The absence of outliers among the variants demonstrates that the results are not driven in any particular direction by any one dataset or methodology, or any of the hypothetical systematics explored. For example, variants V08, V10, and V13 demonstrate that the central value of {\Hcst} is not dominated by Cepheids (V08), parallaxes mainly from {\it Gaia} (V10), or SNe~Ia (V13).  
Figure~\ref{fig:whisker} also readily identifies differences in constraining power brought by various methods: the largest increases in the error on \Hcst{} are seen in V08 (no Cepheids), V13 (no SNe\,Ia), and V16 (no HST). This underlines the crucial precision provided by Cepheids, SNe\,Ia, and {HST} observations, whereas the insensitivity of the central value on \Hcst{} demonstrates that neither dominate the result.
The exploration of such variants establishes robustness of the baseline result and demonstrates that the {\DN} is enhanced by the joint consideration of complementary methodologies. It further underlines that the different methodologies reinforce and strengthen each other, resulting in improved accuracy thanks to the DN approach.  

\subsubsection {Adding one class at a time}

Variants V01 to V07 each add one category of objects or methodology to the baseline solution; these include JAGB, Miras, the Fundamental Plane of elliptical galaxies, empirically and astrophysically calibrated SNe~II, Tully-Fisher, and the SMC as an additional anchor.  For all cases except the Tully-Fisher relation, the impact on the solution is negligible, with the central value of $ H_0 $ changing by 0.2 {\Hunit} or less ($25\%$ of the $1\sigma$ uncertainty), and the uncertainty remaining constant or decreasing slightly.  Including astrophysically calibrated SNe~II (V05) leads to the most significant improvement to the uncertainty, by $8.5\%$ to $ \Herrorepm \Hunit $; however, this method being relatively new, its systematic uncertainties are not yet fully quantified \citep{2025AA...702A..41V}.  

Including the Tully-Fisher relation raises $ H_0 $ by almost 0.5 {\Hunit} ($\sim 60\%$ of the baseline error), and yields a reduced $ \chi^2 $ significantly larger than unity (\redchisqtf). The increase in $ H_0 $ is consistent with the results of the recalibration of the Tully-Fisher relation by \cite{2024arXiv241208449S}.  Inspection of the residuals shows that the anomalous value of $ \chi^2 $ is caused by the dispersion in predicted magnitude for Tully-Fisher calibrators, whose empirical scatter significantly exceeds the assumed intrinsic dispersion, particularly at the bright end. Uncertainties in host distances are subdominant and unlikely to play a role in this comparison. We conclude that the internal dispersion currently assumed for the Tully-Fisher relation is likely underestimated, and recommend avoiding the inclusion of Tully-Fisher results until its intrinsic dispersion can be reevaluated.

 \subsubsection {Leave-one-out variants\label{sec:leave-one-out}}

Variants V08-V18 each remove one class of measurements or a subset of data from the solution.  Most variants do not change the central value of {\Hcst} much, with the exception of V08 (no Cepheids), which drops it by almost 1 {\Hunit} to {\Hvaluenoceph} {\Hunit}, and V12 (no NGC 4258), which decreases it to {\Hvaluenoftfe} {\Hunit}. 
Comparing V08 and the corresponding custom baseline (V08B) clearly identifies that the shift in the no-Cepheids variant (V08) is driven by the implicit exclusion of 20 calibrator SNe\,Ia rather than by the Cepheids themselves. Analogous effects on \Hcst{} induced by subsampling SNe\,Ia calibrators have been extensively discussed in \citet{2024ApJ...977..120R}. Accounting for this effect leaves a difference of merely $0.352$\,\Hunit\ ($44\%$ of the baseline error) due to excluding Cepheids.

Some variants yield significantly increased uncertainties on \Hcst{}. Excluding Cepheids (V08) increases the uncertainty from {\Herrorbase} to {\Herrornoceph} {\Hunit}, and comparison with the uncertainty of the custom baseline V08B (\Herrorcustbasenoceph) reveals that this increase is truly driven by the exclusion of Cepheids rather than the implicitly removed 20 SNe\,Ia calibrators. 
More modest increases occur for the variants without NGC$\,$4258 as anchor, or when TRGB measurements are excluded. This is simply a reflection of the reduced constraining power caused by the exclusion of a broad class of measurements. Excluding either the Milky Way or the LMC as anchors has less impact, as each exclusion leaves two anchors for Cepheids, and does not affect the TRGB, which is only calibrated via NGC$\,$4258. The most dramatic impact is associated with the exclusion of all SNe~Ia; this increases the uncertainty to {\Herrornosnia} {\Hunit}, more than a factor of 2 above the baseline, illustrating the strong constraining power of SNe~Ia, which remain the most powerful way to measure the local value of the Hubble constant. 
When SNe Ia are excluded from the baseline, the primary constraint on {\Hcst} is based on the SBF measurement. Even in this case, the central value of {\Hcst} is changed only little, although one naturally finds a much larger uncertainty. As an illustration, the solution including all other available data but excluding SNe~Ia in the Hubble flow---formally not a variant, since it does not satisfy the criteria established during the Workshop---yields $ \Hvaluehfnosnia \pm \Herrorhfnosnia \, \Hunit $; despite the higher uncertainty, this value is still $\Hsigmahfnosnia \, \sigma $ away from the CMB value in flat $\Lambda$CDM.

The two variants V16 and V17 explore the impact of excluding all calibrator  data from either HST or JWST, testing potential broad systematics, such as calibration errors, that affect either facility. Note that all primary distance indicators as included used space-based data from at least one of the two facilities.
Excluding HST also implies excluding the Milky Way and the LMC as anchors, since we do not yet have a calibration for Cepheids in those systems based on JWST data.  The solutions with either no HST or no JWST data are not fully disjoint since they share a common anchor, NGC 4258, and many of the Hubble Flow tracers, and since the set of calibrators overlap, carrying with them any intrinsic variance in their properties.  However, many of the sources of uncertainty in distance estimates are different and independent. The resulting {\Hcst} values bracket the baseline, as could be expected, with a difference in line with their respective uncertainties: without HST, $ \Hvaluenohst \pm \Herrornohst \, \Hunit $; and without JWST, $ \Hvaluenojwst \pm \Herrornojwst \, \Hunit $.  Not surprisingly, the uncertainty in the result increases more if HST is excluded, since HST results use three anchors (only NGC 4258 is available for JWST) and more HST observations are available, yet both subsets remain close to the baseline with a difference below $ 1 \, \Hunit $. The fact that the inclusion of JWST observations favors a higher value of \Hcst{} corroborates the conclusion based on host-to-host distance comparisons by \citet{2024ApJ...977..120R} that \Hcst{} measurements based purely on HST \citep{2022ApJ...934L...7R} are not biased high by the limited spatial resolution of WFC3/IR (crowding).

\subsubsection {Alternate treatments\label{sec:alternative_medicine}}

Variants V19-V26 show what happens if subsets of data are either excluded or treated differently. For variants V19-V22, we modify the handling of peculiar velocities or redshift selection. In variant V19, we assume that the observed velocities directly represent cosmological redshifts, i.e., we set all peculiar velocity corrections, which are by default derived with the 2M++ model \citep{2015MNRAS.450..317C}, to zero.  The value of {\Hcst} decreases slightly, to $ {\Hvalueusecmb} \pm {\Herrorusecmb} \, \Hunit $, with no meaningful change in the accuracy.
Similarly, restricting the redshift range, either by removing the high redshift end (V20, $ z > 0.06 $) or the low-redshift end (V21, $ 0.3 < z < 1 $), changes the value of {\Hcst} by up to 0.8\,{\Hunit}; the impact on the accuracy is modest.

Variants V22-V27 are noteworthy because they explore different treatments of SNe\,Ia, which drive overall precision (cf. Sect.\,\ref{sec:leave-one-out}). 
Variants V22 and V23 consider different light curve fitters (consistently for both calibrators and tracers), namely SNooPy v2.7 in V22 and BayesSN in V23. Both variants yield slightly higher \Hcst{} values with slightly larger nominal uncertainties. The BayesSN variant V23 furthermore yields a higher reduced $ \chi^2 $. Further detail and comparisons with published articles is provided in App.\,\ref{app:sec:SNIa_literature}, specifically concerning SNooPy in App.\,\ref{app:snoopy}.

Variants V25 and V26 rely exclusively on NIR measurements for SNe~Ia \citep{2023AA...679A..95G}, in the $ H $ and $ J $ band respectively. These also yield slightly lower, yet consistent \Hcst{} results with slightly larger uncertainties due to the smaller number of calibrators and Hubble flow tracers.  Note that we did \textit{not} use SNe~Ia measurements from multiple sources (e.g., Pantheon+ and NIR) simultaneously in any solution; doing so would lead to unreliable, likely incorrect results.  First, the calibration of SN measurements in different samples are inconsistent, as they are based on different filters, fitting processes, and conventions.  More importantly, different measurements of the same SN would share to a large extent the astrophysical variance of the source; therefore we expect such measurements to be significantly correlated in their deviation from the mean, to an extent that; to the best of our knowlegde, has not been sufficiently quantified.  Since the SNe~Ia samples overlap significantly, multiple measurements cannot be included in the same solution in a statistically satisfactory way.  To avoid such issues, we take care not to include multiple measurements of the same SNe~Ia---whether as local calibrators or as Hubble flow objects---in the same solution. Unless otherwise stated, all solutions exclusively consider measurements collected in the Pantheon+ system \citep{2022ApJ...938..113S}.

Another treatment option concerns the covariance between different SNe~Ia in the Hubble flow. Several collections of measurements provide covariances between different objects; these can be due to the effect of standardization parameters, or the survey from which data have been obtained. By default, we include such covariances where available. However, Variant V27 deliberately ignores these covariances and assumes (incorrectly) the stated uncertainties to be independent in order to assess the impact that the SNe\,Ia covariances have on the final result. This results in a miniscule decrease in {\Hcst}, by less than 0.1\,{\Hunit}, and analogously miniscule decrease in the nominal uncertainty. We conclude that covariances between SNe~Ia measurements in the Hubble flow currently have a negligible impact on \Hcst. 

\subsubsection {``Everything'' solution\label{sec:everything_bagel}}

It is naturally interesting to consider the result of including all available methods into the \DN{}. To this end, we constructed variant V99 by including all available measurements.
As noted above, multiple measurements of SNe~Ia cannot be combined in a statistically satisfactory way, so that V99 exclusively considered Pantheon+ SNeIa. V99 also incorporates methods with as yet insufficiently well understood systematics, such as SN~II with EPM and the Tully-Fisher relation. 
These results are provided here for completeness, and we recommend that they \textit{not} be used for further analysis.  The resulting value is $ \Hvalueeverything \pm \Herroreverything \, \Hunit $, which lies $ \Hsigmaeverything \, \sigma $ from the Planck+$\Lambda$CDM solution.  We recognize that this solution may not be fully reliable, as indicated by the relatively large value of reduced $ \chi^2 $, namely $\redchisqeverything $ per degree of freedom; indeed, this solution deviates from the baseline more than several others, primarily because of the TF contribution.  Since the TF relation contributes the most to the excess $\chi^2 $, we also include variant V99a, which is identical to V99 except for the exclusion of TF calibrators and tracers.  This variant results in $ \Hcst = \Hvalueeverythingnotf \pm \Herroreverythingnotf \Hunit $, very close to the baseline with a 12\% smaller uncertainty (and sub-percent precision on \Hcst), has a reasonable reduced $ \chi^2 $ of {\redchisqeverythingnotf}, and lies $ \Hsigmaeverythingnotf \, \sigma $ from the Plank+$\Lambda$CDM solution. 

\begin{figure*}
    \centering
    \includegraphics[width=0.9\linewidth]{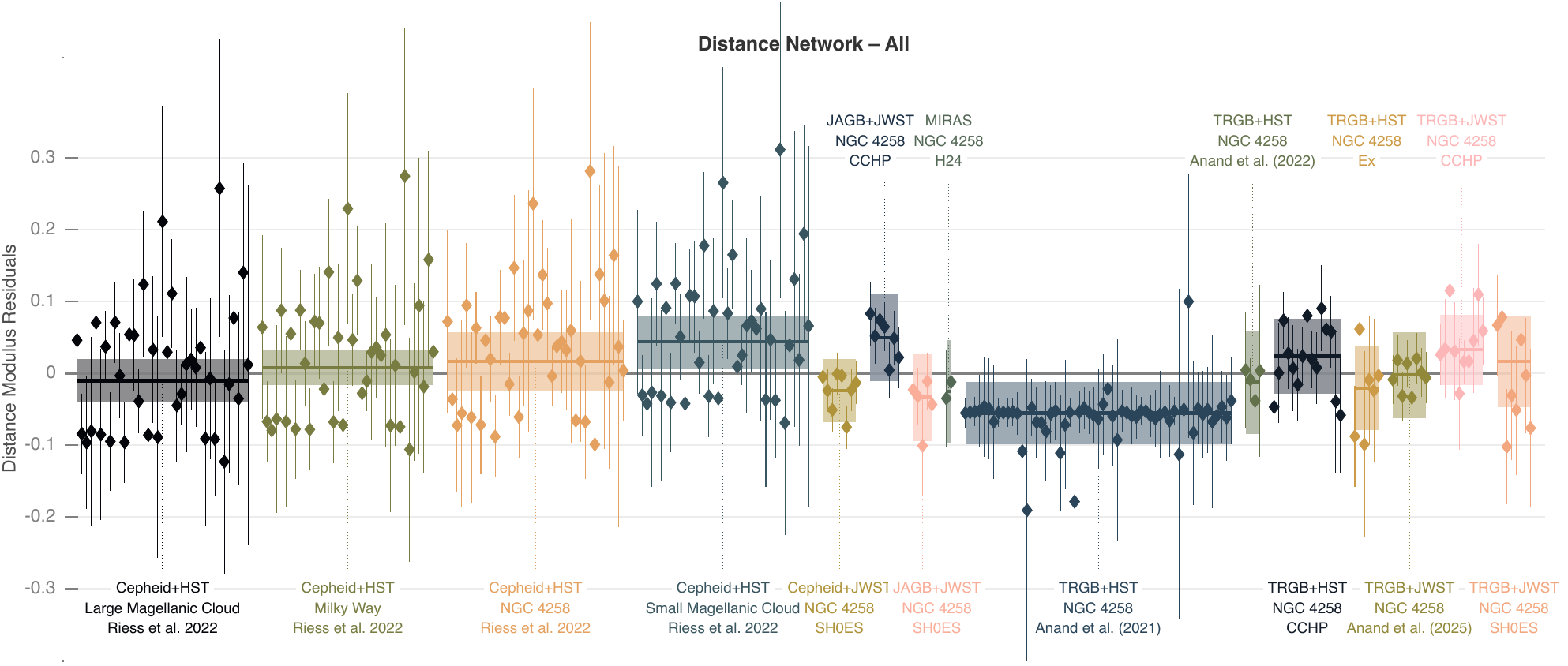}
    \caption{Residuals for host and calibrator distances for the ``everything'' solution.  The groupings, also distinguished by color, correspond to calibrators measured by the same authors with common method and anchor.  The error bars reflect the individual calibrator errors; the shaded areas for each group indicate the combined uncertainty in the anchor distance and its reference value, which are in common for all calibrators in the same grouping.}
    \label{fig:everything_residuals}
\end{figure*}

\begin{figure*}
    \centering
    \includegraphics[width=0.9\linewidth]{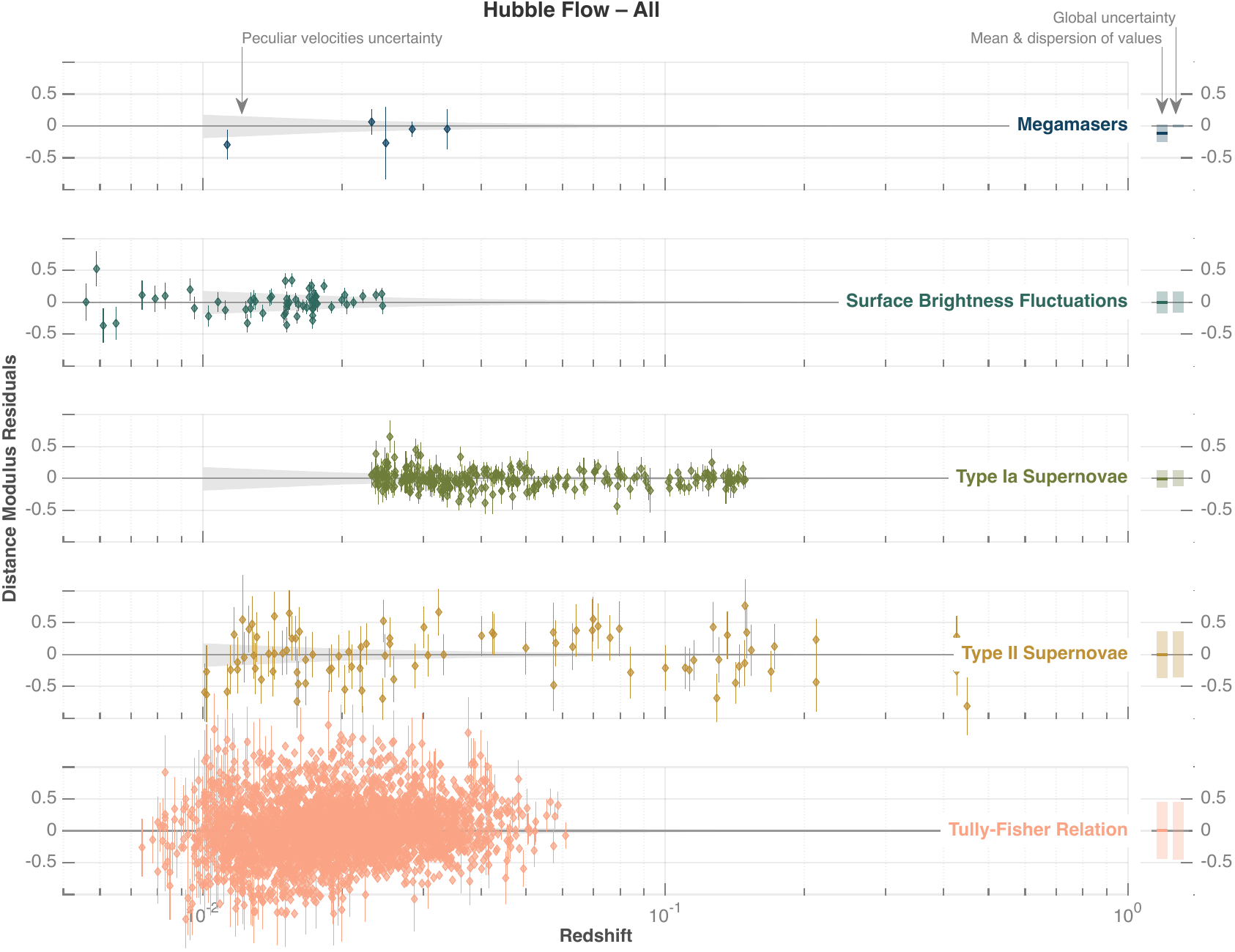}
    \caption{Residuals in $ \log(H_0) $ vs.\ redshift for objects in the Hubble Flow for the ``everything'' solution.  Each subplot refers to sources in the same group, measured by the same method.  Error bars for individual sources are representative of the expected scatter across sources in the same category, and do not include calibration uncertainties for each class.  The shaded bars at the right show the mean and dispersion for each category.}
    \label{fig:everything_hf}
\end{figure*}

\begin{figure}
    \centering
    \includegraphics[width=0.9\linewidth]{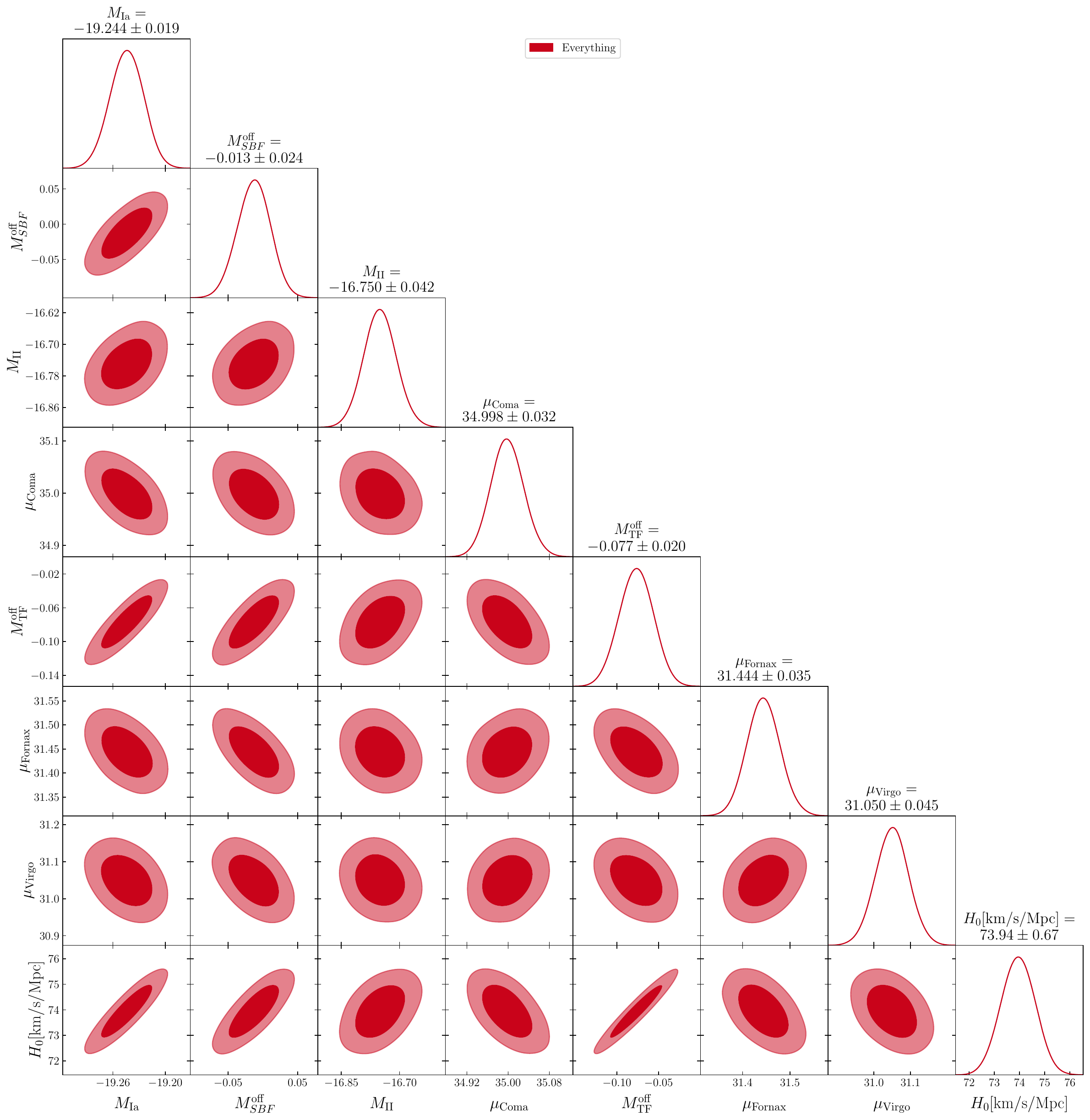}
    \caption{Corner plot illustrating the main optimization parameters for the all-inclusive solution, using the naming convention of Appendix~\ref{sec:optimization_parameters}.  Parameters related to individual host distances are not included. Note that this solution is not an alternative result to the baseline solution, but rather a variant to check the impact of further data on our result, see also Section~\ref{sec:variants}. We show it here to highlight the existing correlations between the various (most important) solution parameters. Deviations from the analytically calculated result are due to the numerical precision of a finite-length chain.}
    \label{fig:everything_corner}
\end{figure}

The everything solution is useful to understanding how elements in the baseline solution compare to the elements not included therein, as well as any covariances among them. Three figures illustrate these points. Figure\,\ref{fig:everything_residuals} shows the fit residuals of host and calibrator distances for V99 in analogy with Fig.\,\ref{fig:baseline_residuals} for the baseline. Figure\,\ref{fig:everything_hf} illustrates the corresponding Hubble flow residuals for V99. A corner plot illustrating covariance among the main global fit parameters of the \DN{} for V99 is shown in Fig.\,\ref{fig:everything_corner}. 

\subsection{Consistency Checks\label{sec:consistency_both}}
We demonstrated the statistical consistency of the baseline solution by building alternate paths through the \DN{} capable of determining \Hcst{}. Statistically independent paths are discussed in Sect.\,\ref{sec:orthogonal_paths}, statistically not independent paths in Sect.\,\ref{sec:consistency_checks}. Replications of \Hcst{} results from the literature are presented in Appendix~\ref{sec:comparisons}.

\subsubsection{Orthogonal paths}\label{sec:orthogonal_paths}

\begin{figure}
    \centering
    \includegraphics[width=0.79\linewidth]{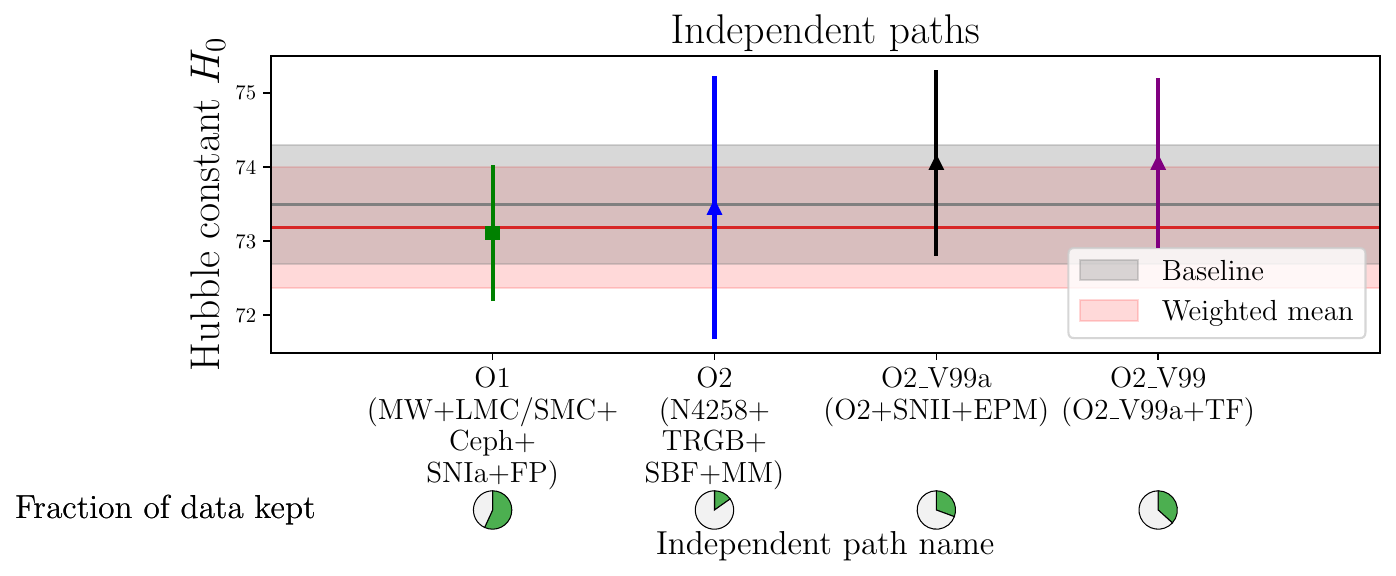}
    \caption{The two paths starting with ``O'' are proper orthogonal paths and can be taken as independent measurements of the Hubble constant. Their weighted mean (red) agrees well with our baseline result despite the latter including all cross-correlations. The shortcuts of the data sets are as defined in the text. The pie charts  ``fraction of data kept'' represent the fraction of statistical power remaining in the path. It is computed as the fraction of the variance of {\Hcst} for the given result (in this way multiple datasets with the same constraining power in $H_0$ will be assigned equal percentages). We show the percentages relative to the baseline result (V00).}
    \label{fig:paths_orth}
\end{figure}

Orthogonal paths strike statistically fully independent paths across the \DN{}. In creating these paths, we sought to identify configurations capable of achieving similar constraining power.
We identified the following two\footnote{In principle, more orthogonal paths could be formulated (e.g., by separating out specific one-step methods), but such combinations of paths are typically not balanced in their constraining power.} orthogonal paths and two extensions of the latter:
\begin{itemize}
    \item \textbf{O1:(MW+LMC+SMC+Ceph+SNIa+FP)} employs the Magellanic Clouds and MW parallaxes as anchors, calibrates SNeIa using Cepheids, and adds in the FP calibrated in Coma. This yields $\Hcst = 73.11 \pm 0.92 \Hunit$ with a $\chi^2=53.5$ and a reduced $\chi^2$ of 0.922.
    \item \textbf{O2:(N4258+TRGB+SBF+MM)} Uses N4258 as anchor, TRGB as calibrators, and SBF as tracers, in addition to megamasers. This corresponds to a modified V07 (baseline+SMC) from which all measurements correlated with O1 have been removed. This yields $\Hcst = 73.45 \pm 1.78 \Hunit$, with a $\chi^2=11.2$, and a reduced  $\chi^2$ of 0.4853. The low reduced $\chi^2$ indicates that the measurement uncertainties used are overestimated.
    \item \textbf{O2\_V99a:(O2+SNII+EPM)} Modified O2 to which both astrophysically and empirically calibrated SNe\,II have been added, i.e., corresponds to V99a without elements in O1. We find $\Hcst = 74.06 \pm 1.25 \Hunit$, with a $\chi^2=28.76$, and a reduced  $\chi^2$ of 0.5752. Here too the reduced $\chi^2$ is a little low.
    \item \textbf{O2\_V99:(O2\_V99a+TF)} Modified O2 with Tully-Fisher relation added, i.e., corresponds to V99 without the elements in O1. We find $\Hcst = 74.93 \pm 1.14 \Hunit$, with a $\chi^2=186.6631$, and a reduced  $\chi^2$ of 1.4359. The very significant increase in reduced $\chi^2$ emphasizes the underestimated uncertainties of the Tully-Fisher results, cf. V99.
\end{itemize} 
Fig.\,\ref{fig:paths_orth} illustrates the results of these orthogonal paths compared to the baseline result. Both O1 and O2 are highly compatible with each other, deviating by only $0.17\sigma$ despite representing fully independent measurements. As previously shown in the context of V13, the significantly higher uncertainty of O2 is due to the exclusion of SNe\,Ia. 
The orthogonal nature of O1 and O2 allows to compute their weighted average, $H_0^\mathrm{O1+O2} = (73.18\pm 0.82)\Hunit$, which agrees to within $0.3\,$\Hunit{} of the baseline. The slightly offset central value and reduced precision arise due to fewer internal cross-calibrations---their combination is missing for example any N4258-anchored calibrations of SNe~Ia. This underlines a key advantage of the covariance-combined approach taken by the H0DN collaboration compared to independent pursuits of \Hcst{}\,.

The two extensions to O2 considered, O2\_V99a and VO2\_V99, add elements from the ``everything'' variants (V99a, V99) to improve precision. Both extensions yield \Hcst{} values marginally higher than O2 and fully compatible with the respective ``everything'' variants V99 and V99a. The weighted average $H_0^\mathrm{O1+O2\_V99a} = (73.45\pm 0.74)\Hunit$ agrees to within $< 0.5\sigma$ with V99a and features a very similar uncertainty. Finally, the weighted average including the Tully-Fisher relation, $H_0^\mathrm{O1+O2\_V99} = (73.77\pm 0.71)\Hunit$, agrees to within $0.4$\,\Hunit{} with V99, with also very similar uncertainty.

Overall these tests confirm both the reliability of the main results as well as the consistent setup of the network. We note that in the case in which the baseline is split into two orthogonal paths we have a mild domination of the uncertainty by the SNe~Ia, whereas in the extended cases the uncertainties are relatively balanced.

\subsubsection{Paths based on subsets of data} \label{sec:consistency_checks}

\begin{figure}
    \centering
    \includegraphics[width=0.79\linewidth]{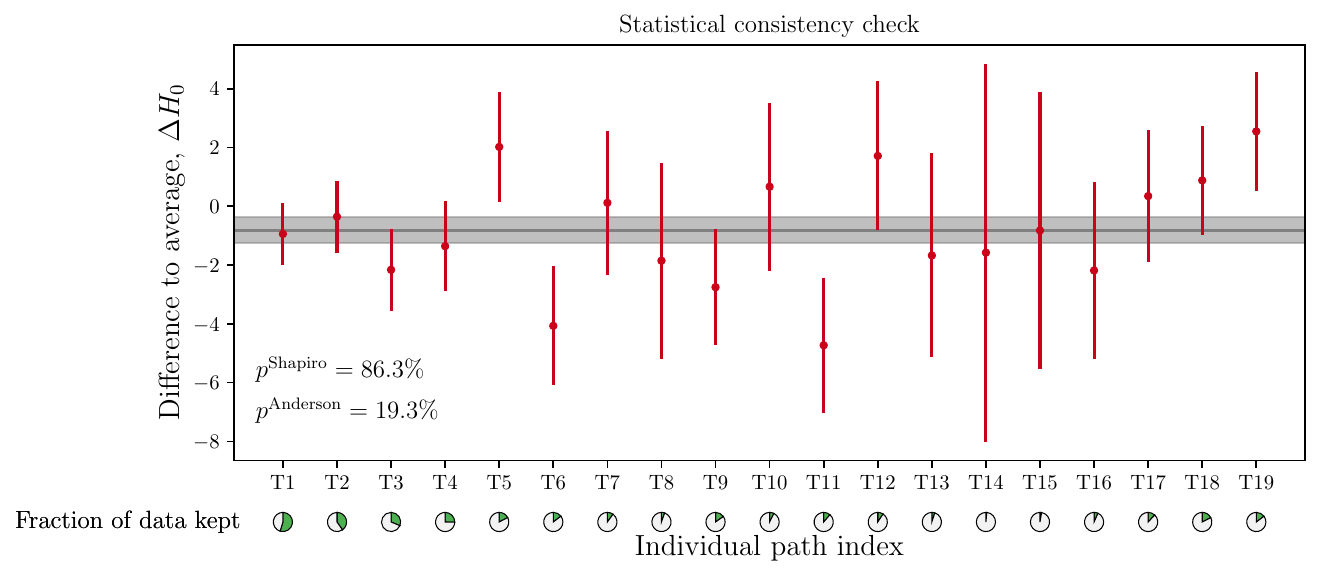} 
    \caption{The statistical consistency of various individual paths through the network. These represent a set of paths that are highly correlated, and often only take a tiny part of the network into account. They are used for a statistical consistency check and should not be interpreted as valid variants in the sense of section \ref{sec:select_variants}. We also show the simple weighted mean of these paths in grey purely for visual purposes, reminding that due to their correlations (and selection effects) it need not necessarily agree with the baseline result. For details on the pie charts, see Fig.~\ref{fig:paths_orth}.}
    \label{fig:paths_consistency}
\end{figure}

In order to verify the internal consistency of different data sets and methods relevant to determining the Hubble constant, 
we show below the $H_0$ values obtained for a set of ``paths'' through the network, each defined by a unique combination of anchors, calibrators, and tracers.\footnote{In practice we construct these paths as minimal combinations, except where this might lead to ambiguities. First we use each of the possible calibrator sets (e.g. Cepheids measured with HST, or Miras, or JAGB, etc.) together with the SNe~Ia as Hubble flow tracers (13 combinations), then using the SNe~II (either just with the minimal calibrator set from \citet{2020MNRAS.496.3402D} or with all available calibrators) (2 combinations), the megamasers alone, the SBF anchored using any hosts anchored by NGC$\,$4258, and the SNII tailored~EPM method (each 1 path), as well as the FP method anchored only by the SBF value (to remain minimal). In total that makes 19 paths through the network.} These paths are not all statistically independent by definition, since they might share, for example, a common anchor or a common set of tracer objects. However, we expect the corresponding distribution of $H_0$ values to align at least as well as for independent variables, since we expect the shared data to produce positive, not negative, correlations. In this case, the corresponding test statistics can be regarded as lower bounds on the true values that would be obtained if the full correlation structure were accounted for (i.e., in plain English: ``if the points sharing the same data are incompatible, that's even worse than points not sharing the same data being incompatible''). We find that the Anderson-Darling test \textit{cannot} reject the null hypothesis of the paths stemming from a Gaussian distribution at 5\% confidence level, and we find $p=86.3\%$ for the Shapiro test, both very well within acceptable bounds. We also find no visually strong $(>2\sigma)$ outliers among the paths shown in Figure~\ref{fig:paths_consistency}. Note that since these paths are constructed always using minimal paths through the network, they explicitly do not qualify as `variants' in the sense defined above. We also note that if we were to include the Tully-Fisher case in the generation of these paths, the results worsen quite a bit, to a $p \simeq 10\%$ for the Anderson-Darling test and a $p\simeq 5\%$ for the Shapiro test, showing once again the high discrepancy of the Tully-Fisher based results within the {\DN}.

%% file: Tables/vartable.tex
\newcommand{\ptt}[1]{\parbox[t]{4cm}{\raggedright #1}}
\begin{table*}
\caption{$ H_0 $ results based on the baseline, variants, and orthogonal paths \label{tab:variants}}
    \begin{center}
        \small \begin{tabular}{lccrrrl}
\hline\hline
\# & $ H_0 $ & 1--$\sigma$ &  \multicolumn{1}{c}{$ \chi^2 $} & $ N_{\rm dof} $ & Reduced & 
Description \\
           & \multicolumn{2}{c}{ \Hunit } & & &\multicolumn{1}{c}{$\chi^2$} \\
\hline
  V00 & 73.499 & 0.809 & 117.5597 & 119 & 0.98790 &   Baseline  \\
  V01 & 73.493 & 0.807 & 125.3081 & 131 & 0.95655 &   Baseline + JAGB \\
  V02 & 73.508 & 0.809 & 117.7560 & 121 & 0.97319 &   Baseline + Miras \\
  V03 & 73.604 & 0.798 & 129.4831 & 133 & 0.97356 &   Baseline + DESI FP calibrated to Coma \\
  V04 & 73.515 & 0.805 & 126.9616 & 129 & 0.98420 &   Baseline + empirically calibrated SNe II \\
  V05 & 73.720 & 0.740 & 123.0193 & 133 & 0.92496 &   Baseline + SNe II with Expanding Photosphere method \\
  V06 & 73.957 & 0.787 & 285.4146 & 194 & 1.47121 &   Baseline + Tully-Fisher \\
  V07 & 73.314 & 0.789 & 118.8900 & 120 & 0.99075 &   Baseline + SMC \\
  V08 & 72.509 & 1.296 &  65.8536 &  69 & 0.95440 &   Baseline without Cepheids \\
 V08B & 72.873 & 0.874 &  95.6518 &  98 & 0.97604 &   Custom baseline for V08 \\
  V09 & 73.526 & 0.878 &  55.2382 &  56 & 0.98640 &   Baseline without TRGB, SBF \\
 V09B & 73.738 & 0.827 &  94.2652 & 106 & 0.88929 &   Custom baseline for V09 \\
  V10 & 73.609 & 0.929 & 117.2767 & 118 & 0.99387 &   Baseline without Gaia parallaxes, MW \\
  V11 & 73.337 & 0.853 & 116.9115 & 118 & 0.99078 &   Baseline without DEB, LMC, SMC \\
  V12 & 73.084 & 0.920 &  41.7194 &  47 & 0.88765 &   Baseline without NGC 4258 \\
 V12B & 73.632 & 0.861 &  84.8933 &  87 & 0.97578 &   Custom baseline for V12 \\
  V13 & 73.434 & 1.795 &  11.1578 &  23 & 0.48512 &   Baseline without SNe Ia \\
  V14 & 73.325 & 0.830 & 106.9393 & 100 & 1.06939 &   Baseline without SBF \\
  V15 & 73.633 & 0.841 & 115.6990 & 114 & 1.01490 &   Baseline without masers in the Hubble flow \\
  V16 & 73.754 & 1.325 &  56.6229 &  55 & 1.02951 &   Exclude HST data, LMC, MW \\
 V16B & 74.024 & 0.924 &  83.3998 &  85 & 0.98117 &   Custom baseline for V16 \\
  V17 & 73.002 & 0.856 &  81.5863 &  85 & 0.95984 &   Exclude JWST data \\
  V18 & 73.690 & 0.822 & 105.7303 & 108 & 0.97898 &   Exclude SN 1994D and earlier \\
  V19 & 73.062 & 0.804 & 117.2456 & 119 & 0.98526 &   Baseline with SNe Ia, SBF, Masers in CMB frame \\
  V20 & 73.190 & 0.862 & 106.8923 & 100 & 1.06892 &   Hubble flow SNe Ia with z $>$ 0.06 (avoid local LSS) \\
  V21 & 72.667 & 0.862 & 106.7433 & 100 & 1.06743 &   SNe Ia in redshift range 0.03--0.10  \\
  V22 & 73.350 & 0.963 &  94.4077 & 119 & 0.79334 &   SNe Ia from CSP, all 55 calibrators \\
  V23 & 73.821 & 1.072 &  98.8529 & 114 & 0.86713 &   SNe Ia from BayesSN \\
  V24 & 73.445 & 0.810 & 109.0825 & 119 & 0.91666 &   SNe Ia from PP processed with SALT3 fitter \\
  V25 & 72.459 & 1.099 &  52.9059 &  64 & 0.82665 &   SNe Ia in H-band \\
  V26 & 72.712 & 1.164 &  47.1329 &  67 & 0.70348 &   SNe Ia in J-band \\
  V27 & 73.411 & 0.800 & 117.6522 & 119 & 0.98867 &   Ignore off-diagonal covariance for SNe Ia in Hubble flow \\
  V28 & 73.080 & 0.801 & 135.0255 & 119 & 1.13467 &   No metallicity correction for Cepheid PL \\
  V99 & 73.994 & 0.697 & 321.7222 & 246 & 1.30781 &   Everything available \\
 V99A & 73.656 & 0.713 & 153.8165 & 172 & 0.89428 &   Everything available except TF \\
\hline
 O1 & 73.110 & 0.920 & 53.3538 & 57 & 0.9360 & % -19.265 & 0.7164 & 42 & 
                                  Orthogonal path 1 MW+LMC/SMC+
Ceph+
SNIa+FP \\ 
 O2 & 73.451 & 1.777 & 11.1628 & 23 & 0.4853 & % -- & -- & 0 & 
                                  Orthogonal path 2 N4258+
TRGB+
SBF+MM (V00 equivalent) \\ 
 O2\_V99a & 74.083 & 1.249 & 33.6889 & 27 & 0.6016 & % -- & -- & 0 & 
                                  Orthogonal path 2 O2+SNII+EPM (V99a equivalent) \\
 O2\_V99 & 74.780 & 1.133 & 220.6600 & 151 & 1.4613 & % -- & -- & 0 & 
                                  Orthogonal path 2 O2+SNII+EPM+TF(V99 equivalent) \\
\hline\hline
        \end{tabular}
    \end{center}
\end{table*}

%% file: 06_discussion_conclusions.tex
The Hubble ``tension'' has persisted for more than a decade and significantly shifted focus amongst endeavors to precisely measure \Hcst. Before $\sim$2012, the primary motivation for measuring $H_0$ with increasing precision was to provide a strong prior for constraining $\Lambda$CDM parameters and understanding dark energy \citep[e.g.,][]{Suyu:2012}. As the statistical significance of the discrepancy between early- and late-Universe measurements grew, relentless attention was given to systematics of local distance measurements with an implicit expectation that the tension would be resolved once the responsible systematic error was identified. In other words, there was a non-negligible preference among many to discard direct measurements (even ones with small and detailed uncertainties) based on astronomical techniques in favor of the \Hcst\ value derived from the perceived ``simple physics'' of the early Universe and a specific cosmological model. A narrative of merely two or three direct measurements disagreeing with one extremely precise prediction based on the {\it Planck} CMB and flat $\Lambda$CDM lent credence to this view, despite a plethora of semi-(in)dependent measurements painting similar pictures \citep[e.g.,][and references therein]{CosmoVerse:2025txj}. A statistically rigorous combination of the available measurements was thus needed to resolve whether any one method might be deemed `bad' or unreliable. The present work provides this long overdue combination supported by expert knowledge from all relevant disciplines, and the results presented discredit the expectation of `unknown unknowns' capable of misleading the direct measurement of the local expansion rate using astronomical techniques.

The {\DN} framework introduced here establishes a reproducible and extensible methodology for synthesizing local distance constraints. Its transparency and modularity allow for the straightforward integration of new data, methods, or calibrators, enabling the community to track the evolution of $H_0$ estimates over time. In addition to the primary distance ladder, the network already incorporates several fully independent paths to $H_0$, including the maser-based anchor, the TRGB–SBF route, and SNe~II distances based on astrophysical modeling. These methods rely on distinct physical assumptions and observational systematics, yet yield results consistent with the primary solution, reinforcing the robustness of the empirical determination. While current uncertainties for these paths are larger, their future competitiveness will depend on targeted improvements in calibration, modeling, and sample size. Promising opportunities include searches for SNe~II in SBF or SNe~Ia host galaxies, extending TRGB and Cepheid measurements deeper into the Hubble flow using ELTs, denser tiling of NGC~4258 to refine JAGB and TRGB calibrations, and establishing additional anchors via geometric methods \citep[e.g., using population-II pulsating stars calibrated by parallax, e.g.,][]{2025arXiv250916331L}. Strengthening these independent routes and developing new anchors will be critical for validating and advancing precision cosmology. The approach presented here serves both as a benchmark of current capabilities and as a foundation for future consensus efforts in the determination of $H_0$.

The {\DN} presented in this work provides the most precise direct measurement to date of the Hubble constant, $H_0 = \Hvaluebase \pm \Herrorbase \, \Hunit$, achieving a relative uncertainty of {\Hrelerrorbase}. This value is the result of a comprehensive, covariance-weighted combination of local distance measurements incorporating multiple distance indicators, each rigorously assessed for reliability, consistency, and methodological transparency. By explicitly accounting for shared uncertainties and correlations, the {\DN} formalism surpasses the precision of previous individual or pairwise combinations and reflects a collaborative consensus on data selection, treatment, and analysis strategy.

In constructing the distance network, certain methodological deviations from state-of-the-art analyses were necessary to achieve full covariance treatment across indicators. For example, the use of supernova intercepts for calibration simplifies network integration but does not incorporate the full cross-rung covariance present in datasets such as Pantheon+, as done in \citet{2022ApJ...934L...7R}. These tradeoffs are explicitly documented to ensure transparency and reproducibility. We are publicly releasing the full {\DN} code upon acceptance of this paper at \url{https://github.com/StefCas789/H0DN}. We are planning to additionally publish the code in the Astrophysics Source Code Library\footnote{\url{https://ascl.net}} -- the corresponding DOI will be updated and linked here in a future version of the paper.

A central feature of our approach is the construction of a baseline solution exclusively based on top-ranked calibrators and tracers, as determined by agreement among all authors as representatives of their respective communities. This was complemented by a suite of predefined variants, decided prior to conducting the analysis, which explore the sensitivity of the result to alternative assumptions, datasets, or methodological choices. The resulting spread in $H_0$ values across all variants is significantly smaller than the statistical uncertainty of any individual variant, confirming that no single dataset or technique dominates the result and that the network solution is not driven by any particular methodology or systematic effect. This is also confirmed by investigating routes through the network that are designed to not have any elements in common, leading to results that are highly consistent with the baseline analysis.

Further scrutiny of systematics affecting {\DN} elements is both necessary and useful for further improvements to {\Hcst} accuracy. However, reconciling the {\DN}-based direct {\Hcst} measurement with the expectation from $\Lambda$CDM would require an extremely unlikely alignment of systematics conspiring to overestimate {\Hcst}. The diversified portfolio of methods and datasets incorporated into the {\DN} effectively protects against significant bias and demonstrates that the Hubble tension cannot be eliminated by invoking as-yet unidentified systematics in any single method or dataset.

Our result differs by more than $6\sigma$ from the value predicted by the $\Lambda$CDM model calibrated to \textit{Planck} CMB observations, by more than $7\sigma$ from that calibrated to SPT+ACT CMB measurements, which are given as  $\Hcst = \Hvaluecmb \pm \Herrorcmb \Hunit $ from Planck+SPT+ACT \citep[Eq.~(54)~of][]{2025arXiv250620707C}, and by about $5\sigma$ from the combination of BBN and DESI BAO constraints, which are $\Hcst = 68.51 \pm 0.58 \Hunit$   \citep[Tab.~V~of][]{DESI:2025zgx}.

Since the {\DN} provides a conclusive direct \Hcst\ measurement with demonstrated control of systematics, the resulting significant discrepancy retires the aforementioned expectation that identifying a simple systematic could resolve the Hubble constant tension. 
In this vein, we advocate that the empirically determined value of $H_0$ from the {\DN} be used as a prior in cosmological analyses. 
This marks a significant shift in perspective: rather than serving solely to constrain dark energy models, as envisioned a decade ago, the improved accuracy of $H_0$ now exposes a broader inconsistency within the standard cosmological framework and strengthens the case for new physics or a deeper reassessment of early-Universe inferences. The evolving role of $H_0$ has already reshaped our understanding of precision cosmology, and further surprises may lie ahead.

%% file: 08_acknowledgments.tex
\begin{acknowledgements}

The authors would like to express their sincere gratitude to the ISSI for their hospitality and organization, and especially for a critical role in enabling the scientific discussion and interactions that made this work possible.  We are grateful to Fabio Crameri for assistance with visualizations (Figs.~4, 5, 8, and 9).

This research has received support from the European Research Council (ERC) under the European Union's Horizon 2020 research and innovation programme (Grant Agreement No. 947660). RIA is funded by the Swiss National Science Foundation through an Eccellenza Professorial Fellowship (award PCEFP2\_194638).
The research leading to these results has received funding from the European Research Council (ERC) under the European Union's Horizon 2020 research and innovation program (project UniverScale, grant agreement 951549).
PK acknowledges support from the Polish-French Marie Sk{\l}odowska-Curie and Pierre Curie Science Prize awarded by the Foundation for Polish Science.
PK acknowledges financial support from the French Agence Nationale de la Recherche (ANR), under grant ANR-23-CE31-0009-01 (Unlock-pfactor).
AB thanks the funding from the Anusandhan National Research Foundation (ANRF), Government of India, under the Prime Minister Early Career Research Grant scheme (ANRF/ECRG/2024/000675/PMS). 
LV and HGM acknowledge support of Spanish MINECO under project PID2022-141125NB-I00 MCIN/AEI, and  Center of Excellence Maria de Maeztu Grant CEX2024-001451-M funded by MICIU/AEI/10.13039/501100011033.
LG acknowledges financial support from AGAUR, CSIC, MCIN and AEI 10.13039/501100011033 under projects PID2023-151307NB-I00, PIE 20215AT016, CEX2020-001058-M, ILINK23001, COOPB2304, and 2021-SGR-01270.
EDV is supported by a Royal Society Dorothy Hodgkin Research Fellowship. NS acknowledges support from the Excellence Cluster ORIGINS which is funded by the Deutsche Forschungsgemeinschaft (DFG, German Research Foundation) under Germany’s Excellence
Strategy - EXC-2094 - 390783311, as well as the funding through a Fraunhofer-Schwarzschild Fellowship at the LMU. 
MC acknowledges support from ASI–INAF grant no. 2024-10-HH.0 (WP8420), the ESO Scientific Visitor Programme, and INAF GO-grant no. 12/2024.
JJ acknowledges support from NASA/STScI through grants HST-GO-17436 and JWST-GO-5989.
SC and GA acknowledge support from NASA/STScI through grants HST-GO-17098, JWST-GO-1685 and JWST-GO-2875.
This article is based upon work from COST Action CA21136 ``Addressing observational tensions in cosmology with systematics and fundamental physics'' (CosmoVerse), supported by COST (European Cooperation in Science and Technology). This research was supported by the International Space Science Institute (ISSI) in Bern/Beijing through ISSI/ISSI-BJ International Team project ID $\#$24-603 – ``EXPANDING Universe'' (EXploiting Precision AstroNomical Distance INdicators in the Gaia Universe). This research was supported by the Munich Institute for Astro-, Particle and BioPhysics (MIAPbP) which is funded by the Deutsche Forschungsgemeinschaft (DFG, German Research Foundation) under Germany's Excellence Strategy – EXC-2094 – 390783311.

\end{acknowledgements}

\textbf{Statement of individuals' relevant expertise}\\
\label{expertise}
% \riacomm{Please fill in below}\\
\noindent Casertano--SN Ia, Cepheids, TRGB, network analysis \\
Anand--TRGB, HST, JWST, TF \\
Anderson--Cepheids, TRGB, star clusters, systematics of stellar standard candles \\
Beaton--TRGB, AGB, stellar populations \\
Bhardwaj--Cepheids, Miras, metallicity, clusters\\
Blakeslee--SBF\\
Boubel--TF relation \\
Breuval--Cepheids, metallicity, clusters\\
Brout--SN Ia, Pantheon$+$\\
Cantiello--SBF\\
Cruz Reyes--Cepheids, clusters\\
Cs\"ornyei--SN II\\
de Jaeger--SN II\\
Dhawan--SN Ia, BayesSN\\
Di Valentino--cosmological analyses\\
Galbany--SN Ia, near-IR, SNooPy, SN II.\\
Gil-Mar\'in--cosmological analyses\\
Graczyk--DEBs\\
Huang--Miras\\
Jensen--SBF\\
Kervella--Cepheids, metallicity, DEBs\\
Leibundgut--SN Ia, SN II\\
Lengen--Clusters\\
Li--JAGB, TRGB\\
Macri--Cepheids, Miras\\
Nota--Organizer\\
\"{O}z\"{u}lker--cosmological analyses\\
Pesce--Masers, Radio observations\\
Riess--SN Ia, Cepheids, HST, JWST\\
Romaniello--Cepheids, metallicity\\
Said-TF, FP, peculiar velocities\\
Sch\"oneberg--cosmological analyses, BAO+BBN, CMB, $H_0$ tension\\
Scolnic--SN Ia, Pantheon $+$\\
Sicignano--Cepheids\\
Skowron--DEBs, reddening maps\\
Uddin--SN Ia cosmology, Hubble constant analysis, CSP\\
Verde--cosmological analyses\\

%% file: Appendices/appendix.tex
\input Appendices/AA_data.tex
\input Appendices/AB_equations.tex
\input Appendices/equation_arrays.tex

\input Appendices/AC_comparisons.tex

\input Appendices/AD_sensitivity_to_anchors.tex
\input Appendices/AE_reproducibility.tex
\input Appendices/AF_distance_table.tex

%% file: Appendices/AA_data.tex
\section{Data} \label{sec:data_appendix}

This section aims to discuss the various data sets employed within this work, giving for each one an overview of the dataset, a description of the source of the data, the methodological details of how the data is obtained and analyzed, possible systematic corrections to be applied, the most important assumptions going into the dataset or method, and the connection to the network. Note that this section primarily describes the data itself---for an overview over the respective working principles, see Section~\ref{sec:data_summary}, and for an overview of the implementation within the {\DN} set of equations, see Appendix~\ref{app:equations}.

\subsection{Geometric distances} \label{sec:anchors}

\subsubsection{Distances to Galactic Cepheids from trigonometric parallaxes}

\paragraph{Overview.} 
Trigonometric parallaxes of Galactic Cepheids were measured with high signal-to-noise by the ESA \textit{Gaia} mission \citep{2016A&A...595A...1G}. The \textit{Gaia} spacecraft was in orbit around the Lagrange point L2 and traveled together with the Earth around the Sun once per year. Given {\it Gaia}'s distance from the Earth and Earth's distance to the Sun, {\it Gaia}'s large baseline made it extremely sensitive and allowed it to measure, with observations of the same star collected over a few years, the apparent shift (parallax) of stars that were within $d < 10$ kpc.

\paragraph{Data Sources.}
Parallaxes of Cepheids measured by the ESA {\it Gaia} mission are published as part of the early third data release~\citep[GEDR3,][]{2021A&A...649A...1G,2021A&A...649A...2L}. Seven additional narrow-angle parallaxes of Milky Way Cepheids measured by spatially scanning {\it HST}~\citep{2014ApJ...785..161R,2016ApJ...825...11C} were added from~\citet{2018ApJ...855..136R}. 

\paragraph{Methodological Details.}
``Field'' Cepheids identify classical Cepheids in the Milky Way with direct trigonometric parallax measurements. Cepheids in clusters have also been the focus of multiple studies: since {\it Gaia} provides parallaxes for all individual members, averaging them yields a value nearly three times more precise than for field Cepheids~\citep{2020A&A...643A.115B, 2022ApJ...938...36R, 2023A&A...672A..85C, 2024AJ....168...34W}. 

\paragraph{Systematic Corrections.}
Well documented systematics of GEDR3 parallaxes were corrected using the formalism by \citet[L21]{2021A&A...649A...4L}, which considers source magnitude, color, and on-sky position (sine of ecliptic latitude). It is also well documented that additional corrections (known as a residual parallax offset $\varpi_{\rm corr}$) are required at a sample level to ensure unbiased parallaxes~\citep[e.g.,][]{2023A&A...680A.105K,2022AJ....163..149W,2021ApJ...909..200B}.~\citet{2021ApJ...908L...6R} simultaneously solved for the fiducial absolute magnitude of Cepheids and for $\varpi_{\rm corr}$ for a sample of 75 field Cepheids observed photometrically with {\it HST} in the SH0ES Wesenheit index (a color-corrected magnitude, $m_H^W$) and found $\varpi_{\rm corr}=-14\,\mu$arcsec. For cluster members (used to infer parallaxes of cluster Cepheids),~\citet[R22]{2022ApJ...934L...7R} find no evidence for a residual parallax offset: the corrections from~\citet{2021A&A...649A...4L} effectively mitigate the parallax bias, as cluster members fall within the color and magnitude range of the sources used in its calibration.  The cluster parallax uncertainty is a combination of the statistical uncertainty and the small-scale angular covariance, with the latter being the dominant term. Table 1 of~\citet{2022ApJ...938...36R} presents the mean parallaxes of 17 Cluster Cepheids from GEDR3 also observed photometrically with HST.

\paragraph{Assumptions.}
For field Cepheids, we assume that the parallax as measured by GEDR3 has a normal distribution with central value equal to the reciprocal of the true distance plus the fixed offset $\varpi_{\rm corr} $ discussed above, and width given by the nominal EDR3 parallax uncertainty for each source.  The calibration of Milky Way Cepheids is obtained from the aggregation of all the measured parallax values, including parallax, photometric, and systematic uncertainties, and marginalized over the solution of the offset value.  The uncertainty in the offset $\varpi_{\rm corr} $ thus determined is therefore included in the final calibration uncertainty.  
For cluster Cepheids, the distance to each cluster is estimated from the aggregation of the parallax measurements for cluster stars; as noted above, cluster stars are generally within the applicability parameters of the standard offset determined in~\citet{2021A&A...649A...4L}.  However, it is important to note that parallax uncertainties within a small angular region are significantly correlated; this correlation is included in the estimate of the combined uncertainty in~\citet{2022ApJ...938...36R}.  Thanks to the combination of multiple sources, individual cluster parallaxes are typically a factor of three better than the nominal uncertainty for a single bright Cepheid.

\paragraph{Connection to the Network.}
Trigonometric parallaxes from {\it Gaia} GEDR3 are used as an anchor of the distance ladder and calibrate the Cepheid period-luminosity relation within the Milky Way.

\subsubsection{Distance to NGC 4258 using Megamasers}
\label{sec:NGC4258MM}

\paragraph{Overview.}
NGC 4258 is a nearby Seyfert 2 galaxy that is well-known for hosting the archetypal circumnuclear water megamaser disk system~\citep{1988IAUS..129..231C,1995ApJ...440..619G}.  VLBI mapping and spectral monitoring of the water masers in this system reveal that they reside in an edge-on disk and appear to be executing nearly perfectly Keplerian orbits within the centermost $\sim$1\,pc of the galaxy, where the gravitational potential is dominated by the central SMBH~\citep{1995Natur.373..127M,1999Natur.400..539H}.  By modeling these maser orbits across $\sim$3.5 years of VLBI monitoring data~\citep[collected by][]{2007ApJ...659.1040A,2008ApJ...672..800H,2013ApJ...775...13H},~\citet{2019ApJ...886L..27R} determined a distance to NGC 4258 of $7.576$\,Mpc, with a statistical uncertainty of $0.082$\,Mpc and a systematic uncertainty of $0.076$\,Mpc.  For the analyses presented in this paper, we adopt a distance modulus of $\mu_{\text{N4258}} = 29.397 \pm 0.032$ for this system \citep{2019ApJ...886L..27R}.

\paragraph{Data Sources.}
The NGC 4258 data used by~\citet{2019ApJ...886L..27R} come from 18 epochs of VLBI monitoring with the Very Long Baseline Array, spanning 3.5 years and described by~\citet{2008ApJ...672..800H} and~\citet{2013ApJ...775...13H}.

\paragraph{Methodological Details.}
The orbital motion of the masers is determined using a combination of spectral and spatial information from the VLBI monitoring, which yields constraints on the sky-plane components of the masers' positions as well as the line-of-sight components of their velocities and accelerations.  The orbital parameters for all masers are modeled alongside global parameters describing the disk morphology, the mass and location of the central black hole, its line-of-sight velocity, and the distance to the system.

\paragraph{Systematic Corrections.}
None.

\paragraph{Assumptions.}
It is assumed that the masers execute Keplerian orbits within a thin disk and that their orbital motion is entirely determined by the gravity of an enclosed point mass. Additionally, the analysis assumes no significant evolution of fundamental physical constants (e.g., $G$, $c$, $\alpha$) across space, time, or environment.

\paragraph{Connection to the Network.}
The distance to NGC 4258 is used as an anchor in the network, where it calibrates the Cepheid variable, TRGB, JAGB, and Mira variable methods.

\subsubsection{Distances to the Magellanic Clouds from detached eclipsing binaries}
\label{sec:MagClDEB}

\paragraph{Overview.}
The eclipsing binary (EB) method is a spectrophotometric method that provides near-geometrical distances to detached EB stars. The method serves both as a standard ruler, via absolute radii of EB components, and as a standard candle, by means of the empirical surface brightness -- color relations (SBCR). The method is applied to two closest galaxies: 
the Large Magellanic Cloud~\citep{2019Natur.567..200P}, giving a distance of 49.59 kpc, and the Small Magellanic Cloud~\citep{2020ApJ...904...13G}, giving a distance of 62.43 kpc. Both galaxies serve as geometrical anchors of the extragalactic distance ladder. For the analysis in the present work we adopt distance moduli of $18.477\pm 0.004$ (stat.)$ \pm 0.026$ (sys.) mag and $18.977 \pm 0.016 $ (stat.)$ \pm 0.028$ (sys.) mag for the LMC and the SMC, respectively.

\paragraph{Data Sources.}
Calibration of SBCR was based on near-infrared interferometry measurements of angular diameters of 40 nearby red clump stars~\citep{2019A&A...632A..31G}. Their optical photometry came from compilation by~\citep{1997A&AS..124..349M} and near-infrared magnitudes came after~\citep{2012MNRAS.419.1637L}. Twenty detached EBs containing non-active late-type giant stars were analysed in the LMC~\citep{2018ApJ...860....1G} and 15 similar systems in the SMC~\citep{2020ApJ...904...13G}. Data used in both papers comprised 20 years of ground-based optical photometry from the OGLE project~\citep{1997AcA....47..319U}, ground-based NIR photometry secured with SOFI instrument on NTT telescope in La Silla Observatory and long-term spectroscopic follow-up with 6 and 8 meter class telescopes in Las Campanas Observatory and Paranal Observatory carried out by the Araucaria project~\citep[e.g.][]{2009ApJ...697..862P}.

\paragraph{Methodological Details.}
Light curves and radial velocity curves were analysed simultaneously with the updated Wilson-Devinney code~\citep{1971ApJ...166..605W} in order to derive masses and radii of eclipsing binary components with a precision better than 3 
percent. Interstellar extinction corrected  photometric indices were used to derive angular diameters of components utilizing the SBCR calibrated by~\citep{2019Natur.567..200P}. The geometric distance to a particular EB results from dividing a radius of a component by its angular diameter. Reddenings were determined by three independent methods using reddening maps~\citep{1998ApJ...500..525S,2011AJ....141..158H}, the sodium doublet~\citep{1997A&A...318..269M} and utilizing extinction-free photometric indices derived from analysis of decomposed spectra of components. 

\paragraph{Systematic Corrections.}
None.

\paragraph{Assumptions.}
The binary stars are assumed to follow Roche-lobe geometry in the modeling. No significant dependence of the SBCR on metallicity is assumed for G- and K-type giant stars. A standard interstellar extinction curve is used~\citep{2007ApJ...663..320F} with $R_V = 3.1$. Grey extinction is assumed to be negligible.

\paragraph{Connection to the Network.}
Distances to the Magellanic Clouds are used to calibrate the period-luminosity relation (Leavitt Law) for the classical Cepheids within these galaxies. 

\subsection{Stellar standard candles}

\subsubsection{Cepheid Variables}\label{ssec:cepheid}

\paragraph{Overview.}
Cepheid type variables are yellow supergiants in the instability strip that follow a period-luminosity relation, also called the Leavitt Law~\citep{1912HarCi.173....1L}. They are considered primary distance indicators and are commonly used to calibrate Type Ia supernovae (though in this paper their use extends to other secondary distance indicators).

\paragraph{Data Sources.}
Distances measured from Cepheids are adopted from the following sources:
\begin{itemize}
\item HST---Cepheids---SH0ES: All Cepheids in the four anchor galaxies were observed in the same three filters ($F160W$, $F555W$, $F814W$) with the Wide Field Camera 3 (WFC3) on the Hubble Space Telescope (HST). A first study including 7 Milky Way field Cepheids with spatial scanning parallaxes and photometry were presented by~\citet{2018ApJ...855..136R}. Later, the photometry for a sample of 75 Milky Way Cepheids with accurate GEDR3 parallaxes was obtained using the same spatial scanning mode to avoid saturation and is listed in Table 1 from~\citet{2021ApJ...908L...6R}.  The photometry for an additional 17 Milky Way Cepheids located in open clusters can be found in Table 2 from~\citet{2022ApJ...938...36R}. In the LMC, 70 Cepheids were measured in Table 2 from~\citet{2019ApJ...876...85R} in the same three filters. The SMC has a non-negligible depth along the line of sight, therefore only Cepheids in the core regions were observed: a sample of 88 Cepheids with HST photometry is given in Table 2 from~\citet{2024ApJ...973...30B}. Finally, in the maser host galaxy NGC 4258, HST photometry of 669 Cepheids in the optical ($F555W$, $F814W$) comes from Table 2 from~\citet{2022ApJ...940...64Y} and in the infrared ($F160W$) from~\citet{2022ApJ...934L...7R}. HST photometry for Cepheids in 37 SN~Ia host galaxies on the second rung are from~\citet{2022ApJ...934L...7R}.
\item JWST---Cepheids---SH0ES: Due to the brightness of Cepheids and the high sensitivity of JWST, the most practical and valuable set of measurements involves following up known Cepheids (discovered in the optical by HST) with JWST NIRCAM in the hosts of SNe~Ia and the maser host NGC 4258. Table A2 of \citet{2024ApJ...977..120R} presents distances to 8 SN Ia hosts measured following this approach.  See~\citet{2023ApJ...956L..18R} for details of how these observations were made.
\end{itemize}

\paragraph{Methodological Details.}
Cepheid pulsation periods are measured from HST optical broad band light curves (in host galaxies) and from well covered ground based light curves in the Milky Way and Magellanic Clouds.  The brightness of each Cepheid is measured from optical and IR observations using HST or JWST, and epoch-corrected using the estimated amplitude and phase of the light curve obtained from optical data.  Periods are known at the $10^{-4}$ day level for Cepheids in the MW and Magellanic Clouds and to 1--2\% precision for those determined with HST in more distant hosts. A linear relation is fitted between the logarithm of the pulsation period and Cepheid brightness, which gives the Period-Luminosity slope and intercept. Intercept differences between the period-luminosity relations in two galaxies give their relative distance.

\paragraph{Systematic Corrections.}
The intercept of the period-luminosity relation depends on the metallicity of Cepheids. Metallicity effects were incorporated by differences in oxygen abundances. Milky Way, LMC, and SMC oxygen abundances of Cepheids were measured using high-resolution spectroscopy~\citep{2022A&A...658A..29R, 2023ApJ...955L..13B, 2024ApJ...973...30B}. Oxygen abundances of Cepheids in host galaxies were estimated using oxygen abundance gradients determined via the $R_{23}$ strong-line method as described by \citet[][Sect.\,3.5]{2022ApJ...934L...7R} and the galactocentric distances of each Cepheid. The period-luminosity metallicity dependence $Z_W=-0.217\pm0.046$ mag/dex was adopted from the global fit to the distance ladder in~\citet{2022ApJ...934L...7R}. Differences among direct measurements of the metallicity intercept impact $\gamma$ in the recent literature\footnote{We distinguish here between $Z_W$ as a nuisance parameter included in the distance ladder, and $\gamma$ as the direct measurement of the metallicity effect on the Leavitt law intercept determined by spectroscopy and geometric distances alone.}~\citep[e.g.,][]{2022ApJ...939...89B, 2023ApJ...955L..13B, 2024A&A...690A.246T, 2024A&A...683A.234B, 2025arXiv250522512K, 2025arXiv250715936B} were discussed in detail and determined to be explained by specifics in methodology, without relevance for the distance network due to the limited range of $\mathrm{[O/H]}$ among SN-host Cepheids~\citep[see Fig.\,21 from][]{2022ApJ...934L...7R}. A variant using $Z_W=0$ and excluding low-metallicity anchors (LMC and SMC) was introduced to assess sensitivity to uncertainties in $Z_W$ and possible mismatches between oxygen abundances determined using optical spectra and $R_{23}$.
As described in \citet{2022ApJ...934L...7R}, statistical corrections for flux contributions due to cluster Cepheids occurring in hosts \citep{2018ApJ...861...36A} were included for HST Cepheids as a background effect. Systematic corrections for time dilation were applied to measured periods of Cepheids in hosts \citep{2019A&A...631A.165A,2022A&A...658A.148A}.

\paragraph{Assumptions.}

It is assumed that the period-luminosity relation of Cepheids does not depend on host galaxy properties other than metallicity. While the period-luminosity relation can be qualitatively explained by stellar models, it is empirically calibrated and does not rely on specific assumptions about stellar physics. The period–luminosity relation is assumed to be linear over the period range $0.4 < \log P < 2$ (with period $P$ in days), and the slope of the period-luminosity relation is considered to be independent of metallicity. There might be a mild metallicity dependence for the slope but it has not been detected to be significant at more than $1\sigma$ level \citep{2024A&A...690A.246T}.

\paragraph{Connection to the Network.}
Geometric distances of Cepheids in anchor galaxies are used to calibrate the period-luminosity relation: GEDR3 parallaxes in the Milky Way, DEBs distances in the LMC and SMC, and the maser distance to NGC 4258. Cepheids in hosts are then used to calibrate the respective secondary distance indicators. 

\subsubsection{Tip of the Red Giant Branch} \label{trgb-data}

\paragraph{Overview.}
The tip of the red giant branch, as a distance indicator, relies on a well-known astrophysical phenomenon: the onset of helium burning in the electron-degenerate helium core of an old, low-mass star~\citep{2017A&A...606A..33S}. The ignition in degenerate conditions results in a runaway process which leads to the helium flash. The core mass at helium ignition varies negligibly over a wide range of stellar population ages, and in certain passbands, metallicity effects are also small or can be standardized \citep{2017A&A...606A..33S, 2018SSRv..214..113B, 2019ApJ...880...63M}. 

\paragraph{Data Sources.}
We use TRGB distances derived from a number of sources and teams as follows: 
\begin{itemize}
\item JWST -- TRGB -- SH0ES --
The SH0ES {JWST} TRGB distances are drawn from the eight hosts of 10 SN~Ia in Table 2 of~\citep{2024ApJ...976..177L}, which are based on observations from {JWST} Cycles 1 \& 2 GO-1685 and GO-2875 (PI: A. Riess). These observations were conducted in the {JWST} {F090W} and {F150W} filters. TRGB reference magnitudes were measured using a Sobel-filter based approach, anchored to the NGC 4258 measurement from~\citet{2024ApJ...966...89A}, also using data from {JWST} Cycle 1 GO-1685.
\item JWST -- TRGB -- CCHP --
The CCHP JWST TRGB distances are measured for a sample of 10 galaxies, plus the geometric anchor NGC~4258~\citep{2025arXiv250311769H, 2025ApJ...985..203F} based on observations from JWST GO–1995~\citep{2021jwst.prop.1995F}. These measurements are made in near-infrared filters: F115W is used as the primary filter, and F356W/F444W are used to measure colors. Given the notable tilt in the TRGB in these near-infrared filters, the TRGB slope is taken into account when performing the measurements.
\item JWST -- TRGB -- TRGB-SBF Project -- The TRGB-SBF Project has measured distances to fourteen nearby elliptical galaxies with JWST/NIRCam \citep{2024ApJ...973...83A, 2025ApJ...982...26A}. These TRGB measurements are anchored to NGC 4258 using JWST observations in the  F090W and F150W filters \citep{2024ApJ...966...89A}. The measurements are performed with maximum-likelihood algorithms which incorporate results from artificial star experiments, although edge-detection measurements with a Sobel filter gives very similar results, on average.
\item HST -- TRGB -- CCHP --
The tip measurements are from~\citet{2019ApJ...882...34F}, but updated with results from~\citet{2021ApJ...915...34H} that added a measurement for NGC\,5643 and determined a direct measurement for NGC\,1404.
We note that the tip measurements from~\citet{2019ApJ...882...34F} are a mix of values derived from new observations collected for the CCHP program and those derived by~\citet{2017ApJ...836...74J} and by~\citet{2019ApJ...882...34F} a single LMC-based TRGB calibration is used. 
The HST data are from the following programs: 
    GO-13691~\citep{2014hst..prop13691F}, 
    GO-15642~\citep{2018hst..prop15642F},
    GO-9351~\citep{2002hst..prop.9351R},
    GO-10497~\citep{2005hst..prop10497R}, and
    GO-10802~\citep{2006hst..prop10802R}.
For the baseline, we elect to use a single geometric anchor, NGC\,4258, for which we take the calibration from~\citet{2021ApJ...906..125J} that assumes no color-term and adopts ${M}_{{\rm{F}}814{\rm{W}}}^{{\rm{TRGB}}}=-4.050\pm 0.028$ (stat) $\pm$ 0.048 (sys) mag.
Thus, where needed we remove the TRGB zero point used in the respective paper and then adopt the NGC\,4258 result. 
\item HST -- TRGB -- EDD -- 
The TRGB measurements from \citet{2022ApJ...932...15A} provide an independent check on the CCHP TRGB measurements, using the same underlying HST data for their Type Ia supernova hosts. The photometry is performed with distinct software, and the TRGB measurements are performed with a model-fitting approach, as opposed to the edge-detection algorithms employed by the CCHP. NGC~4258 remains the sole geometric anchor, but the TRGB absolute magnitude adopted here uses an added color calibration from~\citet{2007ApJ...661..815R} to account for metallicity effects.
\item HST -- TRGB -- Li et al -- We retrieve the distances from~\citet{2025arXiv250408921L}, who adopted a Sobel filter approach to measure the TRGB in five SNe~Ia host galaxies observed by HST, as well as the result from \citet{2022ApJ...934..185D} for NGC~7814 (which is referenced within \citet{2025arXiv250408921L} and not contained in the other references above).
\end{itemize}

\paragraph{Methodological Details.}
TRGB magnitudes are identified using either edge-detection methods (e.g., Sobel filters) (cf. \citealt{2024ApJ...963L..43A} for methodological considerations) or model-fitting to the luminosity function of RGB stars. Color-based calibrations are applied to account for metallicity effects, particularly in the near-infrared. All measurements are anchored to the geometric distance to NGC 4258, with differing treatments of slope and zero point depending on filter set and methodology.

\paragraph{Systematic Corrections.}
Some teams adopt a color-based metallicity calibration to standardize the TRGB magnitude. When a different zero point was originally assumed, it is replaced by the NGC 4258-based calibration to ensure consistency.

\paragraph{Assumptions.} The different authors adopt distinct ways to deal with the metallicity dependence of the TRGB absolute magnitude. The general consistency between distances from the various works imply that the effect of these assumptions may be minor. 

\paragraph{Connection to the Network.} The TRGB distances are geometrically anchored to NGC 4258 in both the native HST and JWST systems. These TRGB distances then provide the ability to calibrate other secondary distance indicators, such as for example the magnitudes of Type Ia supernovae or Surface Brightness Fluctuations.

\subsubsection{Mira distances}

\paragraph{Overview.}
Mira variables are fundamentally-pulsating Asymptotic Giant Branch stars that, similar to Cepheids, follow a period-luminosity Relation.  Despite not being as widely-used or as well-understood as Cepheids, Miras still have two major advantages as independent calibrators of Type Ia SNe: 1) they have larger near-infrared amplitudes than Cepheids, allowing them to be characterized and detected using only near-infrared observations and 2) they are ubiquitous since they have low-to-intermediate masses (stars with initial masses ranging from $0.8 M_\odot < M < 8 M_\odot$ are expected to undergo a Mira phase of evolution), allowing them to be detected in galaxies not reachable by Cepheids. Miras are as bright as Cepheids in the near-infrared and brighter than all of the other commonly-used stellar distances indicators such as Tip of the Red Giant Branch, J-region Asymptotic Giant Branch, and RR Lyrae. 

\paragraph{Data Sources.}
Data and distance measurements to NGC 4258, NGC 1559, and M101 using Miras originate from~\citet{2018ApJ...857...67H, 2020ApJ...889....5H, 2024ApJ...963...83H} respectively. All three galaxies were observed with HST WFC3/IR, primarily in the \emph{F160W} bandpass. NGC 4258 and NGC 1559 also have \emph{F125W} and \emph{F110W} observations used to obtain colors. 

\paragraph{Methodological Details.}
Mira periods are determined using at least 10--12 epochs of photometry spaced unevenly over a minimum baseline of one year. In addition to applying cuts to remove non-variable objects, candidate O-rich Miras are selected on the basis of their peak-to-trough amplitude ($0.4$ mag $ \lesssim A \lesssim 0.8$ mag), period of variability ($200 \lesssim  P \lesssim 400$ days), and approximate color.  These cuts are intended to limit the contamination from Carbon-rich Miras and other types of variables and Asymptotic Giant Branch stars. In each galaxy, the lower period bound is set by an empirically-determined completeness limit and magnitude of contamination from Carbon-rich Miras is modeled empirically using a mixture model based on the OGLE (Optical Gravitational Lensing Experiment) Large Magellanic Cloud observations of long-period variables. Although Miras in Milky Way star clusters and the LMC have also been used as anchors in the literature~\citep[e.g.][]{2025ApJ...990...63B}, these were not adopted here to mitigate systematic uncertainties related to ground-based to HST photometric transformations. 

\paragraph{Systematic Corrections.} None.

\paragraph{Assumptions.} 
It is assumed that Oxygen-rich Miras with periods below approximately 400 days follow a linear period-luminosity Relation. The slope and zeropoint of this relation, which are empirically calibrated, as well as the amplitudes of the Miras themselves, are not expected to evolve as a function of the host environment. See~\citet{2024ApJ...963...83H} for further discussion.

\paragraph{Connection to the Network.} The zeropoint of the Mira period-luminosity Relation is calibrated using the water megamaser distance to NGC 4258. This relation is then used to determine Mira distances to NGC 1559 (host of SN 2005df) and M101 (host of SN 2011fe) and calibrate secondary distance indicators in those hosts.

\subsubsection{J-region AGB method}

\paragraph{Overview.} The J-region asymptotic giant branch (JAGB) method is a statistical method that uses evolved stars that are redder than the red giant branch in color-magnitude diagrams.
The stars are identified using a color and magnitude selection and the application of a statistic such as mean, median, or mode. 

\paragraph{Data Sources.}
We use JAGB distances derived from the following sources:
\begin{itemize}
\item JWST -- JAGB -- SH0ES --
We draw our data for the SH0ES JAGB host galaxy distance measurements from~\citet{2024ApJ...966...20L} and~\citep{2025arXiv250205259L}, which are based on observations from both {JWST} Cycle 1 \& 2 programs GO-1685 and GO-2785 (PI: A. Riess). These observations were obtained in the {JWST} NIRCAM \emph{F150W} and \emph{F277W} filters. 
\item JWST -- JAGB -- CCHP --
We use distances from~\citet{2025ApJ...985..203F} for the CCHP JAGB distances. These distances are based on {JWST} NIRCAM observations from program GO-1995 (PI: W. Freedman) in the \emph{F115W}, \emph{F356W}, and \emph{F444W} filters.
\end{itemize}

\paragraph{Methodological Details.} The JAGB reference magnitude can be measured using the mode, mean, median, or a model fit, as currently used in the literature. To account for systematic differences in selecting any one particular statistic,~\citep{2024ApJ...966...20L} adopted the middle, or median, of the JAGB reference magnitudes across several measurement variants and use the spread in JAGB magnitude across these variants as a systematic uncertainty. The CCHP distances use exclusively the mode.

\paragraph{Systematic Corrections.} For the SH0ES dataset, crowding corrections were applied using artificial star tests, as described in~\citet{2024ApJ...966...20L} and are measured in the `outer disk,' where the crowding bias is less than 0.05~mag in both filters used for the measurement. The values from the CCHP dataset do not apply crowding corrections but seeks to minimize the effects of crowding by also making the measurement in the outer disk. 

\paragraph{Assumptions.} The JAGB method, as  currently described in the literature, assumes that the population of stars in the J-region are homogeneous such that the mean magnitude (or alternatively the median or mode) for all stars in this region remain constant. However, this may not be the case, as noted by~\citet{2020MNRAS.495.2858R, 2021MNRAS.501..933P,2023MNRAS.522..195P, 2024ApJ...966...20L, 2025arXiv250205259L}; such effects can manifest via asymmetry of the J-region luminosity function, which causes difference between the JAGB magnitude measured using different statistics. Systematics from this asymmetry can be quantified through the systematic uncertainty adopted via measurement variants. 

\paragraph{Connection to the Network.} 
The JAGB is calibrated using the water maser distance in NGC 4258 and is used to further calibrate other secondary distance indicators in the respective hosts.

\subsection{Supernovae}
\subsubsection{Type-Ia Supernovae}

\paragraph{Overview.}
Type Ia supernovae (SNe~Ia) are thermonuclear explosions of white dwarf stars in binary systems. They are among the best-established standardizable candles in cosmology due to their consistent intrinsic luminosities after empirical corrections. This allows them to serve as precise distance indicators at cosmological scales. In the local distance ladder, SNe~Ia constitute the third rung, calibrated by nearby objects with independent distance measurements.

\paragraph{Data Sources.}
We use SNe~Ia distances derived from the following sources:
\begin{itemize}
    \item \textit{Pantheon+ (optical):} 
    The Pantheon+ sample of 1701 light curves of 1550 spectroscopically confirmed SNe~Ia compiled from 18 different surveys. The sample itself and light curves are detailed in~\cite{2022ApJ...938..113S}, is cross-calibrated in~\cite{2022ApJ...938..111B}, and host-galaxy assignments and redshifts are updated in~\cite{2022PASA...39...46C}. Pantheon+ utilize the SALT2 SN Ia model from~\cite{2021MNRAS.504.4111T} and apply bias corrections to the observed light-curve quantities following~\cite{p21,p22}. The entire analysis pipeline for Pantheon+ is automated in~\cite{Hinton:2020yik} which facilitated the systematic uncertainties described in~\cite{2022ApJ...938..110B} that are utilized for this work. 
   
    \item \textit{CSP-I\&II (optical+NIR):} The Carnegie Supernova Project (CSP; \citealt{2006PASP..118....2H}) obtained high-quality light-curves of SNe~Ia in $uBgVriYJH$ bands from two observing campaigns between 2004 and 2015. The first observing campaign (CSP-I) ran from 2004 to 2009, and the second campaign (CSP-II) ran from 2011 to 2015. Light-curves from CSP-I are published in~\cite{2017AJ....154..211K}, and spectroscopic analysis is described in~\cite{2013ApJ...773...53F}. While CSP-I followed up SNe~Ia mostly from targeted search, CSP-II followed SNe~Ia that are discovered from rolling search, primarily from La Silla Quest \citep{2013PASP..125..683B}. CSP-II is described in~\cite{2019PASP..131a4002H} and in~\cite{2019PASP..131a4001P}. Spectra of SNe~Ia are published in~\cite{2024ApJ...967...20M}, and individual light-curves will be published (N. Suntzeff et al. 2026, in preparation) soon. An extensive analysis of Hubble constant using various distance calibrators is presented in~\cite{2024ApJ...970...72U}, where light-curve fitting parameters obtained from SNooPy \citep{2011AJ....141...19B} $\texttt{max\_model}$ method are available. 
    
    \item \textit{Template-independent distances in the NIR:} A compilation of all publicly available SNe~Ia with NIR observations was presented in~\cite{2023AA...679A..95G}. These were selected for having pre-maximum data so the peak magnitude could be derived by simple Gaussian Process fitting. The sample include 19 SNe~Ia in galaxies with Cepheid-based distances from SH0ES, and 55 SNe~Ia in the Hubble flow ($z> 0.01$).
    From those, all 19 SNe~Ia in the calibrator sample have $J$-band light curves, however only 16 have light-curves with enough quality in the $H$ band. Similarly, for the Hubble-flow sample, while 52 SNe~Ia have good $J$-band light-curves, 40 have $H$-band light-curves that permit the determination of the peak-brightness. 
    For the $H_0$ determination, an intrinsic scatter of $\sigma_{int}$ of 0.125 mag was included in quadrature to the observed peak magnitude uncertainty. 
    From the baseline analysis presented in~\cite{2023AA...679A..95G}, in this paper we removed the term accounting for the SH0ES ladder in the systematic uncertainty budget (0.51 km s$^{-1}$ Mpc$^{-1}$), so the total systematic uncertainty here is 1.35 km s$^{-1}$ Mpc$^{-1}$ instead of the 1.44 km s$^{-1}$ Mpc$^{-1}$ used in~\cite{2023AA...679A..95G}.
    
    \item \textit{BayeSN (optical+NIR):} An inference of the SN~Ia ``distance,'' i.e. a normalised peak magnitude, based on optical and NIR data (the latter where available), was presented in~\citep{2023MNRAS.524..235D}. The sample included 42 SNe~Ia with Cepheid distances and 18 with TRGB distances. The SN~Ia data is compiled from the literature (individual SN references are in Tables 1 and 2). To restrict the sample to the original training set of BayeSN, we used 67 Hubble flow SNe~Ia. The uncertainties include the fit errors on each objects and the intrinsic scatter was fitted in the MCMC code used to infer $H_0$. When comparing optical only to optical + NIR, the final $H_0$ error improved by $\sim 10\%$. For the case with a $z_{\rm min} = 0.023$, the error in $H_0$ was 1.135 km\,s$^{-1}$\,Mpc$^{-1}$.
    
    \item \textit{SNe~Ia in Coma:} A sample of 13 SNe~Ia located in the Coma Cluster was recently compiled by~\citep{2025ApJ...979L...9S}, using data from multiple modern surveys including ATLAS and YSE. These SNe~Ia were selected to be hosted by passive galaxies in the Coma cluster and are located at effectively the same redshift, providing a local standard-candle ensemble. Light curves are calibrated following the Pantheon+ framework.
\end{itemize}

\paragraph{Systematic Corrections.}
Systematic uncertainties include photometric calibration, light-curve model assumptions, and corrections for selection biases. These are treated internally within each analysis and propagated into the final uncertainties on $H_0$\,.

\paragraph{Assumptions.} SNe~Ia are assumed to be standardizable candles, and the standardization is empirically-calibrated. Therefore, it is assumed that there is no evolution of SNe~Ia properties  as function of time/distance and no dependence on properties of the host such as mass or metallicity  beyond what is modeled in the standardization, see e.g., ~\cite{2023ApJ...951...22J} for a discussion. 

\paragraph{Connection to Network.} SNe~Ia are used as tracers of cosmic expansion in the distance network.

\subsubsection{Type-II supernovae---standard candle method} \label{sec:data:snii_SC}

\paragraph{Overview.}
Type II supernova luminosities can be standardized using the Standard Candle Method (SCM; \citealt{2002ApJ...566L..63H}), an empirical technique based on correlations with photospheric velocity and color. This allows SNe II to be used as distance indicators beyond the local Universe. 

\paragraph{Data Sources.}
The Type-II supernova sample used to derive distances through the SCM is almost the same as that in~\citet{2022MNRAS.514.4620D}. This sample consists of 89 objects with $z>$0.01 from various surveys, including the Carnegie Supernova Project-I (CSP-I; \citealt{2024A&A...692A..95A}), the Lick Observatory Supernova Survey (LOSS; \citealt{2019MNRAS.490.2799D}), the Sloan Digital Sky Survey-II SN Survey (SDSS-II; \citealt{2010ApJ...708..661D}), the Supernova Legacy Survey (SNLS; \citealt{2006ApJ...645..841N}), the Dark Energy Survey Supernova Program (DES-SN; \citealt{2020MNRAS.495.4860D}), and the Subaru Hyper Suprime-Cam Survey (SSP-HSC; \citealt{2017MNRAS.472.4233D}).  Filters used include multiple optical bands (e.g., $B$, $V$, $R$, $I$, $g$, $r$, $i$).
We exclude a total of four SNe~II with Cepheid distances from the list in~\citet{2022MNRAS.514.4620D}. The distances to three supernovae (SN 1999em, SN 1999gi, and SN 2012aw) were based on Cepheid measurements using optical images. These objects were removed for consistency with our baseline, for which all the Cepheid distances were derived using NIR images. Furthermore, SN 2005ay (NGC 3938) was also removed from the sample because it used the Cepheid distance to NGC 3982---both galaxies are part of the Ursa Major groups but are located in different subgroups (South and North, respectively). We add SN 2023ixf~\citep{2025arXiv250313974Z} which exploded in M101, a galaxy for which a Cepheid distance is available~\citep{2022ApJ...934L...7R}.
Finally, the four galaxies NGC 628, NGC 5194, NGC 6946, and NGC 7793 \citep[5~SNe II used by][]{2022MNRAS.514.4620D} with TRGB measurements were re-scaled to the distances from~\citet{2022ApJ...932...15A}, which uses NGC 4258 as the TRGB zeropoint.
Therefore, to calibrate the SN~II absolute magnitude, we use a total of ten calibrators: four objects with Cepheid measurement, five SNe~II with TRGB distances, and one in NGC 4258 with geometric maser distance.

\paragraph{Methodological Details.}
We follow the methodology of~\citet{2020MNRAS.495.4860D} to derive SN velocities, magnitudes, and colors. Expansion velocities are obtained using H$_\beta$ absorption via cross-correlation between observed spectra and an SN~II template library. Light curves are modeled using hierarchical Gaussian Processes with the George Python library, allowing accurate interpolation of magnitudes and colors at fixed epochs (43 days post-explosion).

\paragraph{Systematic Corrections.}
Magnitude corrections include Milky Way extinction, K-correction, and S-correction \citep{2020MNRAS.496.3402D,2022MNRAS.514.4620D}. CMB redshifts are taken from the NASA/IPAC Extragalactic Database and corrected for peculiar velocities using the model of Carrick et al. (2015). A residual peculiar-velocity uncertainty of 250 km s$^{-1}$ is included.

\paragraph{Assumptions.}
SNeII are assumed to be standardizable candles, and the standardization is empirically-calibrated. Therefore it is assumed that there is no significant evolution of the indicator with redshift and no environment-dependent bias (e.g., host mass, metallicity).

\paragraph{Connection to the Network.}
In the distance ladder, SNe~II are used as a tracer in the Hubble flow as secondary distance indicators calibrated using primary distance indicators.

\subsubsection{Type-II supernovae---spectral modeling-based approach}

\paragraph{Overview.}
Type II supernovae also offer a unique opportunity for a single-step distance estimation without the use of primary distance indicators. This path makes use of the tailored Expanding Photosphere Method (tailored EPM, \citealt{2005A&A...439..671D, 2020A&A...633A..88V, 2025AA...702A..41V}), which is a significantly improved version of the classical EPM~\citep{1974ApJ...193...27K, 1994ApJ...432...42S}. The method employs spectral modeling, which allows for the precise estimation of physical parameters of the supernova at each epoch. The latest implementation of the fitting and the distance estimation is described in~\citet{2025AA...702A..41V} and~\citet{2023A&A...678A..44C}.

\paragraph{Data Sources.}
The currently published set of Type II supernovae with astrophysical modeling distances consists of 19 supernovae: ten with $z > 0.01$ which were used for the recent $H_0$ determination of~\citet{2025AA...702A..41V}, four SNe sharing two host galaxies at $z \sim 0.01$ from the SN sibling consistency check of~\citet{2023AA...672A.129C}, and five more supernovae from three additional nearby galaxies from~\citet{2023AA...672A.129C} and~\citet{2023A&A...678A..44C}. The spectra modeled in~\citet{2025AA...702A..41V} were collected by the PESSTO survey~\citep{2015A&A...579A..40S} at the New Technology Telescope of the European Southern Observatory, while~\citet{2023AA...672A.129C} used data collected across the literature, including data from~\citet{2006PASP..118....2H,2006MNRAS.372.1315S,2011ApJ...736...76R,2019ApJ...876...19S,2022ApJ...930...34T}. The observed spectra have been recalibrated to the photometry through the estimation of synthetic colours to rule out potential contaminations. The exact procedure of the analysis and the applied steps are described in~\citet{2025AA...702A..41V}.

\paragraph{Methodological Details.}
The main improvement of this technique over the traditionally used classical EPM is the use of radiative transfer modeling, which solves the outstanding issue of continuum flux dilution and the associated correction factors, which increased the systematic errors of this method in the past~\citep[for details, see][]{2005A&A...439..671D}. The radiative transfer models used for the physical parameter estimation were derived using the radiative transfer code TARDIS~\citep{2014MNRAS.440..387K, 2019A&A...621A..29V}. These models are fit to the spectra on a $\chi^2$ basis employing the emulator and fitter described in~\citet{2020A&A...633A..88V}. The estimated physical parameters are combined with the photometry to derive the absolute flux, total reddening, and the distance of the supernova following the procedure described in~\citet{2025AA...702A..41V}.

\paragraph{Systematic Corrections.}
None.

\paragraph{Assumptions.}
Since the method is based on an astrophysical model, a relatively large number of assumptions are required for modeling the physics of SNe II. The current estimation makes use of the TARDIS radiative transfer code, which assumes a spherical supernova, without time-dependent effects, but including bound-bound, bound-free transitions, and Thomson scattering as the sources of the opacity. The full setup is described by~\citet{2019A&A...621A..29V}. These assumptions hold well for a fully or almost fully ionized ejecta, and the results agree well with codes that include time-dependent effects. 
A further modeling assumption is made for the density profile of the supernova ejecta, modeled with a power-law profile, which was shown to provide adequate results in the past, but is a potential simplification.
Finally, while the method allows for the estimation of the reddening, the total-to-selective extinction ratio $R_V$ has to be assumed for the distance estimation. For the modeling, an $R_V$ of 3.1 was assumed. A more comprehensive list of assumptions and their potential impacts is available in the discussion section of~\citep{2025AA...702A..41V}.

\paragraph{Connection to the Network.}
The SN~II tailored EPM method can be used to directly determine distances to objects in the Hubble flow.  Currently, there is no overlap between the host galaxies used for the distance network and the SNe II modeling approach, however, they could serve as a calibrator in the future. An overlap between the network and the modeling of SNe II can be created by either finding new SNe~II in host galaxies, by estimating further precise standard candle distances to hosts of modelable SNe II, or by observing siblings of SNe~Ia and SNe II in the same host galaxy.  All of these options will open new avenues in the distance network and are being actively pursued.

\paragraph{Comments}
Given the assumptions used in the spectral modeling, not all Type II supernovae can be modeled in the current framework: the radiative transfer models can only fit spectra reliably in a specific epoch range, and only for sufficiently normal Type II supernovae. These can be selected in advance based on the appearance of the spectra; however, a pre-selection is needed to limit exposure to systematic effects. 

\subsection{Surface Brightness Fluctuations}

\paragraph{Overview.}
The Surface Brightness Fluctuation (SBF) technique is a statistical distance indicator that measures pixel-to-pixel variance in the unresolved stellar populations of early-type galaxies. The variance of fluctuations, normalized to the galaxy surface brightness, scales inversely with distance squared, making the galaxy appear smoother as distance increases~\citep{1988AJ.....96..807T}.
The SBF signal, which is dominated by red giant branch stars, can be reliably  measured in nearby calibration galaxies and as a tracer of universal expansion well into the Hubble flow ($\sim$100 Mpc with HST, and potentially several times farther with JWST).

\paragraph{Data Sources.}
In 2024, JWST/NIRCam observed 14 giant elliptical galaxies to simultaneously measure both their TRGB brightnesses and SBF magnitudes~\citep{2024ApJ...973...83A,2025ApJ...982...26A}, enabling a high-precision absolute calibration of the near-IR SBF distance zero point. The TRGB distances were calibrated using the geometric distance to the megamaser galaxy NGC\,4258~\citep{2024ApJ...966...89A}. This approach provides a zero point for the WFC3/IR F110W SBF distance scale that is entirely independent of the Cepheid distance scale.  
The new TRGB-based calibration was applied to SBF measurements of 61 galaxies from~\citet{2021ApJS..255...21J} and~\citet{2021ApJ...911...65B}, using updated optical $({g-z})$ color measurements, to determine the distances used here \citep{2025ApJ...987...87J}.\footnote{\url{https://github.com/jjensen-uvu/sbf_distances_2021}} The distant galaxy SBF sample spans distances up to 100 Mpc. 
In summary, the key measurements for the adopted SBF-based distances in this study come from:
\begin{itemize}
    \item JWST/NIRCam imaging of 14 elliptical galaxies used for both SBF and TRGB measurements ~\citep{2024ApJ...973...83A,2025ApJ...982...26A}. Eight of these galaxies have HST WFC3/IR SBF data, and the full set provided a distance calibration for the Fornax and Virgo clusters that connects the JWST TRGB calibration to the HST IR SBF sample of \citet{2015ApJ...808...91J}.
    \item JWST/NIRCam TRGB distance to the anchor galaxy NGC\,4258~\citep{2024ApJ...966...89A}.
    \item HST/WFC3-IR imaging of 61 early-type galaxies in the F110W filter~\citep{2021ApJS..255...21J,2021ApJ...911...65B}.
\end{itemize}

\paragraph{Methodological Details.}
SBF distances are derived from the amplitude of spatial brightness fluctuations in galaxies, quantified by fitting the Fourier spatial power spectrum with a scaled power spectrum of the image point-spread function. 
Accurate SBF relative distances are obtained using an empirical calibration adjusted for stellar population age and metallicity variations based on optical colors. 
The recent JWST/TRGB calibration of the absolute SBF magnitude, $\overline{M}$, as a function of the observed $({g - z})$ color has replaced the earlier Cepheid-based calibration, which had been used in almost all previous SBF-based distance estimates.

\paragraph{Systematic Corrections.}
The absolute fluctuation magnitude calibration depends on galaxy color, and for this study, a new calibration with updated optical colors and slope was derived \citep{2025ApJ...987...87J}. The SBF zero point is anchored to TRGB distances that were calibrated using the geometric distance to NGC\,4258. To convert distances into Hubble-flow velocities, peculiar velocity corrections were applied based on group assignments from both the Cosmic Flows 3 database~\citep{2019MNRAS.488.5438G} and the 2M++ flow model~\citep{2015MNRAS.450..317C}, following the approach of~\citet{2021ApJ...911...65B} and~\citet{2025ApJ...982...26A}. Standard corrections for foreground extinction were also applied using the prescription from~\citet{2011ApJ...737..103S}.

\paragraph{Assumptions.} 
The main assumptions behind the SBF method are that stellar population variations between galaxies are adequately accounted for by color corrections (i.e., $\overline{M}$ vs.\ $({g-z})$ color calibration), and — for the analysis presented in this work — that the TRGB stars in the spiral megamaser galaxy NGC\,4258 at a given color are representative of those in the 14 giant elliptical galaxies used to calibrate SBF.

\paragraph{Connection to the Network.}
SBF distances are calibrated by TRGB distances from JWST/NIRCam, which are anchored to the geometric distance to NGC\,4258. 
Additionally, SBF can be used to calibrate other distance indicators, thereby adding another rung to the cosmic distance ladder. The published SBF distance to one galaxy in the Coma cluster \citep{2021ApJS..255...21J} was used to provide a calibration of the Fundamental Plane (FP) method used to measure distances to thousands of elliptical galaxies~\citep{2025MNRAS.539.3627S}. SBF has also been used to provide an alternate calibration of type Ia SNe~\citep{2023ApJ...953...35G}. However, this latter path was not included in the baseline network for this study.

\paragraph{Comments and Known Issues.}
While SBF is robust for early-type galaxies, its use is limited in spiral galaxies where star formation, clumpy dust, intermediate age asymptotic giant branch and young massive stars can all affect the RGB-based SBF fluctuation signal useful for determining the distance. The accuracy of SBF distances also depends on high-quality imaging and precise knowledge of the point spread function. The SBF method is potentially sensitive to calibration systematics from $(i)$ the uncertainty in the slope of the stellar population correction arising from the limited number of calibrators and their color range, and $(ii)$ the systematic uncertainty in the zero point calibration from TRGB measured in a single anchor galaxy (NGC\,4258), and the potential differences between the TRGB brightness in that spiral galaxy and in the elliptical galaxies used to calibrate SBF.

%%%% FUNDAMENTAL PLANE %%%%
\subsection{The Fundamental Plane calibrated in Coma\label{app:data:coma}}
\paragraph{Overview.}
The Fundamental Plane (FP) is a three-dimensional scaling relation for early-type galaxies that connects their effective radius, surface brightness, and central velocity dispersion, working as a secondary distance indicator that can measure galaxy distances with $~20-25\%$ precision~\citep{1987ApJ...313...59D,1987ApJ...313...42D}. The FP relation uses two observables of elliptical galaxies that don't require any knowledge of distances, the mean surface brightness ($I_e$) and the central velocity dispersion ($\sigma_0$), to infer the physical effective radius ($R_e$), which requires distances to measure. Comparing the physical radius to the angular radius ($\theta_e$) allows for the measurement of distances. In the distance ladder, the FP acts as a secondary distance indicator: it requires calibration through absolute distance measurements to nearby galaxy clusters, after which it can trace distances to more distant early-type galaxies in the Hubble flow, making it valuable for measuring peculiar velocities and constraining cosmological parameters like $H_0$~\citep{2025MNRAS.539.3627S,2025ApJ...979L...9S} and the growth rate of cosmic structure~\citep{2020MNRAS.494.3275A, 2020MNRAS.497.1275S, 2023MNRAS.518.2436T}.

\paragraph{Data Sources.}
Building the FP relation requires combining two complementary datasets: photometric and spectroscopic observations. The photometric component draws from the DESI Legacy Imaging Surveys~\citep{2019AJ....157..168D}, which provide multi-band imaging across the DESI footprint, while the spectroscopic data come from the Early Data Release of the DESI survey~\citep{2024AJ....168...58D}. The FP analysis included 4191 early-type galaxies within the redshift range  $0 < z < 0.1$, with central velocity dispersion measurements from DESI spectra~\citep{2025MNRAS.539.3627S}. Photometric parameters, including angular effective radius and mean surface brightness, were obtained from the DESI Legacy Imaging Surveys Data Release 9.

\paragraph{Methodological Details.}
The DESI FP methodology employed a maximum likelihood approach to fit a three-dimensional Gaussian model in the parameter space defined by $\log R_e = a \log \sigma_0 + b \log I_e + c$. Central velocity dispersions were measured from DESI spectra using pPXF~\citep{2023MNRAS.526.3273C} with stellar templates from the Indo-U.S. Coudé Feed Spectral Library~\citep{2004ApJS..152..251V}. Photometric parameters were derived from DESI Legacy Survey imaging. 

\paragraph{Systematic Corrections.}
Several systematic corrections were applied to ensure robust distance measurements. For the DESI FP analysis, a photometric zero-point correction of $+0.0234$ mag was applied to northern $r$-band magnitudes to account for systematic differences between BASS and DECaLS surveys~\citep{2025MNRAS.539.3627S}. Velocity dispersion measurements were compared with SDSS pPXF values to assess potential systematic offsets, which were incorporated into the total systematic error budget rather than applied as corrections. Aperture corrections were applied using the~\citet{1995MNRAS.276.1341J} formula to convert to standard aperture sizes. A redshift cut of $z > 0.023$ was implemented to minimise peculiar velocity contamination, and peculiar velocity corrections were applied using  pvhub\footnote{\url{https://github.com/KSaid-1/pvhub}} velocity field maps. 

\paragraph{Assumptions.}
Several key assumptions are made in the application of these distance measurement methods. For the Fundamental Plane analysis, the primary assumptions include: (1) negligible evolution of the FP relation over the redshift range $0.0 < z < 0.1$, with only a modest evolution correction of $0.85z$ applied to account for stellar population aging; (2) universality of the FP relation across different galaxy environments, from field galaxies to dense cluster cores; and (3) adoption of a flat $\Lambda$CDM cosmology with fixed deceleration parameter $q_0 = -0.55$ and jerk parameter $j_0 = 1$, fitting only for the Hubble constant $H_0$ due to the limited redshift range of the data~\citep{2025MNRAS.539.3627S}. 

\paragraph{Connection to the Network.}
Within the distance ladder, the Fundamental Plane operates as a tracer that requires calibration from other distance indicators. The precise linkage is further discussed in Appendix~\ref{sec:eq_special:FP}, and is based primarily on SNe~Ia distances, with a sub-leading fixed SBF calibration also employed.  In the future it might be feasible to extend the FP approach to a full tracer, allowing for consistent modeling of expansion parameters (like $q_0$, $j_0$) and peculiar velocities within the context of the {\DN}. For now, we only rescale the {\Hcst} values of \citet{2025MNRAS.539.3627S,2025ApJ...979L...9S}, see Appendix~\ref{sec:eq_special:FP}.

%%%% TULLY-FISHER %%%%
\subsection{Tully-Fisher relation}

\paragraph{Overview.}
The Tully-Fisher (TF) relation is an empirical scaling relation connecting the rotational velocity of spiral galaxies to their intrinsic luminosity, serving as a secondary distance indicator in the cosmic distance ladder. First used as a distance indicator by~\citet{1977A&A....54..661T}, the relation enables distance measurements to spiral galaxies by comparing their observed apparent magnitude to absolute magnitudes predicted from measured HI line widths. The TF relation is a secondary distance indicator that requires calibration from primary distance indicators to establish its absolute zero-point. The current largest TF catalogue is Cosmicflows-4~\citep{2020ApJ...902..145K} containing $\sim 10\,000$ spiral galaxies with HI line widths, redshifts, and photometry.

\paragraph{Data Sources.}
The primary dataset for TF measurements is Cosmicflows-4 (CF4; \citealt{2020ApJ...902..145K}), which represents the largest TF catalogue to date with approximately 10\,000 spiral galaxies distributed across the full sky, out of which we only use around 3\,400 with complete infrared photometry (as recommended by P. Boubel 2025, priv. comm.). The CF4 sample combines data from multiple surveys, incorporating HI line widths from various sources, optical photometry from the SDSS in the $i-$band, and infrared photometry from WISE in the W1-band. 

\paragraph{Methodological Details.}
The Tully-Fisher model assumed for this dataset is typically a linear relation with a quadratic term introduced at the bright end and an intrinsic scatter that decreases linearly with rotation velocity / luminosity~\citep{2020ApJ...902..145K,2024MNRAS.531...84B}. In the {\DN} implementation, the distances of the calibrators are solved for explicitly in the same way as all other calibrators.

\paragraph{Systematic Corrections.}
The TF analysis applies corrections for galaxy inclination. Magnitude selection functions are modeled as a Gaussian drop-off beyond magnitude limits to account for high flux limit effects. Peculiar velocity corrections are applied through a simultaneous fit of velocity field parameters, including bulk flow and external dipole components. 

\paragraph{Assumptions.}
The TF relation assumes that spiral galaxies follow a universal relation between rotational velocity and luminosity with no significant evolution over the redshift range used ($z<0.1$). The method assumes that HI line widths provide accurate tracers of rotational velocity after inclination corrections, and that the intrinsic scatter in the relation is primarily astrophysical rather than systematic. The peculiar velocity model assumes that large-scale motions can be described by a linear velocity field with bulk flow and external bulk flow components. 

\paragraph{Connection to the Network.}
The TF relation serves as a tracer of the Hubble flow in the {\DN}, requiring calibration from primary distance indicators.

%%% MEGAMASERS
\subsection{Megamasers in the Hubble flow}

\paragraph{Overview.}
AGN accretion disk megamasers similar to those in NGC 4258 are also observed in more distant galaxies, where they can be used to make single-step measurements of the Hubble constant~\citep{2007IAUS..242..399B}. ~\citet{2020ApJ...891L...1P} combined and updated these distance measurement.

\paragraph{Data Sources.}
The Megamaser Cosmology Project (MCP) has carried out an extensive survey of several thousand AGN~\citep{2015IAUGA..2255730B,2020ApJ...892...18K}, identifying a number of megamaser systems that are suitable for precise ($\lesssim$10\% uncertainty) geometric distance measurements.  VLBI mapping and spectral monitoring observations conducted by the MCP have to date yielded such measurements for five megamaser-hosting galaxies: UGC 3789~\citep{2009ApJ...695..287R,2010ApJ...718..657B,2013ApJ...767..154R}, NGC 6264~\citep{2013ApJ...767..155K}, NGC 6323~\citep{2015ApJ...800...26K}, NGC 5765b~\citep{2016ApJ...817..128G}, and CGCG 074-064~\citep{2020ApJ...890..118P}.  For the analyses presented in this paper, we adopt the distance and velocity measurements from~\citet{2020ApJ...891L...1P}, with the distance measurements reprocessed as constraints on distance modulus.  These galaxies span a range of distances from $\sim$50--130\,Mpc, and the MCP observations have utilized the Robert C. Byrd Green Bank Telescope, the Karl G. Jansky Very Large Array, and the Very Long Baseline Array.

\paragraph{Methodological Details.}
For each megamaser-hosting galaxy, the orbital motion of the masers is determined using a combination of spectral monitoring and VLBI mapping, which yield constraints on the sky-plane components of the masers' positions, as well as the line-of-sight components of their velocities and accelerations.  The orbital parameters for all masers are modeled alongside global parameters describing the disk morphology, the mass and location of the central black hole, its line-of-sight velocity, and the distance to the system.  The distance and velocity are combined to constrain {\Hcst}\,.

\paragraph{Systematic Corrections.}
A variety of peculiar velocity corrections have been tested for the megamaser sample, as detailed in~\citet{2020ApJ...891L...1P}. For the analyses presented in this paper, we adopt the 2M++ peculiar velocity correction as primary; one of the variant solutions use the velocity in the CMB frame, without peculiar velocity corrections. 

\paragraph{Assumptions.}
The analysis assumes that the masers execute Keplerian orbits within a thin disk and that their orbital motion is entirely governed by the gravitational influence of an enclosed point mass. It also assumes that fundamental physical constants (e.g., $G$, $c$, $\alpha$) do not vary across space, time, or environment.

\paragraph{Connection to the Network.}
The megamasers are used directly as tracers in the Hubble flow and do not require further calibration.

%% file: Appendices/AB_equations.tex
\section{The system of equations and the covariance matrix}
\label{app:equations}

The system of equations that constrains the parameters of the {\DN} consists of several different types of equations (some involving only individual sources, others global), constraining a number of optimization parameters. Within the approximations of the present analysis, all equations are linear (or linearized) in these parameters, and all probability distributions are Gaussian in magnitude, distance modulus, or $ \log_{10}(\Hcst) $. The optimization procedure is therefore appropriate to a generalized linear least squares problem, which---within the stated assumptions---provides the Best Linear Unbiased Estimator (BLUE) for the solution \citep{aitken1935}. Given the linearized nature of the equations, different methods of solving the equations generally return the same parameter constraints---computing Bayesian credible intervals using MCMC sampling methods or constructing Frequentist confidence intervals would give the same answers.

Critically, we do include several off-diagonal terms in the covariance matrix of all equations, reflecting the fact that their uncertainties are correlated. The solution process also returns the covariance matrix for the optimized parameters, including the Hubble constant (more precisely, its logarithm).

\subsection {Linearity and Gaussianity}

In our approach, we adopt equations that are linear in distance modulus and linearized in the logarithm of quantities non-linearly related to distances, such as redshifts and $ H_0 $.  We also assume that the probability distribution function of all measured quantities is Gaussian in this description, and therefore is fully described by a (non-diagonal) covariance matrix. Unfortunately, many original papers do not provide a full PDF for the quantities they measure, and they implicitly describe the errors as Gaussian; even if asymmetric confidence intervals are given, the lack of a full description of the PDF makes it challenging to go beyond the Gaussian approximation. However, most measured quantities have small enough fractional errors that the Gaussian approximation is accurate, especially in the aggregate; those with large fractional errors (and typically more non-Gaussian profiles) contribute only little to the final results. Finally, we note that most published data are already processed, incorporating uncertainties e.g. due to the underlying anchor. Therefore, we make sure to include the appropriate (see below) covariances for the published data according to how they were obtained from the original measurements. The most direct validation we have for all of the above approximations is our ability to recover the original results for all the examples considered in Appendix~\ref{sec:comparisons}. We note that the linearization naturally implies that results going out to many $>5\sigma$ should be interpreted with this caveat in mind. However, only slight changes could be introduced by these approximations, and we believe that our qualitative statements remain true even in more careful implementations.

\subsection{Solution procedure}

Given the measured data vector $\mathbf{Y}$ with its associated (non-diagonal) covariance matrix $\mathbf{C}$ we can construct a likelihood function as follows
\begin{equation}
    -2 \ln p(\mathbf{Y}|\mathbf{X}) = \chi^2 = (\mathbf{Y - A X})^T \mathbf {C}^{-1} (\mathbf{Y - A X}).\label{eq:chisq_definition}
\end{equation}
where $p(\mathbf{Y}|\mathbf{X})$ is the likelihood, and our linear model has the theory prediction of the data as $\mathbf{Y}^\mathrm{theory} = \mathbf{A} \mathbf{X}$ with some coefficient matrix $\mathbf{A}$ that we determine below. We immediately recognize that this is a formulation of a generalized least squares problem, as indicated in the introduction of this Section \ref{app:equations}. For a problem of this kind, it is well known that Bayesian approaches with flat (unbounded) priors will generally give the same results as Frequentist minimum likelihood estimation approaches. Within this work we will mostly follow the Frequentist view, but we stress that, due to the properties of multivariate Gaussian distributions, our conclusions apply equivalently to the Bayesian approach.

The generalized least squares problem seeks to minimize a quadratic form $Q$ in the residuals $ \mathbf {Y - A X} $:
\begin{equation}
    Q = (\mathbf {Y-AX})^T \mathbf {W} (\mathbf {Y-AX})
\end{equation}
for some positive-definite, symmetric weight matrix $ W $; here $ ()^T $ indicates the transpose.  By the Gauss-Markov theorem, the result of this minimization has the lowest variance if the weight matrix $ \mathbf {W} $ is the inverse of the covariance matrix $ \mathbf C $ of the data \citep{aitken1935}; in this case the quadratic form corresponds to the usual $ \chi^2 $ in Equation~(\ref{eq:chisq_definition}).

However, a difficulty can arise in that the set of these equations is generally not linearly independent. The reason is that the same measured quantities can be combined in different ways to obtain constraints (i.e., equations); if the number of constraints exceeds the number of independent measurements, some of these constraints must be equivalent to a combination of others.  For example, consider the case of Cepheid-related distance constraints based on different anchors. If we have, say, 30 host galaxies with measured intercepts of the Leavitt law, and 3 calibrations based on different anchors, we have a total of 33 independently measured quantities: the apparent magnitude of the intercept in the 30 host galaxies, and the calibrated absolute magnitude of the intercept in the three anchors. However, these result in 90 separate equations for the host distances, one for each \textit{combination} of host and anchor (see also Section \ref{sec:eq_host} below); these equations are by necessity linearly dependent. Other dependencies are more subtle, and result from using the same host distance constraint for different types of calibrators. As a result, the covariance matrix $ \mathbf {C} $, as constructed, is generally singular, as evidenced by extremely small (numerically consistent with zero) elements in the diagonal matrix of its singular-value decomposition. In our baseline case, we have 255 equations, but only 183 distinct  measurements, resulting in 72 superflous/overconstrained equations.

To address the issue, we replace the standard matrix inversion for the covariance matrix with its Moore-Penrose inverse \citep{penrose1956}, also called pseudoinverse, computed, e.g., via the {\tt python} routine {\tt scipy.linalg.pinv} (see, e.g., \citealt{2022zndo....595738G}).  In this implementation, the Moore-Penrose inverse is computed via singular value decomposition, but the singular values that are numerically consistent with zero are ignored, i.e., their inverse is set to zero.  Other languages have similar implementations, and this computation can also be implemented using LAPACK routines (see, e.g., \citealt{2021ascl.soft04020L}). The effect of using the Moore-Penrose inverse of the covariance matrix in Equation~\ref{eq:chisq_definition} is to ignore linearly-dependent equations in the computation of the $ \chi^2 $, and reduce the problem to only linearly independent equations; this obviates the need to carry out a detailed analysis of all constraints. 

We stress, however, that this numerical solution is completely equivalent to analytically making sure that only linearly independent equations are used. However, the current approach has several advantages, such as readability of the code and simplicity of the ensuing system of equations which also simplifies future modifications.

\subsection {Equations and covariance matrix}

Above we have described generally how the problem is set up. Within this section we aim to describe our specific setup of the equations. For this, we introduce a few additional pieces of notation. We minimize the residuals
\begin{equation}
    \sum_{j=1}^{N_\mathrm{par}} A_{i_\mathrm{eq} j} X_j - Y_{i_\mathrm{eq}}
    \label {eq:coefficient_form}
\end{equation}
where $ Y_i $ are the $ N_\mathrm{eq} $ data values, corresponding also to the number of equations to be solved, with $ i_\mathrm{eq} = 1, \dots, N_\mathrm{eq} $ the data indices, and $ X_j $ are the $ N_\mathrm{par} $ optimization parameters, and $ j = 1, \dots, N_\mathrm{par} $ the parameter indices.  We note that the matrix $ \mathbf {A} $ is very sparse; each equation typically involves only one or two parameters. For convenience below we write the equations in the form $\mathbf{A} \mathbf{X} = \mathbf{Y}$, understanding that we are actually implementing the least square solution to the corresponding residuals of that equation.

The form of each equation depends on which data and which parameters are involved. We have four main types of equations: host distance equations; host-group equations; calibrator equations; and Hubble flow equations. The host distance equations (Section~\ref{sec:eq_host}) relate the distance of each host to several published measurements, which are typically not independent; their correlations are accounted for in the covariance matrix, as discussed in the following. Host-group equations (Section~\ref{sec:eq_groups}) relate the distance of each host in a group to the common distance of that group (or cluster).  Calibrator equations (Section~\ref{sec:eq_calib}) relate the absolute calibration of one of the secondary distance indicators---such as SNe~Ia, SNe II, Surface brightness fluctuations, Tully-Fisher relation, and so on---to the distance of a host system obtained via the primary calibration method. Hubble Flow equations (Section~\ref{app:sec:eq_hf}) relate the distance (typically a luminosity or angular diameter distance) of a calibrated secondary distance indicator to its redshift, and thus provides the connection to the Hubble constant. Some methods bypass some of these equations; for example, megamaser distances are calibrated directly, and directly provide a Hubble Flow equation without intermediate steps. Similarly, astrophysically calibrated SNe II (using the Expanding Photosphere method) connect directly to the Hubble flow. In such cases, the equations are modified accordingly; for example, Hubble flow equations for directly calibrated distances will not involve a host distance---see also Section \ref{sec:eq_special} for these special cases.

\subsubsection {Optimization parameters}\label{sec:optimization_parameters}
We begin our discussion of the equations by formally writing down the list of all parameters that will be optimized. For this, we quickly revisit the general notation we adopt.

The quantity $ \mu_{\ihost} $ is the distance modulus of the host $ \ihost $.  This is defined on the basis of the geometric luminosity distance alone; the apparent magnitude $ m_{\ihost} $ is expected to follow the simple relation $ m_{\ihost} = M_{\ihost} + \mu_{\ihost} $, with any corrections due to extinction, redshift effects, and other standardization terms absorbed in the definition of the apparent or absolute magnitude, as appropriate. The reference absolute magnitude $ M_{{\rm ref}, \itype} $ for calibrator type $ \itype $ is defined so that for each calibrator $ \icalib $, the observed magnitude $ m_{ \icalib} $ is expressed as 
\begin{equation}
     m_{\icalib} = M_{{\rm ref}, \itype[\icalib]} + \mu_{\icalib} .
\end{equation}
For a calibrator $\icalib$ within a given host $\ihost[\icalib]$, this equation is used to determine the $\mu_{\mathrm{host},\ihost[\icalib]}$.

The parameters to be optimized include the following:

\begin{itemize}
   \item $ \mu_{\rm host, \ihost}  \quad (\ihost = 1, \dots, N_{\rm hosts}) $: the distance modulus for each system treated as a host
   \item $ M_{\rm ref, \itype} \quad (\itype = 1, \dots, N_{\rm types}) $: the reference magnitude for each type of calibrator used in the solution.  
   Calibrator types include SNe~Ia, SNe II (calibrated empirically), 
   SBF, and Tully-Fisher. For the sake of brevity, we synonymously refer to these as $M_{\rm Ia}\equiv M_{\rm ref, SNe~Ia}$, $M_{\rm II}\equiv M_{\rm ref, SNe~II}$, $M^\mathrm{off}_{\rm SBF}\equiv M_{\rm ref, SBF}$,  $M^\mathrm{off}_{\rm TF}\equiv M_{\rm ref, TF}$, respectively. 
   \item $ \mu_{\igroup} $, the distance modulus to any groups ($ \igroup$) included in the host data (see Section~\ref{sec:eq_groups}). 
   \item $ \mu_{\rm Coma} $ if the FP is used in the \DN.
   \item $ \log_{10} (H_0) $.
\end{itemize}

These are collectively indicated as $ X_j $ in Equation~\ref{eq:coefficient_form}. Other parameters---e.g., those describing the metallicity dependence of a class of calibrators, or other standardization parameters---can in principle be included.  However, in the spirit of this study, we generally adopt whatever conventions and standardizations have been used in the original papers, and eschew re-characterizing the behavior of calibrators.

\subsubsection {Host distance equations}
\label{sec:eq_host}

For every measurements of the distance of a nearby host, we include an equation relating the distance modulus for that host with a specific anchor and method. Separate equations are given for the same host measured by different methods, using different anchors, or by different collaborations/groups (source).  A \textit{method} in this context refers to the combination of a type of primary distance indicator (e.g., Cepheids) observed using a given instrument (e.g., HST). The equations are thus characterized by the method used, the anchors they refer to, and the source of the measurement; collectively we refer to these as \textbf{MAS} (Method, Anchor, Source) and denote them with $\imas[\ieq]$ for the $\ieq$-th equation.  Data with different MAS are considered separately, although they can be correlated. We additionally define the host involved in the $\ieq$-th as $\ihost[\ieq]$ and the anchor as $\ianchor [\ieq]$ -- noting that different equations can have the same $\ihost$ and/or $\ianchor$. Finally, for convenience we also define unique combinations of the host, the method, and the source (HMS) and denote them with $\ihms[\ieq]$ for the $\ieq$-th equation -- where a host calibrated by the same group using the same method but with a different anchor can cause the $\ihms$ of different equations to agree. Only the $\imas$ is unique to the $\ieq$-th equation.

The host equation $\ieq$ can then be written as:
\begin {equation}
\mu_{\ihost[\ieq]} = \mu_{\ieq}
\end{equation}
with correspondingly $A_{\ieq,j}=\delta_{j, \ihost[\ieq]}$ for this $\ieq$, and $\delta_{ij} =1$ if $i=j$ and 0 otherwise.  In order to construct the covariance for this set of equations, we note that the uncertainty of each measurement combines the measured uncertainty of the distance indicator (e.g., the TRGB magnitude or the Cepheid intercept) in the host, $ \sigma_{\ihost[\ieq]| \ihms[\ieq]} $; the uncertainty of the same measurement in the anchor, $ \sigma_{\ianchor[\ieq]| \imas[\ieq]}$; and the uncertainty in the distance modulus of the anchor itself, $ \sigma_{\ianchor[\ieq]} $. Therefore the diagonal element of the covariance matrix is:
\begin {equation}
C_{\ieq, \ieq} = \sigma_{\ihost[\ieq]| \ihms[\ieq]}^2 + \sigma_{\ianchor[\ieq]| \imas[\ieq]}^2 + \sigma_{\ianchor[\ieq]}^2
\end{equation}

The covariance between different equations $ \ieq, \jeq $ depends on which elements they have in common. If the anchor is the same, then the covariance includes the anchor distance uncertainty. If the MAS $\imas$ is the same, then the uncertainty of the measurement of the distance indicator in the anchor is in common. If two equations share the same host measured by the same method and source ($\ihms$) then the covariance includes the uncertainty of the distance indicator measurement within the host. Therefore:
\begin{equation}
    C_{\ieq, \jeq} = \delta_{\ianchor[\ieq], \ianchor[\jeq]} \cdot  \sigma_{\ianchor[\ieq]}^2 + \delta_{\imas[\ieq], \imas[\jeq]} \cdot \sigma_{\ianchor[\ieq]| \imas[\ieq]}^2 + \delta_{\ihms[\ieq], \ihms[\jeq]} \cdot  \sigma_{\ihost[\ieq]| \ihms[\ieq]}^2 
\end{equation}

\subsubsection {Host-group equations}
\label {sec:eq_groups}

In some cases, especially for the Surface Brightness Fluctuations method, a distance estimate for a calibrator or host can come from its membership in a group or cluster.  Most of the SBF calibrators are either in Fornax or in Virgo; some of them have direct distance estimates from TRGB, others do not.  The assumption is that all objects within each of those clusters are at approximately the same distance, with a dispersion reflecting the estimated spatial extent of the cluster.

This situation can be handled by introducing an additional class of parameters: the distance to a group or cluster.  In keeping with the distance network concept, this distance is estimated from all members of the group with a direct distance measurement, and is applied to the members that do not have a direct measurement.  For the members with a measurement, their distance is also partially constrained by the other group members.  

The equations representing these constraints involve the parameters of the group distance estimate, $ \mu_\igroup  $, and the distance to each member, $ \mu_\ihost $.  If a host $ \ihost $ is a member of a group $ \igroup$, we write the simple equation:
\begin{equation}
\mu_{\ihost[\ieq]} - \mu_{\igroup[\ieq]} = 0
\end {equation}
with correspondingly $A_{\ieq, j} = \delta_{j, \ihost[\ieq]} - \delta_{j,\igroup[\ieq]}$ and $Y_\ieq=0$ in this case.  The uncertainty associated to this equation is the estimated dispersion $ \sigma_{\ihost, \igroup} $ of the distance modulus of host $ \ihost $ with respect to the distance of the group $ \igroup $. This value can in principle be different for every member of the group; however, currently we assign the same uncertainty $\sigma_\igroup$ to all hosts in the group, and assume the hosts are randomly distributed within that group (no non-diagonal elements in the covariance matrix).  In the current treatment, this type of equation has no covariance with any other equation.

\subsubsection {Calibrator equations}
\label {sec:eq_calib}

For every calibrator $ \icalib $, we include an equation relating the distance modulus of its host and the quantity being calibrated, the reference magnitude $ M_{\rm ref, \itype} $, with the standardized apparent magnitude $ m_{\icalib } $.  This quantity can be the reference magnitude for TF or SBF, the standardized absolute magnitude of a SN Ia or SN II in the appropriate band, or (in the case of BayesSN) a reference magnitude for the SN. These equations have in general negligible covariance in the data, but it is critical that the intrinsic dispersion in each calibrator's distance estimate (e.g., the intrinsic variance in SNe~Ia absolute magnitude, or the internal dispersion of the TF relation) be included in the uncertainty of each source.  These uncertainties are reduced by using more calibrators. \textit{Distance calibration} uncertainties are propagated from the uncertainty of the host distance modulus, which is derived from the distance network, and therefore are not included in the \textit{data} covariance matrix.

The calibrator equations have the form:
\begin{equation}
    \mu_{\ihost} = m_{\icalib}- M_{\rm ref, \itype}
\end{equation}
where $ \mu_{\ihost} $ is the distance modulus parameter of the host $ \ihost $ in which the calibrator $ \icalib $ is located, which will be optimized from the global solution of the generalized least-squares problem. This equation can be recast in the form of Equation~\ref{eq:coefficient_form} as follows:
\begin{equation}
    M_{\rm ref, \itype[\ieq]} + \mu_{\ihost[\ieq]} = m_{\icalib[\ieq]}.
\end{equation}
with correspondingly $A_{\ieq, j} = \delta_{j, \itype[\ieq]} + \delta_{j, \ihost[\ieq]}$.

The variance associated with these quantities includes not only the contribution from the direct measurement process, such as the photometric uncertainty, but also the variance resulting from measurements of the standardization parameters, such as color and light curve width for SNe~Ia, or the velocity width for Tully-Fisher.  Those uncertainties are those reported in the original papers, unless otherwise stated.  Critically, the variance must also include the intrinsic dispersion in that distance indicator---e.g., the scatter in standardized SN luminosities. or the dispersion in the velocity-based luminosity indicator from Tully-Fisher.  This term is not always included explicitly in the  uncertainties provided by the original sources, and in such cases it must be added in quadrature to the stated uncertainties. We assume that these equations have a diagonal covariance structure.

\subsubsection{Hubble flow}\label{app:sec:eq_hf}
In principle each tracer in the Hubble flow would add its own equation, and with the potentially hundreds to thousands of objects in the Hubble flow, this would greatly increase the size of the problem. Luckily, the additional equations are highly similar, and therefore allow a great simplification of the overall problem. Since the equations for this part of the problem are uncorrelated to the remaining system of equations (the covariance matrix is block-diagonal), this sub-problem can be treated independently. Additionally, the similarity of the equations can be expressed mathematically precisely as a rank deficiency of the coefficient matrix. We therefore first analyze such rank-deficient cases in general, and then apply our knowledge to the Hubble flow tracer equations.

If the coefficient matrix $A$ is rank-one, i.e., $A = b a^T$, then we can write
\begin{align}
    \chi^2 & = (A X - Y)^T F (A X - Y) = (X^T a) (b^T F b) (a^T X) - 2 (X^T a) (b^T F Y) + (Y^T F Y) \\ & = (X^T a - y) f (a^T X - y) + \Delta \chi^2~. \label{eq:full_chi2}
\end{align}
This establishes the equivalence between the full problem with coefficient matrix $A$, parameter vector $X$ and data vector $Y$ with inverse covariance matrix $F = C^{-1}$ and a smaller condensed problem with coefficient vector $a$ and a single data point $y$ with inverse covariance $f$. Here $f = (b^T F b)$ and $y = (b^T F Y)/(b^T F b)$ are both scalars -- the new condensed covariance matrix and data point. Additionally $\Delta \chi^2 = (Y^T F Y) - f y^2$ is a parameter-independent offset of the $\chi^2$ function and correspondingly does not impact the overall parameter estimation. This means that instead of solving the full problem with many data points $Y$, one can solve the reduced problem with a single data point $y$. Interestingly the value of the scalar $y$ is the solution to the independent $\chi^2$ optimization problem
\begin{equation}
    \chi^2 = (b \hat{y} - Y)^T F (b \hat{y} - Y) \label{eq:reduced_chi2}
\end{equation}
which has the solution $\mu_{\hat{y}} = y$ and the variance $\sigma^2_{\hat{y}} = f$. As such, we can first solve the reduced problem of Equation~\ref{eq:reduced_chi2}, and then use the mean solution and its variance as inputs for the condensed problem of Equation~\ref{eq:full_chi2}. In summary, a full problem with a rank-deficient coefficient matrix can be decomposed into a separate optimization problem, whose solution can be used to construct only a single additional equation for the original problem.

The equations for tracers in the Hubble flow could generally be expressed as
\begin{equation}
    m_i = M_{\mathrm{ref},\itype} + \mu_i = M_{\mathrm{ref},\itype} - 5 \log_{10}(H_0) + 5 f(z_i)\label{eq:hf_original}
\end{equation}
with $m_i$ the observed standardized magnitude of the $i$-th tracer, $M_{\mathrm{ref},\itype}$ the common reference absolute magnitude for the tracer type $\itype$, and $\mu_i$ the distance modulus of the $i$-th tracer, which we decompose as $\mu_i = -5 \log_{10}(H_0) + 5 f(z_i)$.
Therefore, in our case we can identify two parameters $X = \{M_{\mathrm{ref},\itype}, \log_{10}(H_0)\}$ and the data vector $Y_i = 0.2 m_i - f(z_i)$. The rank deficient coefficient matrix is then $A = b a^T$ with $b = 1_\mathrm{N_\mathrm{tracer}}$ being a vector of ones with length equal to the number of tracers (of a given type) in the Hubble flow, and $a^T = \{-0.2, 1\}$ is a vector of length 2. We thus can use the techniques for a rank-deficient matrix and first solve the $\chi^2$-minimization reduced sub-problem with equations of the type
\begin{equation}
    a_\itype + 5 = f(z_i) - 0.2 m_i \label{eq:hubble_flow}
\end{equation}
where $\hat{y} = a_\itype + 5$ is the so-called Hubble intercept for tracer type $\itype$. For these reduced equations we use in analogy with Equation~\ref{eq:reduced_chi2} the full covariance matrix of the $m_i$ with possible peculiar velocity corrections (see below). For our full problem we can then reduce the additional equation to only
\begin{equation}
    \log_{10}(H_0) - 0.2 M_\itype  = a_\itype + 5\label{eq:intercept}
\end{equation}
with $A_{\ieq, j} = \delta_{j,\log H_0} - 0.2 \delta_{j, \itype[\ieq]}$, where $a_\itype$ is the solution to the Hubble intercept sub-problem and its covariance is used for the full problem according to Equation~\ref{eq:full_chi2}. Solving the equations determining the Hubble flow then boils down to solving the Hubble flow intercept problem of Equation~\ref{eq:hubble_flow}. We quickly note that the condensation procedure can be repeated independently for the independent tracer types, leading to one additional Hubble intercept equation and one independent Hubble intercept minimization problem per tracer type $\itype$; the form of the equation will differ for SBF and TF because of the form of their luminosity calibration.

\subsubsection{The intercept equations}\label{app:sec:intercept}
For the intercept Equations~\ref{eq:hubble_flow} we have two distinct cases. When the tracer objects are standard candles (SNIa, SNII, etc) we need to use the luminosity distance for the distance modulus, and for standard ruler objects we can use the same equations, replacing the luminosity distance with the angular diameter distance. Therefore we only treat the case of standard candles below, noting that since the two are assumed to be related as $D_A = D_L /(1+z)^2$, we only need to insert the correct factors of $(1+z)^2$ at relevant locations (which we explicitly indicate below) to treat both cases.

The more problematic aspect of the Hubble flow intercept equations is that the measured redshifts $z_i$ are impacted by Doppler shifting from the peculiar motion of the tracer objects. Albeit this effect can be minimized by choosing corresponding tracer objects that are distant enough to minimize the relative influence on the velocity, we still aim to consistently account for these effects, allowing us to also use relatively nearby tracers such as megamasers.

We note that $\mu_i \propto 5 \log_{10}(D)$ and $D \propto H_0^{-1}$ (where $D$ can be either the luminosity or angular diameter distance), and therefore $f(z_i)$ is by design independent of $H_0$\,, independently of which precise expansion rate a specific model has. While it would also be possible to parameterize the distance functional $f(z)$ using a specific cosmology, we choose a cosmographic expansion for our baseline results due to its model-independence, and given the low redshift range $z<0.15$ we typically consider for the tracers in the Hubble flow.

In general the expression for $f(z_i)$ must take into account also the corrections to the observed redshift from peculiar velocities (both of the earth as well as the individual tracer objects). We therefore differentiate between the cosmological redshift that a comoving object would have $z_{\mathrm{HD},i}$, the actually observed redshift $z_i$, and the redshift corrected into the heliocentric frame (removing earth's peculiar motion) as $z_{\mathrm{hel},i}$. We discuss these corrections further in Section~\ref{app:sec:pec_vel} below. The expression of the distance modulus can then be written as \citet{2011ApJ...741...67D,2017JCAP...01..038C}
\begin{equation}
    \mu_i = 5\log_{10}(D(z_i)/\mathrm{Mpc})+25 = 5\log_{10}\left(D(z_{\mathrm{HD},i}) \cdot \frac{1+z_{\mathrm{hel},i}}{1+z_{\mathrm{HD},i}}\right) + 25~.\label{eq:distance_modulus}
\end{equation}
Together with Equation~\ref{eq:hf_original} this leads to the expression
\begin{equation}
    f(z_i) = \log_{10}\left(\frac{D(z_i) H_0}{\mathrm{km/s}}\right)+5 \label{eq:hf_f_of_z}
\end{equation}

The distance can further be expressed using a cosmographic expansion within the desired redshift range, namely
\begin{subequations} \label{eq:distance_taylor}
\begin{align}
    D_L(z) & \approx \frac{cz}{H_0} \left[1+\frac{1}{2}(1-q_0) z - \frac{1}{6}(1-q_0 - 3 q_0^2 + j_0) z^2 + \ldots \right] \equiv \frac{v_L(z)}{H_0} \\ 
    D_A(z) & \approx \frac{cz}{H_0} \left[1-\frac{1}{2}(3+q_0) z + \frac{1}{6}(11+7 q_0 + 3 q_0^2 - j_0) z^2 + \ldots \right] \equiv \frac{v_A(z)}{H_0}
\end{align}
\end{subequations}
leading to the corresponding expansions for $f(z_i)$ that can be used for Equation~\ref{eq:intercept} to compute the intercept.
Here we have used the acceleration and jerk parameters $q_0$ and $j_0$ defined respectively as the evaluations today of $q(a)=-\ddot{a}a/\dot{a}^2$ and $j(a) = \dddot{a}a^2/\dot{a}^3$, which can be used to re-express the redshift Taylor expansions of the distance expressions.
According to \citet{2019MNRAS.490.2948D}, this approximation is adequate out to $ z \sim 0.3 $; additional correction for deviations from a flat Universe are negligible.  The use of Equation~\ref{eq:distance_taylor} is the only direct cosmological dependence of our results.  We adopt the fixed values $ q_0 = -0.55 $ and $ j_0 = +1 $, following \citet{2016ApJ...826...56R}; the value for $ q_0 $ differs slightly from that adopted by \citet{2024MNRAS.531...84B} ($q_0 = -0.5275 $).  Changes to $ q_0 $ and $ j_0 $ do not have a significant impact in the resulting value of $ H_0 $; even a change in $ q_0 $ from $ -0.55 $ to $ -0.50 $ would reduce the baseline value of $ H_0 $ by less than $ 0.1 \, \rm km/s/Mpc $.  This description of distances assumes the validity of the cosmological principle, statistical isotropy and homogeneity (effectively a FRWL metric) and the Etherington relation---that is the $(1+z)^2$ scaling for $D_L/D_A$, which of course assumes photon number conservation in General Relativity. We also assume a given level of smoothness of the expansion history within the given redshift range such that the given expansion is applicable.

The variance of each data item $m_i - 5 f(z_i)$ combines: the measurement error for the indicator (typically a photometric error); the variance in the velocity function due to the uncertainty of the redshift (including any uncertainty in the peculiar velocity corrections); and the intrinsic width of the distance indicator, expressed in magnitudes.  These variances are considered uncorrelated and added together. The elements entering the different intercept equations can (and typically do) have correlated errors; for example, uncertainties in the standardization parameter will propagate to all elements $i$ in ways that depend on how their properties were standardized, and sources from the same survey could have survey-systematics in common. However, this information is not always available. We have included covariance terms between SNe~Ia according to the prescriptions of Pantheon+, resulting in a minor impact on the results; the change in $ H_0 $ between including and excluding off-diagonal covariance terms is less than $ 0.1 \, \rm km/s/Mpc $ (see V00 vs V27 in Table~\ref{tab:variants}). In practice, for all secondary distance indicators in use the intrinsic dispersion of the calibration is significantly larger than any systematic covariance terms; only with very large number of sources could these covariances, which are not abated by simply increasing the number of sources, come into play. The uncertainty in the absolute calibration itself is part of the distance network system and is accounted as part of the default system of equations; it need not be included explicitly in the covariance matrix of the separated intercept equations.

\subsubsection{Peculiar velocities}\label{app:sec:pec_vel}

Peculiar velocities are typically on the order of hundreds of km/s. At low redshifts especially (typically below $cz=3000$\,km/s, i.e $z<0.01$), the scatter introduced to the Hubble diagram by peculiar velocities becomes dominant; we generally consider tracers only at $ z > 0.01 $ ($D \gtrsim 40$Mpc), though preferentially at $z>0.023$ ($D \gtrsim 100$Mpc).
 
In the linear regime, a smoothed density field mapped from a galaxy redshift survey can be used to predict the peculiar velocity field, up to a normalization and any influences external to the survey volume (often approximated as a residual bulk flow) \citep{1995PhR...261..271S}. This method has been applied to the 2M++ \citep{2015MNRAS.450..317C} and Two-Micron All-Sky Redshift Survey (2MRS) redshift surveys \citep{2021MNRAS.507.1557L} to predict the density and velocity fields. For the 2M++ model, the normalization parameter and external bulk flow were re-measured by \cite{2020MNRAS.497.1275S} using peculiar velocity measurements from 6dFGS and SDSS, and by \cite{2024MNRAS.531...84B} using peculiar velocity measurements from the Tully-Fisher Cosmicflows-4 catalog. The uncertainty in these model predictions, $\sigma_v$ is dominated  by the residuals between the non-linear observed field and the predicted linear field, and its value spans $\sim$150--250 km/s depending on the specific flow model. 

Peculiar velocity corrections made using these models have previously been successful at reducing scatter in the redshift-distance relation. In this paper, we adopt the 2M++ model for peculiar velocity corrections in the baseline determination of $H_0$\,. Previous studies have found it to be at least as good as other representations. For example, \cite{2023AAS...24142408P} found it produced the smallest uncertainties in cosmological parameters including $H_0$\,. These peculiar velocities are adopted in all intercept equations, as well as for objects directly linked to the Hubble flow.

In all cases we decompose the observed redshift as
\begin{equation}
    1+z_i = (1+z_{\mathrm{HD},i}) \cdot (1+z_\mathrm{earth}) \cdot (1+z_{\mathrm{pec},i})
\end{equation}
where $z_{\mathrm{HD},i}$ would be the redshift if the object was comoving in the Hubble flow and was measured by a comoving observer, $z_\mathrm{earth}$ is the Doppler redshift due to earth's motion with respect to a comoving frame, and $z_{\mathrm{pec},i}$ is the Doppler redshift caused by the peculiar velocity of the given object with regard to a comoving frame, which can be computed by
\begin{equation}
    z_\mathrm{pec} = \sqrt{\frac{1+v_\mathrm{pec}/c}{1-v_\mathrm{pec}/c}}
\end{equation}
Finally we denote $z_\mathrm{hel,i}$ as $z_\mathrm{hel,i} = \frac{1+z_i}{1+z_\mathrm{earth}}-1$.

\subsubsection {Special cases} \label{sec:eq_special}

The previous Sections describe the equations associated with a typical distance ladder: distance estimates for nearby (host) galaxies, calibration constraints, and secondary distance indicators applied to objects in the Hubble flow, leading to $ H_0 $ constraints.  However, there are some distance estimates that do not fit neatly into this three-rung scheme, and are discussed in the following.

\paragraph {Direct distance estimates for sources in the Hubble flow}

If the distance of a source far enough to be in the Hubble flow can be estimated directly, then the equivalent of Equation~\ref{eq:distance_taylor} can be written for that source. In such a case the equation will not involve $M_{\rm ref, \itype} $ --- which for other systems is used to estimate the distance on the basis of the observed apparent magnitude. Rather, the directly measured quantity is the distance $D_i$, typically in the form of either a luminosity or an angular diameter distance. The equation can simply be written as
\begin{equation}
  \log_{10}(H_0) = \log_{10}(v(z_{\mathrm{HD},i}))-\log_{10}(D_i).\label{eq:HF_direct}
\end{equation}
where the coefficient matrix has $A_{\ieq,j} = \delta_{j, \log_{10}(H_0)}$ and $y_\ieq = \log_{10}(v(z_\ieq)) - \log_{10}(D_\ieq)$, where $D_\ieq$ is the object's measured distance and $v(z)$ is either $v_A(z)$ or $v_L(z)$ from Equation~\ref{eq:distance_taylor}, depending on the nature of the object. Note that the redshift needs to be corrected for peculiar velocities, as discussed in Section~\ref{app:sec:pec_vel}.

We include two such categories: megamasers, the distance of which is determined by fitting Keplerian orbital parameters and matching the physical and angular orbital radii (thus yielding an angular diameter distance); and astrophysically-calibrated SNe II, for which the absolute luminosity is estimated from a model of their expanding photosphere, yielding a luminosity distance for each object. In the latter case, all SNe II thus measured share a common modeling uncertainty; however, this term is not clearly defined in the reference material, and is likely small in comparison with the (rather large) internal dispersion of the distance estimate. See Section~\ref{sec:data_appendix} for more details.

\paragraph {The Fundamental Plane of elliptical galaxies and the Coma cluster\label{sec:eq_special:FP}}
Another special case is the measurement of the parameters of the so-called Fundamental Plane of elliptical galaxies, which is a tight relation between size, surface brightness, and velocity dispersion of bright ellipticals. In this case we make use of the analysis first performed in \citet{2025MNRAS.539.3627S} and updated in \citet{2025ApJ...979L...9S}. We simply re-scale the estimate anchored to Coma by the distance to Coma determined self-consistently from the global solution of the network. For this we primarily implement the following equations:
\begin{equation}
    \mu_\mathrm{Coma} + M_{\rm Ia} = m_\ieq
\end{equation}
with Coma distance modulus $\mu_\mathrm{Coma}$, SNIa reference magnitude $M_{\rm Ia}$, and individual supernovae magnitudes $m_\ieq$\,, where we use the following list of 13 supernovae as part of the Coma cluster: [2019bkh, 2020ags, 2021dch, 2021lxb, 2022frn, 2023aakj, 2023czd, 2023epj, 2024ana, 2021oat, 2010ai, 2013ag, ASASSN-15jt]. Their absolute magnitude uncertainties include intrinsic dispersion but we do not model their correlations within Coma or with other calibrators.  Following \citet{2025MNRAS.539.3627S}, we consider only the YSE measurement of SN 2021lxb.

In addition to the SNe~Ia-derived distance we also allow for a constraint on the Coma distance from SBF measurements. However, in this case we do not employ the full \DN\ (yet), and simply use the distance modulus value $\mu_\mathrm{SBF, fixed} = 34.98\pm 0.1345$ (see  \citet{2025MNRAS.539.3627S,2025ApJ...979L...9S} for details):  
\begin{equation}
    \mu_\mathrm{Coma} = \mu_\mathrm{SBF, fixed}
\end{equation}

Finally, the Coma distance is coupled to the Hubble flow via the rescaling equation
\begin{equation}
    \log_{10}(H_0) + 0.2 \mu_\mathrm{Coma} = \log_{10}([H_0D]_\mathrm{FP}) + 5
\end{equation}
with $[H_0D]_\mathrm{FP}$ being the fundamental plane value of $(76.05 \cdot 99.1)$ with a uncertainty assigned to the logarithm of 0.00742, see \citet{2025MNRAS.539.3627S,2025ApJ...979L...9S}. In the future we are planning to directly compute intercept equations also in the case of the FP method.

\paragraph {An {\Hcst} estimate without Hubble flow SNe~Ia}

\cite{2022ApJ...935...83K} published an analysis of measured Cepheid distances to SNe~Ia host galaxies that did not rely on a calibration of SNe~Ia in the Hubble Flow, essentially bypassing secondary distance indicators.  Their solution used a more in-depth analysis of peculiar velocity corrections to mitigate their impact.  While significantly less accurate, their estimate of {\Hcst} had the significant methodological advantage of being independent of any potential systematics in the properties of SNe~Ia.  Since this analysis does not use additional independent data, it is not included in our formalism; however, we note here this approach for methodological completeness.  If direct distances to more distant galaxies are measured in the future, this approach may be worth renewed consideration.

% \textcolor{red}{any validation necessary?}

% \textcolor{red}{Continue: explicit form of the solution.  Separation of chi2.}

% \newpage
% \noindent

%% file: Appendices/equation_arrays.tex
\subsection {The system of equations in matrix form}

The following equatons show formally how the equations for hosts, calibrators, and Hubble flow objects fit into the system of equations for the distance network.  Which equations are included of course depends on the specific variant chosen; for simplicity, we show the case in which Hubble flow objects in the same class are grouped in via their corresponding $ \alpha_\iobj $ value.  

\begin{equation*}
\begin{gathered}
\makebox[\textwidth][l]{%
\hspace*{-3.75em}
\begin{tikzpicture}[baseline=(m.center)]
  % ================= MATRIX A =================
  \matrix (m) [matrix of math nodes,
               nodes in empty cells,
               left delimiter={[}, right delimiter={]},
               row sep=2pt, column sep=12pt,
               ampersand replacement=\&] {
  % ---------------------------------------------------------
  % Column blocks (by index):
  % 1–4 = Host Distances | 5–6 = Groups | 7 = M_Ia | 8 = M_II |
  % 9 = M_offset | 10 = M_offset_SBF | 11 = log10(H0) | 12 = mu_coma
  % ---------------------------------------------------------
  %------------------------------- Rung 2
  % Row R2-1
  1 \& \cdots \& 0 \& 0 % [Host Distances 1–4]
  \& 0 \& 0 % [Groups 5–6]
  \& 0 % [M_Ia 7]
  \& 0 % [M_II 8]
  \& 0 % [M_offset 9]
  \& 0 % [M_offset_SBF 10]
  \& 0 % [log10(H0) 11]
  \& 0 % [mu_coma 12]
  \\
  % Row R2-mid
  \vdots \& \ddots \& \vdots \& \vdots % [Host Distances 1–4]
  \& \vdots \& \vdots % [Groups 5–6]
  \& \vdots % [M_Ia 7]
  \& \vdots % [M_II 8]
  \& \vdots % [M_offset 9]
  \& \vdots % [M_offset_SBF 10]
  \& \vdots % [log10(H0) 11]
  \& \vdots % [mu_coma 12]
  \\
  % Row R2-*
  0 \& \cdots \& 0 \& 1 % [Host Distances 1–4]
  \& 0 \& 0 % [Groups 5–6]
  \& 0 % [M_Ia 7]
  \& 0 % [M_II 8]
  \& 0 % [M_offset 9]
  \& 0 % [M_offset_SBF 10]
  \& 0 % [log10(H0) 11]
  \& 0 % [mu_coma 12]
  \\
  % Row R2-*
  0 \& \cdots \& 0 \& 1 % [Host Distances 1–4]
  \& 0 \& 0 % [Groups 5–6]
  \& 0 % [M_Ia 7]
  \& 0 % [M_II 8]
  \& 0 % [M_offset 9]
  \& 0 % [M_offset_SBF 10]
  \& 0 % [log10(H0) 11]
  \& 0 % [mu_coma 12]
  \\
  \hline
  %------------------------------- Groups 
  % Row G1
  1 \& \cdots \& 0 \& 0 % [Host Distances 1–4]
  \& 1 \& 0 % [Groups 5–6]
  \& 0 % [M_Ia 7]
  \& 0 % [M_II 8]
  \& 0 % [M_offset 9]
  \& 0 % [M_offset_SBF 10]
  \& 0 % [log10(H0) 11]
  \& 0 % [mu_coma 12]
  \\
  % Row G2
  \vdots \& \ddots \& \vdots \& \vdots % [Host Distances 1–4]
  \& \vdots \& \vdots % [Groups 5–6]
  \& \vdots % [M_Ia 7]
  \& \vdots % [M_II 8]
  \& \vdots % [M_offset 9]
  \& \vdots % [M_offset_SBF 10]
  \& \vdots % [log10(H0) 11]
  \& \vdots % [mu_coma 12]
  \\
  % Row G3
  0 \& \cdots \& 0 \& 1 % [Host Distances 1–4]
  \& 0 \& 1 % [Groups 5–6]
  \& 0 % [M_Ia 7]
  \& 0 % [M_II 8]
  \& 0 % [M_offset 9]
  \& 0 % [M_offset_SBF 10]
  \& 0 % [log10(H0) 11]
  \& 0 % [mu_coma 12]
  \\
  % Row G3
  0 \& \cdots \& 1 \& 0 % [Host Distances 1–4]
  \& 0 \& 1 % [Groups 5–6]
  \& 0 % [M_Ia 7]
  \& 0 % [M_II 8]
  \& 0 % [M_offset 9]
  \& 0 % [M_offset_SBF 10]
  \& 0 % [log10(H0) 11]
  \& 0 % [mu_coma 12]
  \\
  \hline
  %------------------------------- Rung 3
  % Row R3-1
  1 \& \cdots \& 0 % [Host Distances 1–3]
  \& 0 \& 0 \& 0 % [Groups 4–6]
  \& 1 % [M_Ia 7]
  \& 0 % [M_II 8]
  \& 0 % [M_offset 9]
  \& 0 % [M_offset_SBF 10]
  \& 0 % [log10(H0) 11]
  \& 0 % [mu_coma 12]
  \\
  % Row R3-mid
  \vdots \& \ddots \& \vdots % [Host Distances 1–3]
  \& \vdots \& \vdots \& \vdots % [Groups 4–6]
  \& \vdots % [M_Ia 7]
  \& 1 % [M_II 8]
  \& \vdots % [M_offset 9]
  \& 0 % [M_offset_SBF 10]
  \& \vdots % [log10(H0) 11]
  \& \vdots % [mu_coma 12]
  \\
  % Row R3-*
  0 \& \cdots \& 0 % [Host Distances 1–3]
  \& 1 \& 0 \& 0 % [Groups 4–6]
  \& 0 % [M_Ia 7]
  \& \vdots % [M_II 8]
  \& 1 % [M_offset 9]
  \& \vdots % [M_offset_SBF 10]
  \& 0 % [log10(H0) 11]
  \& 0 % [mu_coma 12]
  \\
  % Row R3-*
  0 \& \cdots \& 0 % [Host Distances 1–3]
  \& 1 \& 0 \& 0 % [Groups 4–6]
  \& 0 % [M_Ia 7]
  \& 0 % [M_II 8]
  \& 0 % [M_offset 9]
  \& 1 % [M_offset_SBF 10]
  \& 0 % [log10(H0) 11]
  \& 0 % [mu_coma 12]
  \\
  \hline
  %------------------------------- ieq_h0_m0 family
  % Row eq-1 (ieq_h0_m0)
  0 \& \cdots \& 0 % [Host Distances 1–3]
  \& 0 \& 0 \& 0 % [Groups 4–6]
  \& -0.2 % [M_Ia 7]
  \& 0 % [M_II 8]
  \& 0 % [M_offset 9]
  \& 0 % [M_offset_SBF 10]
  \& 1 % [log10(H0) 11]
  \& 0 % [mu_coma 12]
  \\
  % Row eq-2
  0 \& \cdots \& 0 \& 0 \& 0 \& 0 \& 0 \& -0.2 \& 0 \& 0 \& 1 \& 0 \\
  % Row eq-3
  0 \& \cdots \& 0 \& 0 \& 0 \& 0 \& 0 \& 0 \& -0.2 \& 0 \& 1 \& 0 \\
  % Row eq-4
  0 \& \cdots \& 0 \& 0 \& 0 \& 0 \& 0 \& 0 \& 0 \& -0.2 \& 1 \& 0 \\
  \hline
  %------------------------------- Megamaser
  % Row M-1
  0 \& \cdots \& 0 \& 0 \& 0 \& 0 \& 0 \& 0 \& 0 \& 0 \& 1 \& 0 \\
  % Row M-mid
  \vdots \& \ddots \& \vdots \& \vdots \& \vdots \& \vdots \& \vdots \& \vdots \& \vdots \& \vdots \& \vdots \& \vdots \\
  % Row M-*
  0 \& \cdots \& 0 \& 0 \& 0 \& 0 \& 0 \& 0 \& 0 \& 0 \& 1 \& 0 \\
  \hline
  %------------------------------- Coma SN
  % Row C-1
  0 \& \cdots \& 0 \& 0 \& 0 \& 0 \& 1 \& 0 \& 0 \& 0 \& 0 \& 1 \\
  % Row C-mid
  \vdots \& \ddots \& \vdots \& \vdots \& \vdots \& \vdots \& \vdots \& \vdots \& \vdots \& 0 \& \vdots \& \vdots \\
  % Row C-*
  0 \& \cdots \& 0 \& 0 \& 0 \& 0 \& 1 \& 0 \& 0 \& 0 \& 0 \& 1 \\
  \hline
  %------------------------------- C_SBF
  0 \& \cdots \& 0 \& 0 \& 0 \& 0 \& 0 \& 0 \& 0 \& 0 \& 0 \& 1 \\
  \hline
  %------------------------------- H_{0,C}
  0 \& \cdots \& 0 \& 0 \& 0 \& 0 \& 0 \& 0 \& 0 \& 0 \& 1 \& 0.2 \\
  };

  % A = (flush to left bracket)
  \node[anchor=east] at ([xshift=-6pt]m.west) {$\mathbf{A} =$};

  % ---------- Horizontal row vector X aligned to A's columns ----------
  \newcommand{\Xraise}{2.0em} % adjust here
  \coordinate (XrowY) at ($(m.north)+(0,\Xraise)$);
  \node (X1)  at ($(m-1-1.north |- XrowY)$) {$\mu_1$};
  \node (X2)  at ($(m-1-2.north |- XrowY)$) {$\cdots$};
  \node (X3)  at ($(m-1-3.north |- XrowY)$) {$\cdots$};
  \node (X4)  at ($(m-1-4.north |- XrowY)$) {$\mu_\gamma$};
  \node (X5)  at ($(m-1-5.north |- XrowY)$) {$\mu_{\rm g,1}$};
  \node (X6)  at ($(m-1-6.north |- XrowY)$) {$\mu_{\rm g,2}$};
  \node (X7)  at ($(m-1-7.north |- XrowY)$) {$M_{\rm Ia}$};
  \node (X8)  at ($(m-1-8.north |- XrowY)$) {$M_{\rm II}$};
  \node (X9)  at ($(m-1-9.north |- XrowY)$) {$M^{\rm off}_{\rm TF}$};
  \node (X10) at ($(m-1-10.north |- XrowY)$) {$M^{\rm off}_{\rm SBF}$};
  \node (X11) at ($(m-1-11.north |- XrowY)$) {$\log_{10} H_0$};
  \node (X12) at ($(m-1-12.north |- XrowY)$) {$\mu_{\rm C}$};
  % Brackets and label for X 
  \node[anchor=east] at ([xshift=-8pt]X1.west) {$\mathbf{X} =$};
  \node at ([xshift=-4pt]X1.west) {$\left[\vphantom{\int}\right.$};
  \node at ([xshift= 4pt]X12.east) {$\left.\vphantom{\int}\right]$};

  % Column block labels 
  \node at ($(m-1-2.north)+(0.9em,1.2em)$) {\scriptsize Host Distances};
  \node at ($(m-1-6.north)+(-0.9em,0.8em)$) {\scriptsize Groups};
  \node at ($(m-1-8.north)+(2em,0.8em)$) {\scriptsize ------Absolute magnitudes------};
  \node at ($(m-1-11.north)+(0,0.8em)$) {\scriptsize $H_0$};
  \node at ($(m-1-12.north)+(0,0.8em)$) {\scriptsize $\mu_{\rm coma}$};

% Row block labels 
   \def\matrixcols{12}
   \node[right=12pt of m-2-\matrixcols]  {\scriptsize Hosts};
   \node[right=12pt of m-6-\matrixcols]  {\scriptsize Groups};
   \node[right=12pt of m-10-\matrixcols]  {\scriptsize Calibrators};
   \node[right=12pt of m-14-\matrixcols] {\scriptsize $H_0$, $M_{\rm Calib}$};
   \node[right=12pt of m-18-\matrixcols] {\scriptsize MM/EPM};
   \node[right=12pt of m-21-\matrixcols] {\scriptsize Coma SN};
   \node[right=12pt of m-23-\matrixcols] {\scriptsize ${\rm C}_{\rm SBF}$};
   \node[right=9pt of m-24-\matrixcols] {\scriptsize $H_{0,{\rm C}}$};

  % Optional vertical separators 
  \foreach \i in {4,11}{
    \draw[dashed] ($(m-1-\i.north east)+(0.7em,0)$) -- ($(m.south east -| m-1-\i.north east)+(0.7em,0)$);
  }
    \draw[dashed] ($(m-1-6.north east)+(1em,0)$) -- ($(m.south east -| m-1-6.north east)+(1em,0)$);
    \draw[dashed] ($(m-1-7.north east)+(1em,0)$) -- ($(m.south east -| m-1-7.north east)+(1em,0)$);
    \draw[dashed] ($(m-1-8.north east)+(1.4em,0)$) -- ($(m.south east -| m-1-8.north east)+(1.4em,0)$);
    \draw[dashed] ($(m-1-9.north east)+(1.4em,0)$) -- ($(m.south east -| m-1-9.north east)+(1.4em,0)$);
    \draw[dashed] ($(m-1-10.north east)+(1em,0)$) -- ($(m.south east -| m-1-10.north east)+(1em,0)$);

  % ---------------- Y VECTOR matching matrix row height
  \matrix (yvec) [
    matrix of math nodes, nodes in empty cells,
    left delimiter={[}, right delimiter={]},
    row sep=2.0pt, column sep=12pt,
    anchor=west, right=55pt of m,
    nodes={minimum height=1em, text depth=0.1ex}
  ] {
    \strut\mu_{\rm host,1} \\
    \vdots \\
    \strut\mu_{\rm host,n-1}\\
    \mu_{\rm host,n} \\
    \hline
    0 \\ \vdots \\ 0 \\ 0 \\
    \hline
    \strut m_{0B,1}\\ \vdots \\ \strut m_{0B,k-1} \\ m_{0B,k}\\
    \hline
    \alpha_{\rm Ia}+5 \\ \alpha_{\rm II}+5 \\ \alpha_{\rm TF}+5 \\ \alpha_{\rm SBF}+5 \\
    \hline
    \log_{10}(H_0)_{\rm MM,1} \\ \vdots \\ \log_{10}(H_0)_{{\rm EPM},m} \\
    \hline
    m_{0,{\rm Coma},1} \\ \vdots \\ \strut m_{0,{\rm Coma},p} \\
    \hline
    \strut \mu_{\rm Coma SBF} \\
    \hline
    \log_{10}(H_0)_{\rm C} +0.2\mu_{\rm C}\\
  };
  \node[anchor=west] at ([xshift=10pt]yvec.east) {$=\ \mathbf{Y}$};
\end{tikzpicture}
}
\\[1ex]
\end{gathered}
\end{equation*}

\pagebreak
\noindent
\begin{equation*}
\begin{gathered}
\makebox[\textwidth][l]{%
\hspace*{-3.75em}
\text{$\mathbf{C}$} =
\begin{tikzpicture}[baseline=(c.center)]
  % COVARIANCE MATRIX DEFINITION
  \matrix (c) [matrix of math nodes,
               nodes in empty cells,
               left delimiter={[},
               right delimiter={]},
               row sep=2pt, column sep=2pt,
               ampersand replacement=\&] {
  % =========================================================
  % Column blocks (by index):
  % 1–3 = Hosts | 4–6 = Groups | 7–9 = Calibrators |
  % 10–13 = Intercpets | 14–16 = MM/EPM | 17–19 = Coma SN |
  % 20 = C_SBF | 21 = H0_C
  % =========================================================
  %------------------------------------ Hosts block rows
  % Row H1
  \sigma^2_{\rm tot,1} \& \cdots \& \sigma^2_{\rm HMAS} % [Hosts]
  \& 0 \& \cdots \& 0 % [Groups]
  \& 0 \& \cdots \& 0 % [Calibrators]
  \& 0 \& 0 \& 0 \& 0 % [Intercepts]
  \& 0 \& \cdots \& 0 % [Megamaser]
  \& 0 \& \cdots \& 0 % [Coma SN]
  \& 0 % [C_SBF]
  \& 0 % [H0_C]
  \\
  % Row H2 (elliptical middle of Hosts)
  \vdots \& \ddots \& \vdots % [Hosts]
  \& \vdots \& \ddots \& \vdots % [Groups]
  \& \vdots \& \ddots \& \vdots % [Calibrators]
  \& \vdots \& \vdots \& \vdots \& \vdots % [Intercepts]
  \& \vdots \& \ddots \& \vdots % [Megamaser]
  \& \vdots \& \ddots \& \vdots % [Coma SN]
  \& \vdots % [C_SBF]
  \& \vdots % [H0_C]
  \\
  % Row Hn
  \sigma^2_{\rm HMAS} \& \cdots \& \sigma^2_{\rm tot,n} % [Hosts]
  \& 0 \& \cdots \& 0 % [Groups]
  \& 0 \& \cdots \& 0 % [Calibrators]
  \& 0 \& 0 \& 0 \& 0 % [Intercepts]
  \& 0 \& \cdots \& 0 % [Megamaser]
  \& 0 \& \cdots \& 0 % [Coma SN]
  \& 0 % [C_SBF]
  \& 0 % [H0_C]
  \\
  \hline
  %------------------------------------ Groups block rows 
  % Row G1
  0 \& \cdots \& 0 % [Hosts]
  \& \sigma^2_{\rm g,1} \& \cdots \& 0 % [Groups]
  \& 0 \& \cdots \& 0 % [Calibrators]
  \& 0 \& 0 \& 0 \& 0 % [Intercepts]
  \& 0 \& \cdots \& 0 % [Megamaser]
  \& 0 \& \cdots \& 0 % [Coma SN]
  \& 0 % [C_SBF]
  \& 0 % [H0_C]
  \\
  % Row G2
   \vdots \& \ddots \& \vdots % [Hosts]
  \& \vdots \& \ddots \& \vdots % [Groups]
  \& \vdots \& \ddots \& \vdots % [Calibrators]
  \& \vdots \& \vdots \& \vdots \& \vdots % [Intercepts]
  \& \vdots \& \ddots \& \vdots % [Megamaser]
  \& \vdots \& \ddots \& \vdots % [Coma SN]
  \& \vdots % [C_SBF]
  \& \vdots % [H0_C]
  \\
  % Row G3
  0 \& \cdots \& 0 % [Hosts]
  \& 0 \& \cdots \& \sigma^2_{\rm g,2} % [Groups]
  \& 0 \& \cdots \& 0 % [Calibrators]
  \& 0 \& 0 \& 0 \& 0 % [Intercepts]
  \& 0 \& \cdots \& 0 % [Megamaser]
  \& 0 \& \cdots \& 0 % [Coma SN]
  \& 0 % [C_SBF]
  \& 0 % [H0_C]
  \\
  \hline
  %------------------------------------ Calibrators block rows
  % Row C1
  0 \& \cdots \& 0 % [Hosts]
  \& 0 \& \cdots \& 0 % [Groups]
  \& \sigma^2_{1} \& \cdots \& 0 % [Calibrators]
  \& 0 \& 0 \& 0 \& 0 % [Intercepts]
  \& 0 \& \cdots \& 0 % [Megamaser]
  \& 0 \& \cdots \& 0 % [Coma SN]
  \& 0 % [C_SBF]
  \& 0 % [H0_C]
  \\
  % Row Cmid
  \vdots \& \ddots \& \vdots % [Hosts]
  \& \vdots \& \ddots \& \vdots % [Groups]
  \& \vdots \& \ddots \& \vdots % [Calibrators]
  \& \vdots \& \vdots \& \vdots \& \vdots % [Intercepts]
  \& \vdots \& \ddots \& \vdots % [Megamaser]
  \& \vdots \& \ddots \& \vdots % [Coma SN]
  \& \vdots % [C_SBF]
  \& \vdots % [H0_C]
  \\
  % Row Ck
  0 \& \cdots \& 0 % [Hosts]
  \& 0 \& \cdots \& 0 % [Groups]
  \& 0 \& \cdots \& \sigma^2_{k} % [Calibrators]
  \& 0 \& 0 \& 0 \& 0 % [Intercepts]
  \& 0 \& \cdots \& 0 % [Megamaser]
  \& 0 \& \cdots \& 0 % [Coma SN]
  \& 0 % [C_SBF]
  \& 0 % [H0_C]
  \\
  \hline
  %------------------------------------ Intercepts block rows
  % Row I1 (Ia)
  0 \& \cdots \& 0 % [Hosts]
  \& 0 \& \cdots \& 0 % [Groups]
  \& 0 \& \cdots \& 0 % [Calibrators]
  \& \sigma^2_{\alpha_{\rm Ia}} \& 0 \& 0 \& 0 % [Intercepts]
  \& 0 \& \cdots \& 0 % [Megamaser]
  \& 0 \& \cdots \& 0 % [Coma SN]
  \& 0 % [C_SBF]
  \& 0 % [H0_C]
  \\
  % Row I2 (II)
  0 \& \cdots \& 0 % [Hosts]
  \& 0 \& \cdots \& 0 % [Groups]
  \& 0 \& \cdots \& 0 % [Calibrators]
  \& 0 \& \sigma^2_{\alpha_{\rm II}} \& 0 \& 0 % [Intercepts]
  \& 0 \& \cdots \& 0 % [Megamaser]
  \& 0 \& \cdots \& 0 % [Coma SN]
  \& 0 % [C_SBF]
  \& 0 % [H0_C]
  \\
  % Row I3 (TF)
  0 \& \cdots \& 0 % [Hosts]
  \& 0 \& \cdots \& 0 % [Groups]
  \& 0 \& \cdots \& 0 % [Calibrators]
  \& 0 \& 0 \& \sigma^2_{\alpha_{\rm TF}} \& 0 % [Intercepts]
  \& 0 \& \cdots \& 0 % [Megamaser]
  \& 0 \& \cdots \& 0 % [Coma SN]
  \& 0 % [C_SBF]
  \& 0 % [H0_C]
  \\
  % Row I4 (SBF)
  0 \& \cdots \& 0 % [Hosts]
  \& 0 \& \cdots \& 0 % [Groups]
  \& 0 \& \cdots \& 0 % [Calibrators]
  \& 0 \& 0 \& 0 \& \sigma^2_{\alpha_{\rm SBF}} % [Intercepts]
  \& 0 \& \cdots \& 0 % [Megamaser]
  \& 0 \& \cdots \& 0 % [Coma SN]
  \& 0 % [C_SBF]
  \& 0 % [H0_C]
  \\
  \hline
  %------------------------------------ Megamaser block rows
  % Row M1
  0 \& \cdots \& 0 % [Hosts]
  \& 0 \& \cdots \& 0 % [Groups]
  \& 0 \& \cdots \& 0 % [Calibrators]
  \& 0 \& 0 \& 0 \& 0 % [Intercepts]
  \& \sigma^2_{\rm 1} \& \cdots \& 0 % [Megamaser]
  \& 0 \& \cdots \& 0 % [Coma SN]
  \& 0 % [C_SBF]
  \& 0 % [H0_C]
  \\
  % Row Mmid
  \vdots \& \ddots \& \vdots % [Hosts]
  \& \vdots \& \ddots \& \vdots % [Groups]
  \& \vdots \& \ddots \& \vdots % [Calibrators]
  \& \vdots \& \vdots \& \vdots \& \vdots % [Intercepts]
  \& \vdots \& \ddots \& \vdots % [Megamaser]
  \& \vdots \& \ddots \& \vdots % [Coma SN]
  \& \vdots % [C_SBF]
  \& \vdots % [H0_C]
  \\
  % Row Mm
  0 \& \cdots \& 0 % [Hosts]
  \& 0 \& \cdots \& 0 % [Groups]
  \& 0 \& \cdots \& 0 % [Calibrators]
  \& 0 \& 0 \& 0 \& 0 % [Intercepts]
  \& 0 \& \cdots \& \sigma^2_{\rm m} % [Megamaser]
  \& 0 \& \cdots \& 0 % [Coma SN]
  \& 0 % [C_SBF]
  \& 0 % [H0_C]
  \\
  \hline
  %------------------------------------ Coma SN block rows
  % Row P1
  0 \& \cdots \& 0 % [Hosts]
  \& 0 \& \cdots \& 0 % [Groups]
  \& 0 \& \cdots \& 0 % [Calibrators]
  \& 0 \& 0 \& 0 \& 0 % [Intercepts]
  \& 0 \& \cdots \& 0 % [Megamaser]
  \& \sigma_1^2 \& \cdots \& 0 % [Coma SN]
  \& 0 % [C_SBF]
  \& 0 % [H0_C]
  \\
  % Row Pmid
  \vdots \& \ddots \& \vdots % [Hosts]
  \& \vdots \& \ddots \& \vdots % [Groups]
  \& \vdots \& \ddots \& \vdots % [Calibrators]
  \& \vdots \& \vdots \& \vdots \& \vdots % [Intercepts]
  \& \vdots \& \vdots \& \vdots % [Megamaser]
  \& \vdots \& \ddots \& \vdots % [Coma SN]
  \& \vdots % [C_SBF]
  \& \vdots % [H0_C]
  \\
  % Row Pp
  0 \& \cdots \& 0 % [Hosts]
  \& 0 \& \cdots \& 0 % [Groups]
  \& 0 \& \cdots \& 0 % [Calibrators]
  \& 0 \& 0 \& 0 \& 0 % [Intercepts]
  \& 0 \& \cdots \& 0 % [Megamaser]
  \& 0 \& \cdots \& \sigma_p^2 % [Coma SN]
  \& 0 % [C_SBF]
  \& 0 % [H0_C]
  \\
  \hline
  %------------------------------------ C_SBF scalar row
  % Row CSBF
  0 \& \cdots \& 0 % [Hosts]
  \& 0 \& \cdots \& 0 % [Groups]
  \& 0 \& \cdots \& 0 % [Calibrators]
  \& 0 \& 0 \& 0 \& 0 % [Intercepts]
  \& 0 \& \cdots \& 0 % [Megamaser]
  \& 0 \& \cdots \& 0 % [Coma SN]
  \& \sigma^2_{{\rm C}_{\rm s}} % [C_SBF]
  \& 0 % [H0_C]
  \\
  \hline
  %------------------------------------ H0_C scalar row
  % Row H0C
  0 \& \cdots \& 0 % [Hosts]
  \& 0 \& \cdots \& 0 % [Groups]
  \& 0 \& \cdots \& 0 % [Calibrators]
  \& 0 \& 0 \& 0 \& 0 % [Intercepts]
  \& 0 \& \cdots \& 0 % [Megamaser]
  \& 0 \& \cdots \& 0 % [Coma SN]
  \& 0 % [C_SBF]
  \& \sigma^2_{\rm C}% [H0_C]
  \\
};

  \def\matrixcolsC{21}

  \node[right=12pt of c-2-\matrixcolsC] {\scriptsize Hosts};
  \node[right=12pt of c-5-\matrixcolsC] {\scriptsize Groups};
  \node[right=12pt of c-8-\matrixcolsC] {\scriptsize Calibrators};
  \node[right=11pt of c-12-\matrixcolsC, yshift=10pt] {\scriptsize Intercepts};
  \node[right=12pt of c-15-\matrixcolsC] {\scriptsize MM/EPM};
  \node[right=12pt of c-18-\matrixcolsC] {\scriptsize Coma SN};
  \node[right=11pt of c-20-\matrixcolsC] {\scriptsize ${\rm C}_{\rm SBF}$};
  \node[right=8pt of c-21-\matrixcolsC] {\scriptsize $H_{0,{\rm C}}$};

  % Block labels (top)
  \node[above=6pt of c-1-2] {\scriptsize Hosts};
  \node[above=5pt of c-1-5] {\scriptsize Groups};
  \node[above=6pt of c-1-8] {\scriptsize Calibrators};
  \node[above=3.5pt of c-1-12, xshift=-10pt] {\scriptsize Intercepts};
  \node[above=5pt of c-1-15] {\scriptsize MM/EPM};
  \node[above=6pt of c-1-18] {\scriptsize Coma SN};
  \node[above=3pt of c-1-20] {\scriptsize ${\rm C}_{\rm SBF}$};
  \node[above=2pt of c-1-21] {\scriptsize $H_{0,{\rm C}}$};

  % Vertical lines between blocks
  \draw[dashed] ($(c-1-3.north east)+(0,0)$) -- ($(c-\matrixcolsC-3.north east)+(1em,-1.25em)$);
  \foreach \i in {6,9,13,16,19,20} {
    \draw[dashed] ($(c-1-\i.north east)+(0.5em,0)$) -- ($(c-\matrixcolsC-\i.north east)+(0.5em,-1.25em)$);
  }
\end{tikzpicture}
}
\\[1ex]
\text{\footnotesize System of equations and covariance matrix adopted within this work.}
\end{gathered}
\end{equation*}

%% file: Appendices/AC_comparisons.tex
\section{Literature comparison}\label{sec:comparisons}
Given that the distance network is an extension of the distance ladder framework, it is only natural, and an important test, that we should be able to reproduce various literature results using this framework. A few caveats to this arise from the simplifications we made in order to build the distance network, as well as possibly different treatments of objects in the Hubble flow (e.g., different peculiar velocity treatments). However, these are not expected to cause significant differences ($\lesssim 0.3$ km\,s$^{-1}$Mpc$^{-1}$) in the final determination of \Hcst. 

\subsection{Literature Replication and Comparison for SNe~Ia\label{app:replication}}
Given the dominant role of the SNe~Ia in previous claims of a Hubble tension, we focus in this part of the appendix on the agreement of the SNe~Ia-derived results by exploring a broader range of SN Ia light curve fitters and SN Ia calibration samples than those included in the main variants Section~\ref{sec:variants} that may help explain the origin of differences in $H_0$ compared to literature studies. 
\begin{figure}
    \centering
    \includegraphics[width=0.9\linewidth]{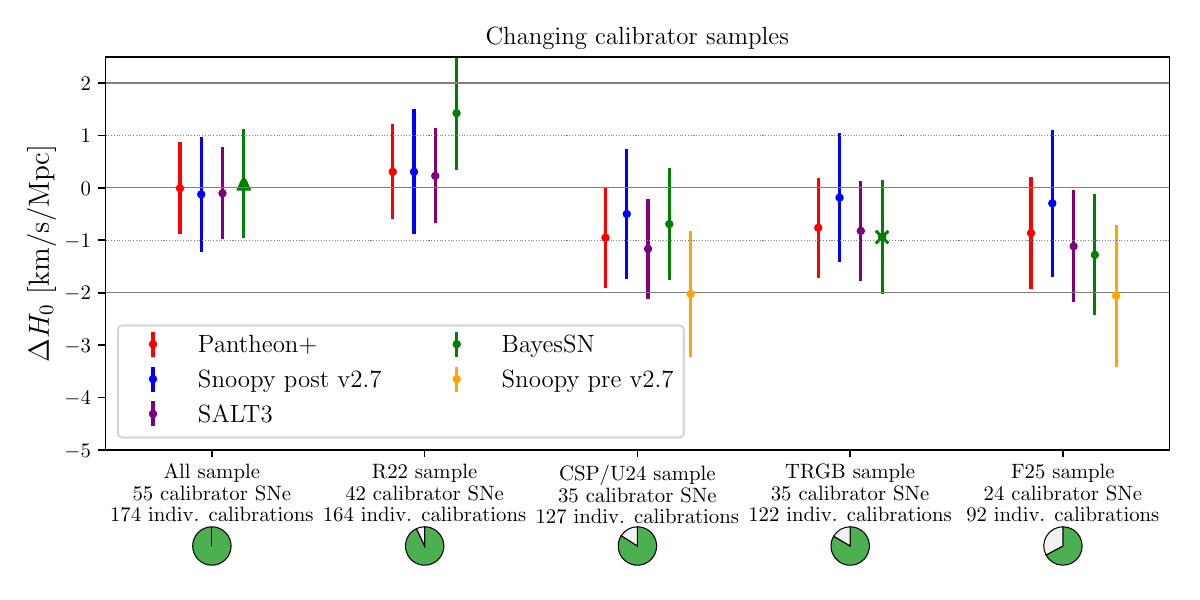}\caption{Comparison of the impact of supernova type Ia calibrator samples (sets of hosts) on the Hubble constant inference $H_0$\,, each one relative to the baseline result, for the various light curve fitting codes. The triangle shows a case where for the bayesSN fitter only 52 out of the 55 hosts was calibrated due to several missing supernovae in the fit, and the cross one where only 32 out of 35 hosts were calibrated. For details on the pie charts, see Fig.~\ref{fig:paths_consistency}. In this case we take 100\% of the pies to be the case involving all supernovae data instead of the baseline.}
    \label{fig:literature_sn1a}
\end{figure}

\begin{table}[h]
\begin{center}
\centering
\begin{tabular}{l c c c c c}
\hline
SN Sample/Calibration  (SN count) & Pantheon+ & SNooPy post v2.7 & SALT3 & BayesSN & SNooPy pre v2.7 \\
\hline
All/All (55) & $73.49 \pm 0.87$ & $73.37 \pm 1.10$ & $73.39 \pm 0.88$ & $(73.58 \pm 1.03)^{*}$ & --- \\
R22/All (42) & $73.80 \pm 0.91$ & $73.80 \pm 1.19$ & $73.72 \pm 0.91$ & $74.92 \pm 1.09$ & --- \\
R22/Cepheids (42) & $73.17 \pm 0.96$ & $73.33 \pm 1.23$ & $73.06 \pm 0.96$ & $74.67 \pm 1.14$ & --- \\
CSP in U24/All (35) & $72.54 \pm 0.95$ & $72.99 \pm 1.23$ & $72.33 \pm 0.95$ & $72.80 \pm 1.08$ & $71.47 \pm 1.20$ \\
CSP in U24/TRGB (27) & $71.61 \pm 1.53$ & $72.62 \pm 1.75$ & $71.50 \pm 1.52$ & $71.44 \pm 1.57$ & $70.99 \pm 1.70$ \\
TRGB HST-JWST/All (35) & $72.73 \pm 0.96$ & $73.31 \pm 1.23$ & $72.67 \pm 0.96$ & $(72.55 \pm 1.09)^{+}$ & --- \\
TRGB HST-JWST/TRGB (35) & $72.18 \pm 1.48$ & $72.66 \pm 1.64$ & $72.13 \pm 1.48$ & $(71.55 \pm 1.55)^{+}$ & --- \\
F25/All (24) & $72.63 \pm 1.06$ & $73.20 \pm 1.40$ & $72.38 \pm 1.06$ & $72.22 \pm 1.16$ & $71.44\pm 1.36$ \\
F25/TRGB (24) & $71.09 \pm 1.63$ & $72.05 \pm 1.85$ & $70.97 \pm 1.63$ & $71.00 \pm 1.66 $& $70.31 \pm 1.80$ \\
\hline
\end{tabular}
\caption{\label{tab:literature_sn1a}Comparison of Hubble constant values (in km\,s$^{-1}$Mpc$^{-1}$) from different supernova sets and calibrators. Note that for the same calibrator sets differences in the uncertainty stem mostly from the number of objects in the Hubble flow. SNooPy pre v2.7 is the one used in U24/F25. \\ {\footnotesize${}^{*}$ In this case only 52 out of the 55 calibrators could be matched with the given set of fitted SNeIa.}\\{\footnotesize${}^{+}$ In this case only 32 out of the 35 calibrators could be matched with the given set of fitted SNeIa.}}
\end{center}
\end{table}

First, we compare our distance network in the case that we only use type Ia supernovae as secondary distance indicators and in the Hubble flow. For this case, we can compare to three references, namely \cite{2022ApJ...934L...7R,2023MNRAS.524..235D,2025ApJ...985..203F}. This comparison is graphically summarized in Figure~\ref{fig:literature_sn1a} and Figure~\ref{fig:literature_sn1a_delta} , and the corresponding numbers can be found in Tables~\ref{tab:literature_sn1a} and \ref{tab:literature_other}. We observe that all comparisons agree between the publication and the \DN{} emulation within less than $0.15\sigma$ on the inferred value of $H_0$\,. In particular, when using only the 42 calibrators of \cite{2022ApJ...934L...7R} for Pantheon+ supernovae and adopting only the Cepheid calibration as in that study, we find $\Hcst = 73.17 \pm 0.96 \Hunit$ where they find $\Hcst=73.15 \pm 0.97\Hunit$ ($0.02\sigma$ difference, 1\% smaller uncertainty). Note that we compared to the results of \cite{2022ApJ...938...36R}, as the {\DN} also includes the cluster Cepheids that tighten the calibration in the anchors. When comparing to the Cepheid-based results of \cite{2023MNRAS.524..235D} based on BayesSN supernovae, we find in this case $\Hcst = 74.67 \pm 1.14 \Hunit$ where they find $\Hcst = 74.82 \pm 1.28 \Hunit$ ($-0.12\sigma$ difference, 11\% smaller uncertainty). In this case, there is an additional caveat that \cite{2023MNRAS.524..235D} only used 41 calibrators whereas with the BayesSN supernova fitter we used 42 Cepheid calibrators in this case. Finally, when comparing to the 24 TRGB-calibrated SN Ia results of \cite{2025ApJ...985..203F} (from HST or JWST), we find $\Hcst = 70.31 \pm 1.80 \Hunit$ where they find $\Hcst = 70.39 \pm 1.22 \mathrm{(stat)} \pm 1.33 \mathrm{(sys)} \pm 0.70 \mathrm{(\sigma_\mathrm{SN})} \Hunit$ ($-0.06\sigma$ difference, 0.3\% smaller uncertainty when comparing the relevant uncertainties in quadrature). We can also compare the values of the calibration in these cases, finding for \cite{2022ApJ...934L...7R} $M_B = -19.260 \pm 0.027$ ($-0.18\sigma$ compared to $M_B = -19.253 \pm 0.027$), for \cite{2023MNRAS.524..235D} $\Delta M_B = 0.039 \pm 0.027$ ($+0.4\sigma$ compared to $\Delta M_B = 0.030 \pm 0.023$, with one additional calibrator), and for \cite{2025ApJ...985..203F} $M_B = -19.186 \pm  0.053$ ($-0.11\sigma$ compared to $M_B = -19.18 \pm 0.04 \mathrm{(stat only)}$).  We thus take as a noteworthy conclusion that the DN can accurately reproduce the individual distance ladder results based on the host distances, standardized SN magnitudes, and geometric calibrations used in these studies.  

\subsection{Understanding differences in \texorpdfstring{$H_0$}{H0} for past SN Ia-based ladders}\label{app:sec:SNIa_literature}

Since we can reproduce the literature results within our framework, we can also investigate the sources of differences
in $H_0$ which may be attributed to specific analysis choices and datasets.  

\subsubsection{Refitting with new, SNooPy v2.7.0\label{app:snoopy}}
An immediate challenge to comparing and extending the SN Ia results that make use of the SNooPy fitter (used in all CSP and CCHP studies) is both the absence of many useful SN calibrators in the \citet{2024ApJ...970...72U} (hereafter U24) the compilation that was used in \cite{2025ApJ...985..203F} (hereafter F25) as well as the update to the SNooPy fitter to version v2.7.0 or later (since Aug 27, 2024)-- with the prior version which we dub ``pre v2.7'' (the prior release was 2.5.3 Jan 29, 2020) that was used in U24 not readily accessible nor recommended for further use by its author.  According to C. Burns who supports SNooPy development (private communication, 2025) current studies are advised to use the latest, v2.7.0.  The most relevant update in v2.7 involves the method used by SNooPy to extrapolate the light curve templates beyond the training sample ($s(B-V) < 0.3$ and $s(B-V) > 1.27$).
Such SN Ia have light curves and inferred luminosities near the extrema of the distributions (often going by the name ``91bg-like'' at the faint and fast side and "91T-like" at the bright and slow/high stretch side).  This has very little affect on the calibrator sample (which are often pre-selected to be normal or near the middle of the distribution, see R22).  However, the Hubble flow sample, and specifically the CSPII sample has a surprising abundance of ``high-stretch'' or 91T-like SNe Ia (U24). We find $18$ supernovae in CSPII and $5$ in CSP I (so 23 in all).  Such SNe Ia will generally have $s(B-V) \geq 1.15$ for which the impact from the version update is quite large.
The prior SNooPy version (pre v2.7) was found to over-correct these, motivating the revision to the fitter. For the updated SNooPy v2.7, prior fits with $s(B-V) \sim 1.15$ are lowered on average to $s(B-V) \sim 1.0$ (the fitted, apparent peak magnitudes are hardly impacted) and the implied reddening in $B-V$ is increased by $\sim$ 0.04 mag. The net change comparing old and new follows as $\Delta m \approx 0.9 \cdot (1.15-1.0)-2.9 \cdot(-0.04) \approx 0.25$ mag (neglecting a small quadratic term in $s(B-V)$ for this illustration) for the high $s(B-V)$ supernovae with the new standardized magnitude being brighter (since a brightness previously attributed to over-luminosity and low extinction is now reflected in the newly, brighter standardized magnitudes). Specific examples which all have $s(B-V)>$1.1 include CSP12G, CSP13Z, CSP13aad, LSQ12agd, LSQ12gef, LSQ12hxx, LSQ13dqh, and PTF14uo which all change by 0.2-0.3 mag in the same, brighter direction.  As 91T's represent $\sim$ 8\% of the CSP sample, we might expect a mean change of $\sim$ 0.02 mag which is similar to the actual change from the new version of $\sim$0.03 mag (which includes error weighting).  This appears as an increase in the Hubble flow intercept of $a_B=0.6843\pm0.0025$ to $a_B=0.6895\pm0.0026$ $(+2\sigma)$. As a result $H_0$ increases by 1.0 km\,s$^{-1}$Mpc$^{-1}$ due only to the change in the measurement of the Hubble flow for CSP from the new SNooPy.

An independent consideration is that it is also necessary to use the newer SNooPy v2.7 to fit additional SN Ia calibrators not included in the U24 compilation produced from pre v2.7 SNooPy.  New SNooPy v2.7 fits were thus made by C. Burns (private communication, 2025) using the versions of the light curves in the Pantheon+ light curve database, most of which are not CSP light curves.  In the process a number of discrepancies were identified from the prior SNooPy fits in U24 and \citet{2019ApJ...882...34F} (hereafter F19).  These were either the use of an older data release when a newer and better one was available (e.g., \texttt{Cfa I} from \citet{1999AJ....117..707R} did not employ host subtraction which was applied and released in \citet{2005ApJ...627..579R,2007ApJ...659...98R}), an inconsistency in a filter function definition (i.e., when older published data were provided after transformation to the Landolt/Bessel system, not the ``natural'' system, e.g., 1998bu and 2002fk)  or the use of a broader set of optical light curves was available. 

These differences tend to pertain to older light curves from \texttt{CfA I} or \texttt{LOSS}, with changes in either direction, as tabulated in Table~\ref{table:choice_data}. 

\begin{table}[h!]
\centering
\caption{Differences in SN~Ia lightcurve data adopted in F19/U24/F25 compared to the best available data.}
\begin{tabular}{lllll}
\hline
\hline
SN & Used in F19/U24/F25 & Best Available & Improvement & Difference $\Delta B_{corr}$\\
\hline
1992A  (Cfa I)      & Suntzeff (1996)   & Suntzeff (1996)    &  fit to UBVRI (not BV) &  $0.08$ \, mag \\
1994ae (Cfa I)      & \citet{1999AJ....117..707R}$^{(a)}$ & \citet{2005ApJ...627..579R}   &  host subtraction, 4x phot cal. &  $0.02$ \, mag \\
1995al (Cfa I)     & \citet{1999AJ....117..707R} & \citet{2009ApJS..183..109R}   &  host subtraction, 4x phot cal. &  $0.29$ \, mag \\
1998bu (Cfa I)      & \citet{1999ApJS..125...73J} & \citet{1999ApJS..125...73J}   &  filter function &  $0.25$ \, mag \\
2002fk (LOSS)      & \citet{2009ApJS..183..109R}  & \citet{2009ApJS..183..109R}    &  filter function &  $-0.06$ \, mag \\
2012cg (LOSS) & \citet{2016ApJ...820...92M}  & \citet{2019MNRAS.490.3882S}    & final reduction  & $-0.28$ \, mag \\ 
2005cf (LOSS) & Open Source Catalogue &  \citet{2019MNRAS.490.3882S} & final reduction & $-0.06$ \, mag \\ 
2005cf (LOSS) &  Open Source Catalogue & \citet{2019MNRAS.490.3882S} & final reduction & $-0.04$ \, mag \\ 
2013dy (LOSS) &  Open Source Catalogue  & \citet{2019MNRAS.490.3882S}  &  final reduction  & $0.35$ \, mag \\ 
2017cbv (LOSS) & Open Source Catalogue & \citet{2019MNRAS.490.3882S} &  final reduction & $-0.10$ \, mag \\
2011fe  & \citet{2012JAVSO..40..872R} & LOSS &  coverage &  $0.04$ \, mag  \\
\hline
~ \\
\end{tabular}
\label{table:choice_data}
\tablecomments{$^{(a)}$ Reference F19 claims that 1994ae used the version of the light curve from  \citet{2005ApJ...627..579R} , but C.~Burns, who performed the fit, sourced the data from the Open Source Catalogue, which instead used \citet{1999AJ....117..707R}.  We further confirm that the value in \citet{2024ApJ...970...72U} matches \citet{1999AJ....117..707R}, not \citet{2005ApJ...627..579R}}
\end{table}

Due to the rarity of NIR light curves for calibrators which span decades and require data before NIR observations were common, to avoid introducing a systematic error (from the difference between optical and NIR calibration) we did not include NIR data for the CSP Hubble flow (in practice we found this difference is negligible). See however \ref{sec:variant_description} for variants that specifically use only NIR information for both calibrators and Hubble flow.

The net change from SNooPy pre-v2.7 to post-v2.7 makes the mean calibrator fainter in $M_B$ by 0.026 mag for the F25 sample of $24$ SNeIa and 0.019 mag for the full in-common sample of $35$ SNeIa. {\bf Overall, compared to prior SNooPy fits, the fits of the current version (SNooPy v2.7) raise $H_0$ for the same SN samples by $\mathbf{\sim1.7\,\mathbf{km \, s^{-1} \, Mpc^{-1}}}$ (e.g., from 70.31 for the F25 sample to 72.05).}  This difference is caused then primarily from improvements to SNooPy and to a lesser degree from better, later versions of the SN Ia calibrator data, see Table~\ref{tab:snoopy}.

\begin{table}[h!]
\centering
\caption{Dependence of $H_0$ on SNooPy Fitter version and SN Sample for TRGB Ladder\label{tab:snoopy}}
\begin{tabular}{lcc}
\hline
Result/difference &  \textbf{$H_0$} & \textbf{Notes} \\
\hline
F25 (SNooPy pre v2.7, F25 sample) & ${\bf 70.39} \pm 1.80$ ($N=24$) & F25 as published (without additional SN calibration error) \\
\,\,\,\,\,\,  \rotatebox[origin=c]{180}{$\Lsh$} No change to SNooPy version or sample & $70.31 \pm 1.80$ ($N=24$) & Reproduced within our network \\
\,\,\,\,\,\,  \rotatebox[origin=c]{180}{$\Lsh$} SNooPy post v2.7, F25 sample & $72.05 \pm 1.85$ ($N=24$) & New SNooPy raises $H_0$ \\
\,\,\,\,\,\,  \rotatebox[origin=c]{180}{$\Lsh$} SNooPy pre v2.7, all available pre v2.7 sample & 71.40 $\pm 1.72$ ($N=28$) & SN in U24 not included in F25 raises $H_0$ \\
SNooPy post v2.7, all available post v2.7 sample &  ${\bf 72.66}\pm 1.64$  ($N=35$) & combination, new SnooPy+all SNe with TRGB \\
\hline
\end{tabular}
\vspace{2mm}
\parbox{0.9\linewidth}{\footnotesize \textit{Note.} Differences with F25 for Same Ladder ,  NGC 4258 $\rightarrow$ HST+JWST TRGB $\rightarrow$ CSP SN Ia w/ SNooPy

SNooPy pre v2.7 is from Uddin+24 and also limited to sample available there.  SNooPy post v2.7 includes current version of SNooPy with refits of best data versions by C. Burns as described in text. There is no difference in peculiar velocity treatment across entries.}
\end{table}

\subsubsection{Differences with Calibrator Sample Size}

Lastly, within the use of every SN fitter, increasing the sample size from the smallest, selected study (the F25 sample of 24 calibrators) to the full available sample of 55 calibrators shows a clear trend towards increasing $H_0$---as evident from Figure~\ref{fig:literature_sn1a} and Table~\ref{tab:literature_sn1a}. It increases by $0.9\Hunit$ for Pantheon+ fitting, by $0.2\Hunit$ for SNooPy post-v2.7 fitting, by $1.0\Hunit$ for Salt 3 fitting, and by $1.4\Hunit$ for BayesSN fitting---a mean of $+0.9 \Hunit$. This difference due to calibrator sample is even greater between the F25 sample and that from R22, about $1.2 \Hunit$ in the mean, agreeing with the findings of \cite{2024ApJ...977..120R}. This is shown in Figure~\ref{fig:literature_sn1a}. It seems reasonable to expect greater convergence in estimates of $H_0$ as sample sizes increase.

\begin{figure}
    \centering
    \includegraphics[width=0.95\linewidth]{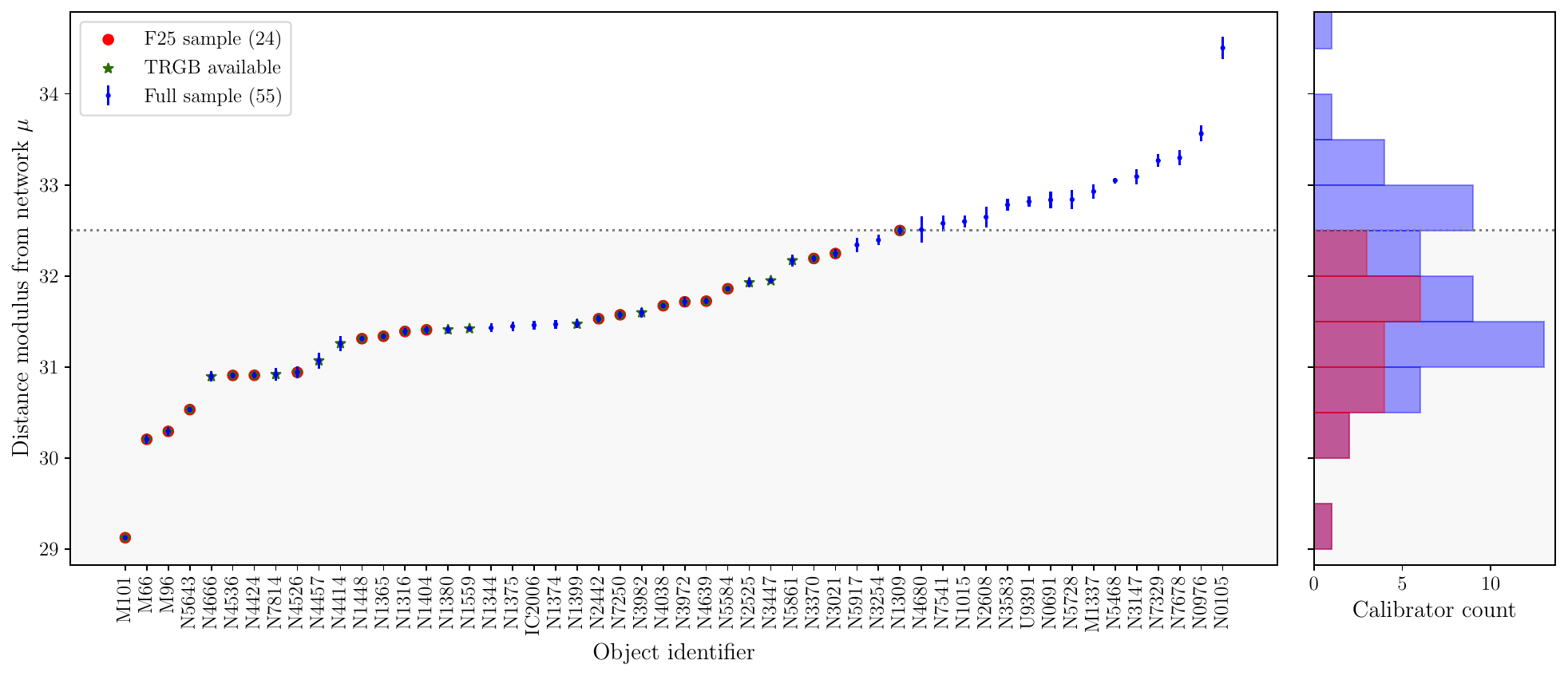}
    \caption{The full network sample (blue) and the F25 sample of calibrator hosts (red) in comparison. The grey region marks the maximal extent of the F25 sample. The green stars mark hosts with TRGB distances available. The left hand side shows an ordered scatterplot of the distance moduli, while the right hand side shows the corresponding histogram. Note that while F25 calibrates 24 SN~Ia, these are located in only 20 hosts (hence only 20 points are marked red). The 11 additional hosts marked with green stars would correspond to potentially 10 additional SNe~Ia (with the 11th host NGC 1399 calibrating SBF, not SNe~Ia). One additional SN~Ia (2021pit) is present in an already F25-calibrated host (N5643) but not within the CSP sample. In total that makes 24+10+1=35 calibrators that are available for SNe~Ia calibrated through TRGB. Even within the sample of the CSP I\&II supernovae only, there would still be 4 additional CSPI\&II calibrators to be gained using TRGB observations of nearby hosts within the range of the F25 sample---with 7 additional TRGB-calibratable SNe~Ia not present within the CSP I\&II samples. The 4 ``missing'' TRGB-F25 calibrators are 2012ht in NGC~3447, 1998aq in NGC~3982, 1992A in NGC~1380, and ASAS14-lp in NGC~4666.}
    \label{fig:mu_sample}
\end{figure}

We note that the F25 sample is not a natural sample selection from the perspective of the \DN{}---it cannot be constructed from simple cuts in redshift/distance or calibrator type. As such, we do not consider it as a variant (according to the methodology laid out in Section~\ref{sec:select_variants}, those should be based on physical assumptions or hypotheses). We analyze it here to provide for a direct comparison with \citet{2025ApJ...985..203F}. We show the distribution in distance of the F25 ($N=24$) and full samples ($N=55$) according to the network in Figure~\ref{fig:mu_sample}---which shows that 37 of the 55 objects would have to be included if a cut was made in distance modulus to retain the full F25 sample. The smaller F25 sample is at a mean distance moduli of $\mu_\mathrm{host} = 31.13 \pm 0.85$, and the remaining sample is at a mean distance moduli of $\mu_\mathrm{host} = 32.01 \pm 0.73$, which are thus a little further away on average (though not very significantly). 

\begin{figure}
    \centering
    \includegraphics[width=0.7\linewidth]{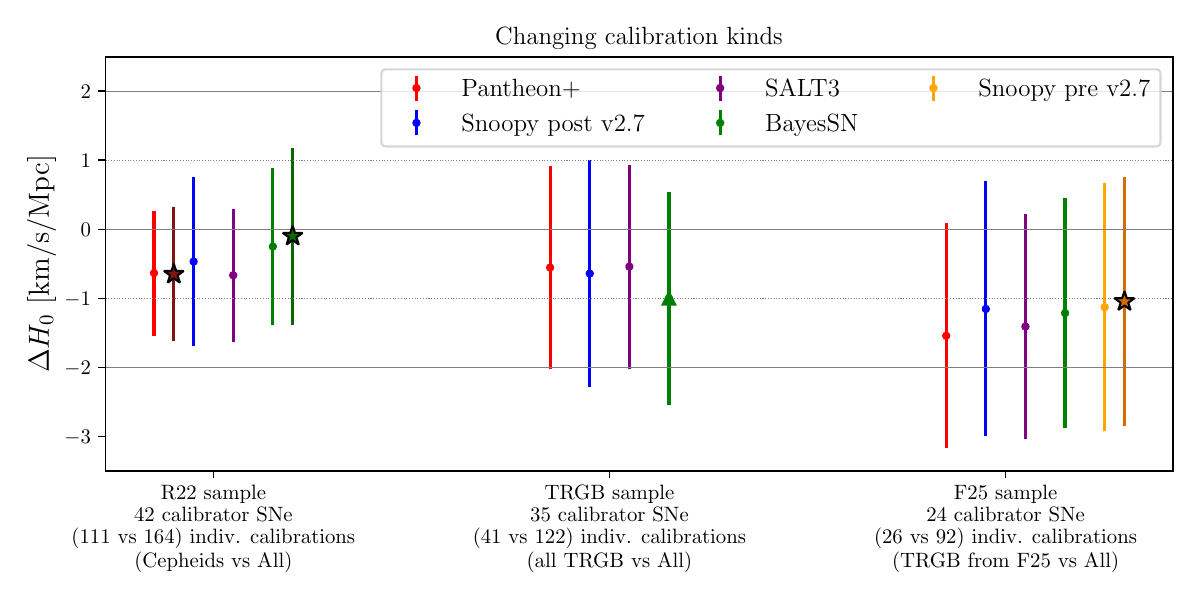}
    \caption{Comparison of $H_0$ between different ways of calibrating a given set of hosts and a gievn SN fitter. It is computed as the result with the hosts calibrated through any means minus the result with the hosts calibrated through a specific calibration type (such as only Cepheids, only TRGB, or only F25 TRGB). The results with the star mark the literature results for the given (reduced) set of calibrating a given set of hosts. The triangle shows a case where for the BayesSN fitter only 32 out of the 35 hosts were calibrated due to several missing supernovae.}
    \label{fig:literature_sn1a_delta}
\end{figure}

For this purpose we can also look at Figure~\ref{fig:literature_sn1a_delta}, which shows the differences between calibrator sets for the same supernovae standardizations and datasets. In each case we take the same supernovae to be used for calibration, but vary with which primary distance indicators the distances to their hosts are calibrated, for example using only HST Cepheids vs using all methods, or using only the TRGB (JWST and HST) vs using all methods, or using only the TRGB calibrations adopted in F25 vs using all methods. These differences are typically compatible with the uncertainty at $\sim 1\sigma$, with shifts around $-1\Hunit$ in central value. We also see that the reduced calibrations agree nicely with the results reported in the literature.

\subsection{Literature comparison for other distance indicators}

\begin{figure}
    \centering
    \includegraphics[width=0.95\linewidth]{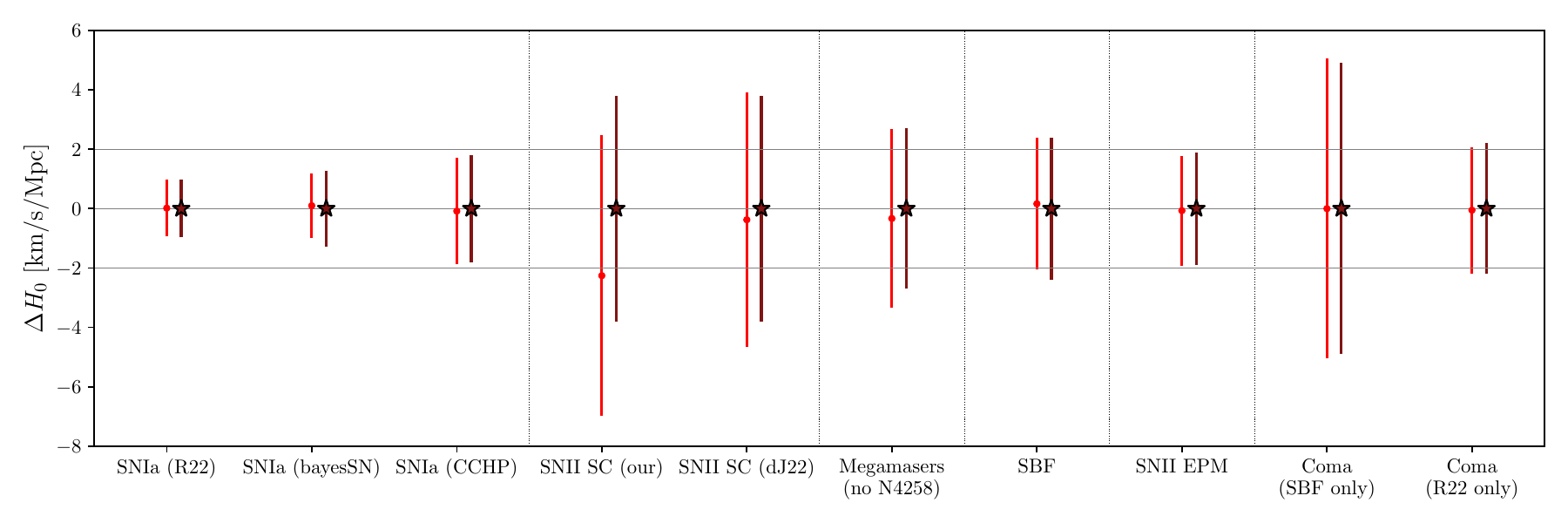}
    \caption{All literature comparisons for the given configuration/method (see x-axis label and main text). The light red results are the results for the given configuration/method coming from the {\DN}, while the dark red results are those available in the literature (differences are always zero but illustrates literature uncertainties), see also Table~\ref{tab:literature_other}.}
    \label{fig:literature_other}
\end{figure}

Next, we compare our distance network to the case where we use other distance indicators. The results can be found in Figure~\ref{fig:literature_other} and Table~\ref{tab:literature_other}. These results are generally very consistent with the literature results, deviations typically staying at $0.1\sigma$ or less, once the same configuration is enforced. 

The SNII standard candle method naively appears to give discrepant results ``out-of-the-box'' in the distance network compared to the results from \citet{2022MNRAS.514.4620D}. However, this is due to the change in calibration described in Appendix~\ref{sec:data:snii_SC}, where we only use 9 out of the 13 possible supernovae as described in that appendix. If we adopt the precise calibration of \citet{2022MNRAS.514.4620D}, we recover their value very closely. For the megamasers of \citet{2020ApJ...891L...1P}, we recover the value without any changes. However, it is important to compare with the correct value, as our baseline employs 2M++ velocity corrections, and does not include N4258 as a non-anchor megamaser. For surface brightness fluctuations (SBF), it is important to correct the literature value for the impact of peculiar velocities manually, which we do by using the difference from peculiar velocities from \citet{2021ApJ...911...65B} (which is $\Delta H_0 = + 0.27 \Hunit$) and applying that to the value of \citet{2025MNRAS.539.3627S}, correcting the central value from $73.9$ to $74.17\Hunit$. We assume no further changes in the uncertainty. For the astrophysical modeling of type II supernovae (using the tailored expanding photosphere method (EPM)), we can directly reproduce the literature value within great agreement. For the Coma-derived value we check both values derived using only the SBF calibration or using only the supernovae from \citet{2022ApJ...934L...7R}. In both cases the result has excellent agreement with the literature value. Finally, we note that we did not manage to recover the literature value of the Tully-Fisher relation, but private communication with P. Boubel revealed that there might have been an issue in their code, leading to an incorrect value.  This only affects \cite{2024MNRAS.531...84B}; future work will be based on the data in earlier publications, which were not available to us at the time of this analysis.

\begin{table}[h]
\caption{Comparison between Recent Literature Results and DN Emulation with the Same Data}
    \centering
    \begin{tabular}{l|l l c}
       Configuration  & Literature & Reference value & Our (DN) value \\ \hline
       SNIa (Pantheon+, R22 sample) & \citet{2022ApJ...938...36R} & $73.17\pm0.96$ & $73.15 \pm 0.97$ \\
       SNIa (BayesSN, R22 sample) & \citet{2023MNRAS.524..235D} & $(74.92 \pm  1.09)^{*}$ & $74.82 \pm 1.28$ \\
       SNIa (SNooPy pre v2.7, F25 sample) & \citet{2025ApJ...985..203F} & $70.31\pm1.80$ & $70.39 \pm 1.80$ \\
       SNII standard candle (SC), our calibration & \citet{2022MNRAS.514.4620D} & $75.4^{+3.8}_{-3.7}$ & $73.14 \pm 4.72$ \\
       SNII standard candle (SC), literature calibration & \citet{2022MNRAS.514.4620D} & $75.4^{+3.8}_{-3.7}$ & $75.02 \pm 4.29$ \\
       Megamasers (no N4258, 2M++ velocities) & \citet{2020ApJ...891L...1P} & $72.1 \pm 2.7$ & $71.77 \pm  3.01$ \\
       SBF (corrected) & \citet{2025ApJ...987...87J} & $74.17 \pm 2.40$ & $74.33 \pm  2.22$ \\
       SNII EPM & \citet{2025AA...702A..41V} & $74.9 \pm 1.9$ & $74.83\pm1.85$ \\
       DESI Fundamental plane (Coma via SBF-TRGB) & \citet{2025MNRAS.539.3627S} & $76.05 \pm  4.90$ & $76.05 \pm 5.05$ \\ 
       DESI Fundamental plane (Coma via SN-Cepheid)& \citet{2025ApJ...979L...9S} & $76.5 \pm 2.2$ & $76.45 \pm  2.13$ \\ 
    \end{tabular}
    \caption{Comparison of literature values with non-SNIa determinations of the Hubble constant from the distance network.\\{\footnotesize${}^*$ In this case only 41 out of the 42 SNe~Ia were fit in the reference.}}
    \label{tab:literature_other}
\end{table}

%% file: Appendices/AD_sensitivity_to_anchors.tex
\section{Sensitivity of Baseline to Distance to NGC 4258}

\begin{figure}
    \centering
    \includegraphics[width=0.95\linewidth]{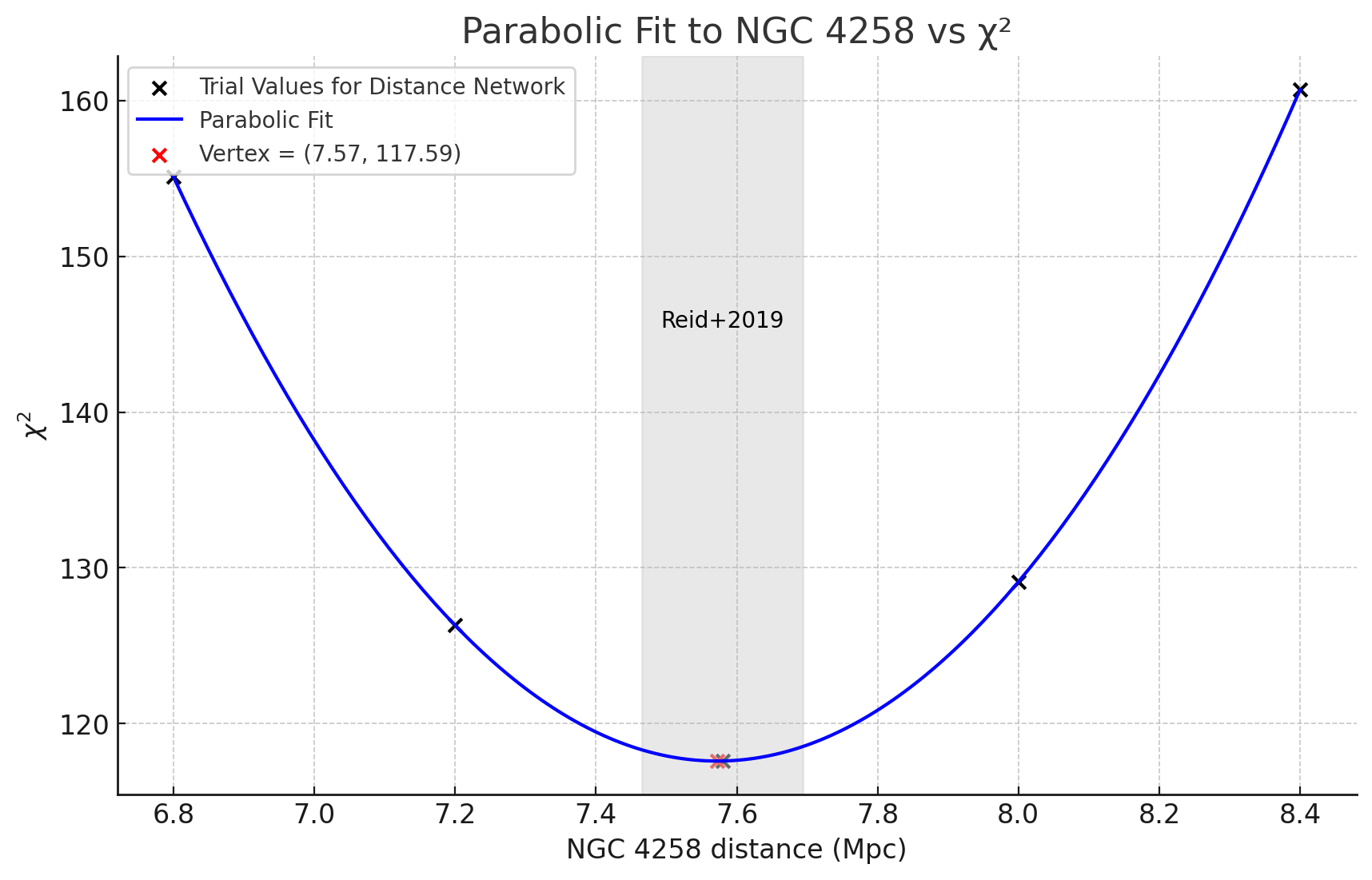}
    \caption{Trial values for the distance to NGC 4258 and associated $\chi^2$ values.  Besides the \citet{2019ApJ...886L..27R} measure at $D=7.58 \pm 0.11$ Mpc (shown as grey band) we also tried increasing or decreasing the distance by $\pm$ 5\% and 10\% (black crosses). The result its a large internal inconsistency in the distance network (due to other geometric anchors) which is highly significant. See also Table~\ref{tab:N4258_sens} for impact to $H_0$. The global minimum (vertex) is marked by a red cross.}
    \label{fig:N4258_trials.png}
\end{figure}

Given the prominent role of NGC 4258 in the distance network, it may be of interest to consider the degree to which the baseline results depend on the accuracy of the distance to NGC 4258.  Such a change is easily evaluated using the distance network by shifting distances with direct trace to NGC 4258 calibration.
For the baseline, the distance to NGC 4258 of $D=7.58 \pm 0.11$ Mpc comes from \citet{2019ApJ...886L..27R}.  In Table~\ref{tab:N4258_sens} we show extreme cases of raising or lowering the distance by $\pm 5\%$ and $\pm 10\%$   (many times the stated uncertainty while retaining the uncertainty).  These would lower/raise $H_0$ by 1.3\% and 2.6\%, respectively, or approximately 25\% of the change in the distance to NGC 4258. The relative size of the change in {\Hcst} compared to that of the anchor distance is mitigated by the availability of the other anchors from Gaia parallaxes and DEBs in the Magellanic Clouds (i.e., it would be a 33\% change in $H_0$ of the change in distance for NGC~4258 if these 3 anchors had equal weight). More importantly, such large shifts are strongly disfavored by the distance network goodness of fit as in either case the $\chi^2$ increases significantly, by 15 for the increased distance and by 38 for the decreased distance as shown in Figure \ref{fig:N4258_trials.png}. This is consistent with the finding of variant V12, which removes NGC 4258 from the baseline resulting in $H_0$=73.1 $\pm 0.92$ {\Hunit}, representing only a small decrease of 0.4 {\Hunit} and an increase in the error by only 15\%. This shows that the literature N4258 distance is compatible with the remainder of the {\DN} and strong deviations are incompatible, and the presence of the N4258 anchor is not crucial for the {\DN}.

We conclude that the Hubble tension is relatively insensitive to the stated accuracy of NGC 4258 alone.

\begin{table}[h!]
\centering
\caption{Sensitivity of baseline to large changes in NGC 4258 distance}
\begin{tabular}{lcccc}
\hline
\textbf{Version} & \textbf{NGC 4258 Distance (Mpc)} & \textbf{$H_0$ (km s$^{-1}$ Mpc$^{-1}$)} & \textbf{Error} & \textbf{$\chi^2$} (Ndof) \\
\hline
baseline & 7.58 & 73.48 & 0.80 & 117.6 (119) \\
\hline
increased 10\% & 8.4 & 71.58 & 0.78 & 160.7 (119) \\
decreased 10\% & 6.8 & 75.43 & 0.83 & 155.1 (119) \\
increased 5\% & 8.0 & 72.53 & 0.79 & 129.1 (119) \\
decreased 5\% & 7.2 & 74.45 & 0.82 & 126.3 (119) \\
\hline
Exclude N4258 & --- & 73.08 & 0.92 & 41.7 (47)  \\
\hline
\end{tabular}
 \label{tab:N4258_sens}
\end{table}

%% file: Appendices/AE_reproducibility.tex
\section{Code availability and reproducing our results}

The analysis code that underlies the results presented here is available in the repository \url{https://github.com/StefCas789/H0DN/}.\footnote{This repository will be made public when the paper is formally published.} The original version of the code is in IDL; a python version preserving all the main functionality is also available.
All the data files needed to run the code as well as corresponding instructions are also available in the repository.

The code is driven by configuration files that determine which data sets are included and other processing parameters. Configuration files corresponding to all the variants presented in Table~\ref{tab:variants} and many of the other tests are provided. Summary results for each run are provided in the form of text files; in addition, the user has the option of saving the main data, equation coefficients, and covariance matrices in FITS files for more detailed perusal.\footnote{The IDL version can also be run using command-line parameters, and has options to provide more detailed output if desired.} It is our intent to update the code and data files with new options and published measurements as they become available.

%% file: Appendices/AF_distance_table.tex
\section{Additional details from the Distance Network: calibration parameters and host/calibrator distances}\label{app:sec:distance_table}

The main result of the {\DN} analysis is of course the value and uncertainty of {\Hcst}, reported in Table~\ref{tab:variants} for all variants, together with the value of the total $ \chi^2 $ to evaluate the quality of the fit.  In addition, the {\DN} process can provide more detailed information on each class of calibrators and on host and the calibrated distances.

\subsection{Calibration parameters}

The {\DN} approach can include several secondary distance indicators whose calibration is determined simultaneously with all other parameters of the solution; each can be used to constrain {\Hcst} via tracers in the Hubble Flow.  For each variant, Table~\ref{tab:extrapars} provides the resulting calibration parameter, the Hubble Flow intercept, and the number of calibrators for each of the main secondary distance indicators: SNe~Ia, SBF, SNe~II, and TF. For SNe~Ia and SN~II, the calibration parameter is generally the reference absolute magnitude of a standardized SN in the appropriate passband; the exception is the BayesSN method, which uses a different definition.  For SBF and TF, the calibration parameter is reported as the \textit{difference} with respect to the published value, as described in Section~\ref{sec:data_appendix}. In all cases, the reported uncertainty is marginalized over all other parameters in the {\DN}.

The Hubble Flow intercept is the parameter $ a_{\itype} $ defined in Equation~\ref{eq:hubble_flow} and discussed in Section~\ref{app:sec:eq_hf}.  It is expressed in units of $ \log(\Hcst) $ (numerically equivalent to 0.2 mag); the reported uncertainty includes the intrinsic and measurement error for the tracers, but does not include the calibration uncertainty.  For some versions of SNe~Ia, the reported uncertainty includes covariance terms between different tracers, as reported in the original sources.

\input{Tables/extrapars.tex}

\subsection{Host and calibrator distances}

One of the secondary results of the {\DN} approach is an optimized set of distance moduli for all hosts and calibrators included in the Network.  These distances are based not only on the multiple measurements included in the {\DN}, but also on the consistency equations for the various classes of calibrators to which they belong.  They represent the values of the distances that, together, produce the best-fitting solution to all of the {\DN} equations.  As a result, the estimated distances for each host will differ slightly depending on which classes of calibrators are included in the solution, even when no additional measurements are introduced.

Table~\ref{tab:host_distances} reports the best distance estimate for each of the hosts (or host/calibrators) included in two solutions, Baseline (V00) and ``Everything,'' the solution including all methods (V99).  Since hosts not tied to any calibrators are excluded from each solution, many of the hosts do not have a distance estimate in the Baseline version.  SN calibrators are not included, since their distances are not separate parameters; they are assumed identical to the distances to their host, which as a consequence are also affected by the SN calibration via the (one or more) SNe they host.  On the other hand, SBF calibrator galaxies in Fornax and Virgo \textit{are} included individually; their distances are allowed to deviate from the group distance, because of depth effects, and are affected by their SBF luminosity measurements via their global calibration.  The quoted distance uncertainties are marginalized over all parameters of the respective {\DN} solution, and therefore are \textit{not} independent.

\startlongtable
\begin{deluxetable}{lcccc}
\tabletypesize{\scriptsize}
\tablecaption{Distance moduli and statistical uncertainties for host and calibrator systems in the {\DN}.\label{tab:host_distances}}
\tablehead{Name & \multicolumn{2}{c}{Baseline solution} & \multicolumn{2}{c}{``Everything'' solution} \\
           & Value & Error & Value & Error}
\startdata
\input{Tables/distance_table.tex}
\enddata
\end{deluxetable}

%% file: Tables/extrapars.tex
\begin{longrotatetable}
\movetabledown=2cm
\begin{deluxetable}{lccccccccrrrr}
\tabletypesize{\tiny}
\tablecaption {Additional fitted parameters for all variants\label{tab:extrapars}}
\tablehead{
\colhead {Variant}  & \multicolumn{4}{c}{Calibration parameter (mag)\tablenotemark{\textup{(}a\textup{)}}}                      &  \multicolumn{4}{c}{Hubble flow intercept\tablenotemark{\textup{(}b\textup{)}}}                    & \multicolumn{4}{c}{Number of calibrators} \\
\colhead{} &  
\multicolumn{1}{c}{SN~Ia}&\multicolumn{1}{c}{SBF}&\multicolumn{1}{c}{SN~II}&\multicolumn{1}{c}{TF}& 
\multicolumn{1}{c}{SN~Ia}&\multicolumn{1}{c}{SBF}&\multicolumn{1}{c}{SN~II}&\multicolumn{1}{c}{TF}& 
\multicolumn{1}{c}{SN~Ia}&\multicolumn{1}{c}{SBF}&\multicolumn{1}{c}{SN~II}&\multicolumn{1}{c}{TF}
}
\startdata
 V00 & $ -19.252 \pm   0.022 $  & $  -0.023 \pm   0.027 $  &          ---             &          ---             & $   0.716 \pm   0.002 $  & $  -3.128 \pm   0.003 $  &          ---             &          ---             &   55 &   14 &    0 &    0 \\
 V01 & $ -19.252 \pm   0.022 $  & $  -0.024 \pm   0.027 $  &          ---             &          ---             & $   0.716 \pm   0.002 $  & $  -3.128 \pm   0.003 $  &          ---             &          ---             &   55 &   14 &    0 &    0 \\
 V02 & $ -19.252 \pm   0.022 $  & $  -0.023 \pm   0.027 $  &          ---             &          ---             & $   0.716 \pm   0.002 $  & $  -3.128 \pm   0.003 $  &          ---             &          ---             &   55 &   14 &    0 &    0 \\
 V03 & $ -19.252 \pm   0.022 $  & $  -0.021 \pm   0.027 $  &          ---             &          ---             & $   0.716 \pm   0.002 $  & $  -3.128 \pm   0.003 $  &          ---             &          ---             &   55 &   14 &    0 &    0 \\
 V04 & $ -19.251 \pm   0.022 $  & $  -0.023 \pm   0.027 $  & $ -16.760 \pm   0.044 $  &          ---             & $   0.716 \pm   0.002 $  & $  -3.128 \pm   0.003 $  & $   0.219 \pm   0.008 $  &          ---             &   55 &   14 &    9 &    0 \\
 V05 & $ -19.246 \pm   0.021 $  & $  -0.017 \pm   0.025 $  &          ---             &          ---             & $   0.716 \pm   0.002 $  & $  -3.128 \pm   0.003 $  &          ---             &          ---             &   55 &   14 &    0 &    0 \\
 V06 & $ -19.241 \pm   0.022 $  & $  -0.012 \pm   0.026 $  &          ---             & $  -0.076 \pm   0.024 $  & $   0.716 \pm   0.002 $  & $  -3.128 \pm   0.003 $  &          ---             & $  -3.116 \pm   0.001 $  &   55 &   14 &    0 &   74 \\
 V07 & $ -19.257 \pm   0.022 $  & $  -0.029 \pm   0.027 $  &          ---             &          ---             & $   0.716 \pm   0.002 $  & $  -3.128 \pm   0.003 $  &          ---             &          ---             &   55 &   14 &    0 &    0 \\
 V08 & $ -19.282 \pm   0.038 $  & $  -0.051 \pm   0.040 $  &          ---             &          ---             & $   0.716 \pm   0.002 $  & $  -3.128 \pm   0.003 $  &          ---             &          ---             &   35 &   14 &    0 &    0 \\
V08B & $ -19.271 \pm   0.025 $  & $  -0.041 \pm   0.029 $  &          ---             &          ---             & $   0.716 \pm   0.002 $  & $  -3.128 \pm   0.003 $  &          ---             &          ---             &   35 &   14 &    0 &    0 \\
 V09 & $ -19.249 \pm   0.024 $  &          ---             &          ---             &          ---             & $   0.716 \pm   0.002 $  &          ---             &          ---             &          ---             &   42 &    0 &    0 &    0 \\
V09B & $ -19.243 \pm   0.023 $  & $  -0.020 \pm   0.028 $  &          ---             &          ---             & $   0.716 \pm   0.002 $  & $  -3.128 \pm   0.003 $  &          ---             &          ---             &   42 &   14 &    0 &    0 \\
 V10 & $ -19.249 \pm   0.026 $  & $  -0.020 \pm   0.030 $  &          ---             &          ---             & $   0.716 \pm   0.002 $  & $  -3.128 \pm   0.003 $  &          ---             &          ---             &   55 &   14 &    0 &    0 \\
 V11 & $ -19.257 \pm   0.024 $  & $  -0.028 \pm   0.028 $  &          ---             &          ---             & $   0.716 \pm   0.002 $  & $  -3.128 \pm   0.003 $  &          ---             &          ---             &   55 &   14 &    0 &    0 \\
 V12 & $ -19.262 \pm   0.026 $  &          ---             &          ---             &          ---             & $   0.716 \pm   0.002 $  &          ---             &          ---             &          ---             &   42 &    0 &    0 &    0 \\
V12B & $ -19.246 \pm   0.024 $  &          ---             &          ---             &          ---             & $   0.716 \pm   0.002 $  &          ---             &          ---             &          ---             &   42 &    0 &    0 &    0 \\
 V13 &          ---             & $  -0.029 \pm   0.052 $  &          ---             &          ---             &          ---             & $  -3.128 \pm   0.003 $  &          ---             &          ---             &    0 &   14 &    0 &    0 \\
 V14 & $ -19.255 \pm   0.023 $  &          ---             &          ---             &          ---             & $   0.716 \pm   0.002 $  &          ---             &          ---             &          ---             &   55 &    0 &    0 &    0 \\
 V15 & $ -19.248 \pm   0.023 $  & $  -0.020 \pm   0.028 $  &          ---             &          ---             & $   0.716 \pm   0.002 $  & $  -3.128 \pm   0.003 $  &          ---             &          ---             &   55 &   14 &    0 &    0 \\
 V16 & $ -19.244 \pm   0.038 $  & $  -0.017 \pm   0.040 $  &          ---             &          ---             & $   0.716 \pm   0.002 $  & $  -3.128 \pm   0.003 $  &          ---             &          ---             &   24 &   14 &    0 &    0 \\
V16B & $ -19.236 \pm   0.026 $  & $  -0.009 \pm   0.030 $  &          ---             &          ---             & $   0.716 \pm   0.002 $  & $  -3.128 \pm   0.003 $  &          ---             &          ---             &   24 &   14 &    0 &    0 \\
 V17 & $ -19.267 \pm   0.024 $  & $  -0.039 \pm   0.029 $  &          ---             &          ---             & $   0.716 \pm   0.002 $  & $  -3.128 \pm   0.003 $  &          ---             &          ---             &   55 &   13 &    0 &    0 \\
 V18 & $ -19.246 \pm   0.023 $  & $  -0.019 \pm   0.027 $  &          ---             &          ---             & $   0.716 \pm   0.002 $  & $  -3.128 \pm   0.003 $  &          ---             &          ---             &   48 &   14 &    0 &    0 \\
 V19 & $ -19.249 \pm   0.022 $  & $  -0.027 \pm   0.027 $  &          ---             &          ---             & $   0.713 \pm   0.002 $  & $  -3.130 \pm   0.003 $  &          ---             &          ---             &   55 &   14 &    0 &    0 \\
 V20 & $ -19.255 \pm   0.023 $  &          ---             &          ---             &          ---             & $   0.716 \pm   0.002 $  &          ---             &          ---             &          ---             &   55 &    0 &    0 &    0 \\
 V21 & $ -19.254 \pm   0.023 $  &          ---             &          ---             &          ---             & $   0.712 \pm   0.003 $  &          ---             &          ---             &          ---             &   55 &    0 &    0 &    0 \\
 V22 & $ -19.122 \pm   0.027 $  & $  -0.030 \pm   0.031 $  &          ---             &          ---             & $   0.689 \pm   0.003 $  & $  -3.128 \pm   0.003 $  &          ---             &          ---             &   55 &   14 &    0 &    0 \\
 V23\tablenotemark{\textup{(}c\textup{)}} &
       $   0.014 \pm   0.029 $ & $  -0.017 \pm   0.033 $  &          ---             &          ---             & $  -3.135 \pm   0.004 $  & $  -3.128 \pm   0.003 $  &          ---             &          ---             &   50 &   14 &    0 &    0 \\
 V24 & $ -19.293 \pm   0.022 $  & $  -0.024 \pm   0.027 $  &          ---             &          ---             & $   0.724 \pm   0.002 $  & $  -3.128 \pm   0.003 $  &          ---             &          ---             &   55 &   14 &    0 &    0 \\
 V25 & $ -18.354 \pm   0.030 $  & $  -0.055 \pm   0.034 $  &          ---             &          ---             & $   0.529 \pm   0.004 $  & $  -3.128 \pm   0.003 $  &          ---             &          ---             &   14 &   14 &    0 &    0 \\
 V26 & $ -18.567 \pm   0.033 $  & $  -0.048 \pm   0.036 $  &          ---             &          ---             & $   0.574 \pm   0.004 $  & $  -3.128 \pm   0.003 $  &          ---             &          ---             &   17 &   14 &    0 &    0 \\
 V27 & $ -19.251 \pm   0.022 $  & $  -0.026 \pm   0.027 $  &          ---             &          ---             & $   0.716 \pm   0.002 $  & $  -3.128 \pm   0.003 $  &          ---             &          ---             &   55 &   14 &    0 &    0 \\
 V28 & $ -19.264 \pm   0.022 $  & $  -0.036 \pm   0.027 $  &          ---             &          ---             & $   0.716 \pm   0.002 $  & $  -3.128 \pm   0.003 $  &          ---             &          ---             &   55 &   14 &    0 &    0 \\
 V99 & $ -19.243 \pm   0.020 $  & $  -0.011 \pm   0.024 $  & $ -16.749 \pm   0.042 $  & $  -0.075 \pm   0.021 $  & $   0.716 \pm   0.002 $  & $  -3.128 \pm   0.003 $  & $   0.219 \pm   0.008 $  & $  -3.116 \pm   0.001 $  &   55 &   14 &    9 &   74 \\
V99A & $ -19.251 \pm   0.020 $  & $  -0.019 \pm   0.025 $  & $ -16.757 \pm   0.042 $  &          ---             & $   0.716 \pm   0.002 $  & $  -3.128 \pm   0.003 $  & $   0.219 \pm   0.008 $  &          ---             &   55 &   14 &    9 &    0 \\
\enddata
\tablenotetext{\textup{(}a\textup{)}}{\quad Absolute magnitude for SNe, offset to the nominal calibration for SBF and TF. Uncertainty is marginalized over all other parameters.}
\tablenotetext{\textup{(}b\textup{)}}{\quad In units of $\log_{10}(\Hcst)$. Uncertainty includes only Hubble Flow terms, and does not include the calibration uncertainty.}
\tablenotetext{\textup{(}c\textup{)}}{\quad Calibration and intercept for SNe~Ia measured with the BayesSN method have a different, internally consistent definition.}
\end{deluxetable}
\end{longrotatetable}

%% file: Tables/distance_table.tex
       IC 2006     &   31.469 &    0.048     &   31.457 &    0.046 \\
          M 33     &    ---   &    ---       &   24.847 &    0.075 \\
          M 64     &    ---   &    ---       &   28.219 &    0.042 \\
          M 66     &   30.200 &    0.045     &   30.202 &    0.044 \\
          M 81     &    ---   &    ---       &   27.901 &    0.081 \\
          M 96     &   30.287 &    0.045     &   30.289 &    0.044 \\
         M 101     &   29.120 &    0.026     &   29.125 &    0.023 \\
         M 104     &    ---   &    ---       &   29.851 &    0.041 \\
      Mrk 1337     &   32.930 &    0.082     &   32.922 &    0.082 \\
        NGC 24     &    ---   &    ---       &   29.315 &    0.036 \\
       NGC 105     &   34.507 &    0.119     &   34.501 &    0.119 \\
       NGC 247     &    ---   &    ---       &   27.854 &    0.036 \\
       NGC 253     &    ---   &    ---       &   27.844 &    0.036 \\
       NGC 300     &    ---   &    ---       &   26.592 &    0.058 \\
       NGC 628     &    ---   &    ---       &   29.897 &    0.122 \\
       NGC 672     &    ---   &    ---       &   29.288 &    0.084 \\
       NGC 691     &   32.836 &    0.089     &   32.821 &    0.085 \\
       NGC 891     &    ---   &    ---       &   30.005 &    0.074 \\
       NGC 976     &   33.566 &    0.087     &   33.553 &    0.085 \\
      NGC 1015     &   32.600 &    0.062     &   32.593 &    0.062 \\
      NGC 1309     &   32.498 &    0.044     &   32.498 &    0.043 \\
      NGC 1316     &   31.385 &    0.042     &   31.386 &    0.041 \\
      NGC 1344     &   31.439 &    0.049     &   31.427 &    0.047 \\
      NGC 1365     &   31.335 &    0.035     &   31.332 &    0.029 \\
      NGC 1374     &   31.479 &    0.049     &   31.466 &    0.048 \\
      NGC 1375     &   31.457 &    0.052     &   31.445 &    0.051 \\
      NGC 1380     &   31.417 &    0.037     &   31.406 &    0.036 \\
      NGC 1399     &   31.504 &    0.039     &   31.493 &    0.037 \\
      NGC 1404     &   31.407 &    0.037     &   31.396 &    0.035 \\
      NGC 1448     &   31.308 &    0.028     &   31.313 &    0.025 \\
      NGC 1559     &   31.423 &    0.030     &   31.422 &    0.026 \\
      NGC 1560     &    ---   &    ---       &   27.539 &    0.148 \\
      NGC 2188     &    ---   &    ---       &   29.618 &    0.050 \\
      NGC 2403     &    ---   &    ---       &   27.528 &    0.058 \\
      NGC 2442     &   31.499 &    0.049     &   31.530 &    0.036 \\
      NGC 2525     &   31.929 &    0.052     &   31.912 &    0.051 \\
      NGC 2608     &   32.648 &    0.115     &   32.560 &    0.111 \\
      NGC 2683     &    ---   &    ---       &   29.979 &    0.057 \\
      NGC 2903     &    ---   &    ---       &   29.817 &    0.049 \\
      NGC 2915     &    ---   &    ---       &   28.191 &    0.076 \\
      NGC 2976     &    ---   &    ---       &   27.822 &    0.050 \\
      NGC 3021     &   32.242 &    0.049     &   32.237 &    0.048 \\
      NGC 3109     &    ---   &    ---       &   25.712 &    0.094 \\
      NGC 3147     &   33.094 &    0.084     &   33.086 &    0.083 \\
      NGC 3254     &   32.396 &    0.057     &   32.410 &    0.055 \\
      NGC 3368     &    ---   &    ---       &   30.245 &    0.036 \\
      NGC 3370     &   32.189 &    0.036     &   32.193 &    0.035 \\
      NGC 3432     &    ---   &    ---       &   29.929 &    0.161 \\
      NGC 3447     &   31.956 &    0.029     &   31.951 &    0.027 \\
      NGC 3583     &   32.783 &    0.064     &   32.782 &    0.063 \\
      NGC 3621     &    ---   &    ---       &   29.224 &    0.042 \\
      NGC 3627     &    ---   &    ---       &   30.259 &    0.042 \\
      NGC 3972     &   31.718 &    0.051     &   31.716 &    0.050 \\
      NGC 3982     &   31.600 &    0.053     &   31.599 &    0.052 \\
      NGC 4038     &   31.672 &    0.030     &   31.673 &    0.029 \\
      NGC 4144     &    ---   &    ---       &   29.194 &    0.042 \\
      NGC 4151     &    ---   &    ---       &   31.036 &    0.070 \\
      NGC 4236     &    ---   &    ---       &   28.298 &    0.050 \\
      NGC 4242     &    ---   &    ---       &   29.403 &    0.076 \\
      NGC 4244     &    ---   &    ---       &   28.184 &    0.075 \\
      NGC 4298     &    ---   &    ---       &   30.842 &    0.167 \\
      NGC 4414     &   31.264 &    0.081     &   31.249 &    0.078 \\
      NGC 4424     &   30.910 &    0.034     &   30.909 &    0.033 \\
      NGC 4455     &    ---   &    ---       &   29.120 &    0.076 \\
      NGC 4457     &   31.076 &    0.086     &   31.074 &    0.086 \\
      NGC 4458     &   31.097 &    0.056     &   31.085 &    0.055 \\
      NGC 4472     &   31.088 &    0.039     &   31.078 &    0.037 \\
      NGC 4489     &   31.002 &    0.056     &   30.990 &    0.055 \\
      NGC 4517     &    ---   &    ---       &   29.663 &    0.135 \\
      NGC 4526     &   30.936 &    0.066     &   30.937 &    0.066 \\
      NGC 4536     &   30.892 &    0.034     &   30.906 &    0.028 \\
      NGC 4552     &   30.978 &    0.041     &   30.968 &    0.040 \\
      NGC 4559     &    ---   &    ---       &   29.733 &    0.042 \\
      NGC 4565     &    ---   &    ---       &   30.415 &    0.036 \\
      NGC 4592     &    ---   &    ---       &   29.749 &    0.050 \\
      NGC 4605     &    ---   &    ---       &   28.723 &    0.036 \\
      NGC 4631     &    ---   &    ---       &   29.333 &    0.036 \\
      NGC 4636     &   31.111 &    0.039     &   31.100 &    0.038 \\
      NGC 4639     &   31.772 &    0.048     &   31.728 &    0.035 \\
      NGC 4649     &   31.071 &    0.040     &   31.060 &    0.038 \\
      NGC 4666     &   30.903 &    0.057     &   30.903 &    0.057 \\
      NGC 4680     &   32.510 &    0.146     &   32.508 &    0.146 \\
      NGC 4697     &   30.347 &    0.041     &   30.337 &    0.039 \\
      NGC 4945     &    ---   &    ---       &   27.765 &    0.058 \\
      NGC 5023     &    ---   &    ---       &   28.926 &    0.058 \\
      NGC 5055     &    ---   &    ---       &   29.773 &    0.036 \\
      NGC 5194     &    ---   &    ---       &   29.619 &    0.059 \\
      NGC 5204     &    ---   &    ---       &   28.443 &    0.216 \\
      NGC 5398     &    ---   &    ---       &   30.320 &    0.185 \\
      NGC 5468     &   33.050 &    0.031     &   33.050 &    0.030 \\
      NGC 5584     &   31.853 &    0.030     &   31.861 &    0.026 \\
      NGC 5643     &   30.535 &    0.027     &   30.533 &    0.023 \\
      NGC 5728     &   32.842 &    0.102     &   32.819 &    0.094 \\
      NGC 5861     &   32.171 &    0.064     &   32.176 &    0.062 \\
      NGC 5907     &    ---   &    ---       &   31.179 &    0.057 \\
      NGC 5917     &   32.343 &    0.077     &   32.339 &    0.077 \\
      NGC 6503     &    ---   &    ---       &   28.997 &    0.066 \\
      NGC 6744     &    ---   &    ---       &   29.861 &    0.066 \\
      NGC 6946     &    ---   &    ---       &   29.206 &    0.089 \\
      NGC 7090     &    ---   &    ---       &   29.887 &    0.058 \\
      NGC 7250     &   31.603 &    0.048     &   31.570 &    0.036 \\
      NGC 7329     &   33.269 &    0.069     &   33.264 &    0.068 \\
      NGC 7541     &   32.579 &    0.089     &   32.577 &    0.089 \\
      NGC 7640     &    ---   &    ---       &   29.652 &    0.066 \\
      NGC 7678     &   33.300 &    0.082     &   33.295 &    0.082 \\
      NGC 7793     &    ---   &    ---       &   27.788 &    0.053 \\
      NGC 7814     &   30.926 &    0.071     &   30.948 &    0.068 \\
      PGC 6574     &    ---   &    ---       &   29.360 &    0.076 \\
      PGC 9962     &    ---   &    ---       &   28.535 &    0.036 \\
     PGC 11139     &    ---   &    ---       &   28.844 &    0.036 \\
     PGC 13163     &    ---   &    ---       &   28.966 &    0.042 \\
     PGC 16957     &    ---   &    ---       &   27.891 &    0.199 \\
     PGC 45084     &    ---   &    ---       &   28.937 &    0.067 \\
     PGC 47847     &    ---   &    ---       &   29.211 &    0.050 \\
     PGC 54392     &    ---   &    ---       &   27.283 &    0.103 \\
     PGC 64181     &    ---   &    ---       &   29.838 &    0.122 \\
     PGC 65603     &    ---   &    ---       &   28.699 &    0.058 \\
      UGC 1281     &    ---   &    ---       &   28.615 &    0.036 \\
      UGC 9391     &   32.818 &    0.056     &   32.817 &    0.056 \\
      UGCA 319     &    ---   &    ---       &   28.853 &    0.050 \\